\documentclass[10pt,aps,pra,twocolumn,showpacs,superscriptaddress]{revtex4-2}
\usepackage[english]{babel} 	
\usepackage[usenames, dvipsnames]{color}
\usepackage{graphicx} 
\usepackage{bm} 
\usepackage{amsmath}
\usepackage{amsthm} 
\usepackage{subfigure}
\usepackage{amssymb} 
\usepackage{times}
\usepackage[pdftex,colorlinks=true,urlcolor= blue,linkcolor=Blue,citecolor=RedViolet]{hyperref}
\usepackage{multirow}
\usepackage{amsmath,amsthm,amssymb,physics,color,enumerate,tabularx,makecell,mathrsfs,float,bbm,stackrel,mathtools,multirow,bm}
\usepackage{natbib}
\usepackage{amsmath}
\usepackage{physics}
\usepackage{xcolor}
\usepackage[small,bf,justification=Justified]{caption}
\usepackage{soul}
\begin{document}

\title{Experimental simulation of daemonic work extraction in open quantum batteries on a digital quantum computer}
\author{Seyed Navid Elyasi}
\affiliation{Department of Physics, University of Kurdistan, Kurdistan, Iran}
\author{Matteo A. C. Rossi}
%\email{matteo.rossi@helsinki.fi}%
%\affiliation{InstituteQ - the Finnish Quantum Institute, Aalto University, Finland}
%\affiliation{QTF Centre of Excellence, Department of Applied Physics, Aalto University, FI-00076 Aalto, Finland}
\affiliation{QTF Centre of Excellence, Department of Physics,
University of Helsinki, FI-00014 Helsinki, Finland.}
\affiliation{Algorithmiq Ltd., Kanavakatu 3C, FI-00160 Helsinki, Finland}
\author{Marco G. Genoni}
\email{marco.genoni@unimi.it}
\affiliation{Dipartimento di Fisica ``{\it Aldo Pontremoli}'', Universit\`{a} degli Studi di Milano, I-20133 Milano, Italy}

\date{\today}

%%%%%%%%%%%%%%%%%%%%%%%%%%%%%%%%%%%%%%%%%%%%%%
%%%%%%%%%%%%%%%%%%%%%%%%%%%%%%%%%%%%%%%%%%%%%%
%%%%%%%%%%%%%%%%%%%%%%%%%%%%%%%%%%%%%%%%%%%%%%
%%%%%%%%%%%%%%%%%%%%%%%%%%%%%%%%%%%%%%%%%%%%%%
%%%%%%%%%%%%%%%%%%%%%%%%%%%%%%%%%%%%%%%%%%%%%%
\begin{abstract}
The possibility of extracting more work from a physical system thanks to the information obtained from measurements has been a topic of fundamental interest in the context of thermodynamics since the formulation of the Maxwell's demon thought experiment. We here consider this problem from the perspective of an open quantum battery interacting with an environment that can be continuously measured. By modeling it via a continuously monitored collisional model, we show how to implement the corresponding dynamics as a quantum circuit, including the final conditional feedback unitary evolution that allows to enhance the amount of work extracted. By exploiting the flexibility of IBM quantum computers and by properly modelling the corresponding quantum circuit, we experimentally simulate the work extraction protocol showing how the obtained experimental values of the daemonic extracted work are close to their theoretical upper bound quantified by the so-called daemonic ergotropy. We also demonstrate how by properly modelling the noise affecting the quantum circuit, one can improve the work extraction protocol by optimizing the corresponding extraction unitary feedback operation.
\end{abstract}

%%%%%%%%%%%%%%%%%%%%%%%%%%%%%%%%%%%%%%%%%%%%%%
%%%%%%%%%%%%%%%%%%%%%%%%%%%%%%%%%%%%%%%%%%%%%%
%%%%%%%%%%%%%%%%%%%%%%%%%%%%%%%%%%%%%%%%%%%%%%
%%%%%%%%%%%%%%%%%%%%%%%%%%%%%%%%%%%%%%%%%%%%%%
%%%%%%%%%%%%%%%%%%%%%%%%%%%%%%%%%%%%%%%%%%%%%%

\maketitle
\section{Introduction}
\label{sec:Introduction}
The role of measurement in thermodynamics has been the subject of both fundamental and practical research since the introduction of the Maxwell daemon's thought experiment. This topic is even more appealing in the context of quantum thermodynamics, whose aim is to understand and possibly extend the laws of classical thermodynamics at the quantum scale~\cite{binderThermodynamicsQuantumRegime2018}. In fact, the act of measuring in quantum mechanics also has a disruptive effect on the quantum state of the system, and for this reason, its role has been studied in great detail both in the context of work extraction quantum protocols inspired by the Maxwell daemon ~\cite{francicaDaemonicErgotropyEnhanced2017,manzanoOptimalWorkExtraction2018,MorronePRApp,stahl2024demonstrationenergyextractiongain} and of quantum measurement engines~\cite{Campisi_2017,ElouardPRL2017,YiPRE2017,Mohammady_2017,BuffoniPRL2019,StevensPRL2022,yanikThermodynamicsQuantumMeasurement2022,JussiauPRR2023,LinpengPRR2024}.  

Here we will focus on work extraction protocols from quantum systems~\cite{allahverdyanMaximalWorkExtraction2004}, a topic that is strictly related to the study on quantum batteries. Quantum batteries (QBs) are defined as energy storage devices, whose charging and discharging protocols are described according to the laws of quantum mechanics~\cite{campaioliQuantumBatteriesReview2018,CampaioliRMP2024}. It has been shown how possible advantages in the charging protocols can be achieved by exploiting collective or purely quantum phenomena~\cite{alickiEntanglementBoostExtractable2013,binderQuantacellPowerfulCharging2015,campaioliEnhancingChargingPower2017,ferraroHighPowerCollectiveCharging2018,andolina_extractable_2019,ZhangPowerfulHarmonicCharging2019,CrescenteChargingEnergyFluctuations2020,JuliaFarreBounds2020,rossiniQuantumAdvantageCharging2020,gyhmQuantumChargingAdvantage2022,seahQuantumSpeedupCollisional2021,salviaQuantumAdvantageCharging2022,landiBatteryChargingCollision2021,mayoCollectiveEffectsQuantum2022,barraEfficiencyFluctuationsQuantum2022,rodriguezOptimalQuantumControl2022,QiMagnonMediatedQuantum2021,CrescenteEnhancingCoherentEnergy2022,mazzonciniOptimalControlMethods2023,rodriguezAIdiscoveryNewCharging2023,Gyhm2024,RazzoliQST2024}. All these studies led to the first proof-of-principle experimental implementations of quantum batteries via superconducting qubits \cite{hu2022optimal} and in an organic microcavity~\cite{quach2022superabsorption}. 

In most of the seminal publications, quantum batteries are considered as closed quantum systems and the interaction with the surrounding environment is ignored; on the other hand the role of noise and decoherence has been investigated, studying the efficiency of charging protocols for open quantum batteries both in the Markovian~\cite{farinaChargermediatedEnergyTransfer2019} and non-Markovian~\cite{morroneChargingQuantumBattery2023} regime. Efforts have then been made to employ control strategies \cite{santosStableAdiabaticQuantum2019,rodriguez2022optimal,gherardini2020stabilizing,mitchisonChargingQuantumBattery2021,yaoOptimalChargingOpen2022a,Ahmadi2024,Lu2024} to improve the charging protocol of quantum batteries both in the closed and open scenarios. In this context, the possibility of enhancing the amount of extractable work by continuously monitoring the environment has been studied in detail~\cite{MorronePRApp}. Continuously monitored quantum systems and quantum feedback control~\cite{wisemanQuantumMeasurementControl2009,AlbarelliPLA2024} are active and fascinating research areas that have been extensively studied, both for their fundamental significance and their practical applications in quantum state engineering.~\cite{wisemanQuantumTheoryOptical1993,dohertyFeedbackControlQuantum1999,thomsenSpinSqueezingQuantum2002,serafiniDeterminationMaximalGaussian2010,genoniOptimalFeedbackControl2013,genoniQuantumCoolingSqueezing2015,brunelliConditionalDynamicsOptomechanical2019,digiovanniUnconditionalMechanicalSqueezing2021,candeloroFeedbackAssistedQuantumSearch2023,isaksenMechanicalCoolingSqueezing2023}, quantum estimation~\cite{gutaOptimalEstimationQubit2008,
tsangOptimalWaveformEstimation2010,tsangQuantumMetrologyOpen2013,
gammelmarkBayesianParameterInference2013,gammelmarkFisherInformationQuantum2014,ralphMultiparameterEstimationQuantum2017,genoniCramErRaoBound2017,albarelliUltimateLimitsQuantum2017,rossiPRL2020,
AmorosBinefa2021,iliasCriticalityEnhancedQuantumSensing2022,yangEfficientInformationRetrieval2022,FallaniPRXQuantum2022,amorosbinefa2024}, and also in the context of quantum thermodynamics~\cite{elouardWorkHeatEntropy2018,manzanoQuantumThermodynamicsContinuous2022,garrahanThermodynamicsQuantumJump2010,garrahanQuantumTrajectoryPhase2011,cilluffoQuantumJumpStatistics2019,belenchiaEntropyProductionContinuously2020,rossiExperimentalAssessmentEntropy2020,landiIrreversibleEntropyProduction2021,LandiPRXQuantum2022,bhandariContinuousMeasurementBoosted2022,yanikThermodynamicsQuantumMeasurement2022,bhandariMeasurementBasedQuantumThermal2023}. Single quantum trajectories of continuously monitored quantum systems has been now experimentally shown in different platforms~\cite{Murch2013,Campagne-Ibarcq2016,Ficheux2017,Minev2019,Wieczorek2015,Rossi2018} and continuous feedback protocols able to cool mechanical oscillators towards their quantum ground state have also been demonstrated~\cite{rossi2018control,Magrini2021,Tebbenjohanns2021}. 

In this work we will simulate a discrete-time continuously monitored quantum system by exploiting the collisional models (CMs) formalism. CMs are a powerful and flexible tool to study open quantum systems~\cite{CiccarelloPra2013,CiccarelloPR2022,Campbell2021_CM,cattaneoBriefJourneyCollision2022a}. In these models both the environment and time are discretized: the environment is indeed represented by a discrete set of auxiliary quantum systems that interact with the system sequentially at discrete times. This, for example, allows one to easily describe non-Markovian evolutions, and they have been employed as a tool also to describe charging protocols of quantum batteries from different perspectives~\cite{seahQuantumSpeedupCollisional2021,salviaQuantumAdvantageCharging2022,landiBatteryChargingCollision2021,mayoCollectiveEffectsQuantum2022,barraEfficiencyFluctuationsQuantum2022}. 

More importantly for our purposes, one can show that via the CM formalism, one can easily derive master equations in Lindblad form describing Markovian open quantum systems~\cite{CiccarelloPR2022,AlbarelliPLA2024} and stochastic master equations describing continuously monitored quantum systems~\cite{AlbarelliPLA2024}. Continuously monitored collisional models (CMCM) have been for example introduced and employed to study entropy production in continuously monitored quantum systems~\cite{LandiPRXQuantum2022,BelenchiaPRA2022}. The main goal of our work is to experimentally simulate a daemonic work extraction in open quantum systems as discussed in~\cite{MorronePRApp}, by mapping it to an experimentally realizable CMCM.  

IBM's cloud quantum platforms serve as highly useful tools enabling the implementation of different quantum protocols. They comprise superconducting qubits based on Josephson junctions in a transmon configuration, offering longer coherence time and lower readout error compared to other counterparts~\cite{burnett2019decoherence}. In the context of QBs, in~\cite{gemmeIBMQuantumPlatforms2022a} the authors utilized IBM quantum computers within the pulse modulation technique using the Qiskit Pulse package to investigate the charging process in terms of time and energy storage. Furthermore, both CMs and CMCMs have been successfully implemented on IBM quantum platform via properly designed quantum circuits in Refs.~\cite{garcia-perezIBMExperienceVersatile2020,cech2023thermodynamics,cattaneoQuantumSimulationDissipative2023}. 

The main goal of our work is to exploit IBM quantum computers, and in particular the recently implemented features of mid-circuit measurements and dynamic circuits ~\cite{javadiabhari2024quantumcomputingqiskit}, in order to experimentally simulate a continuously monitored quantum battery via a CMCM, along with the conditional work extraction operations for each quantum trajectory. This will allow us to experimentally verify the corresponding daemonic enhancement obtainable from the protocol in a physically relevant CMCM. 
\begin{figure}[t]
\includegraphics[width=0.5\textwidth]{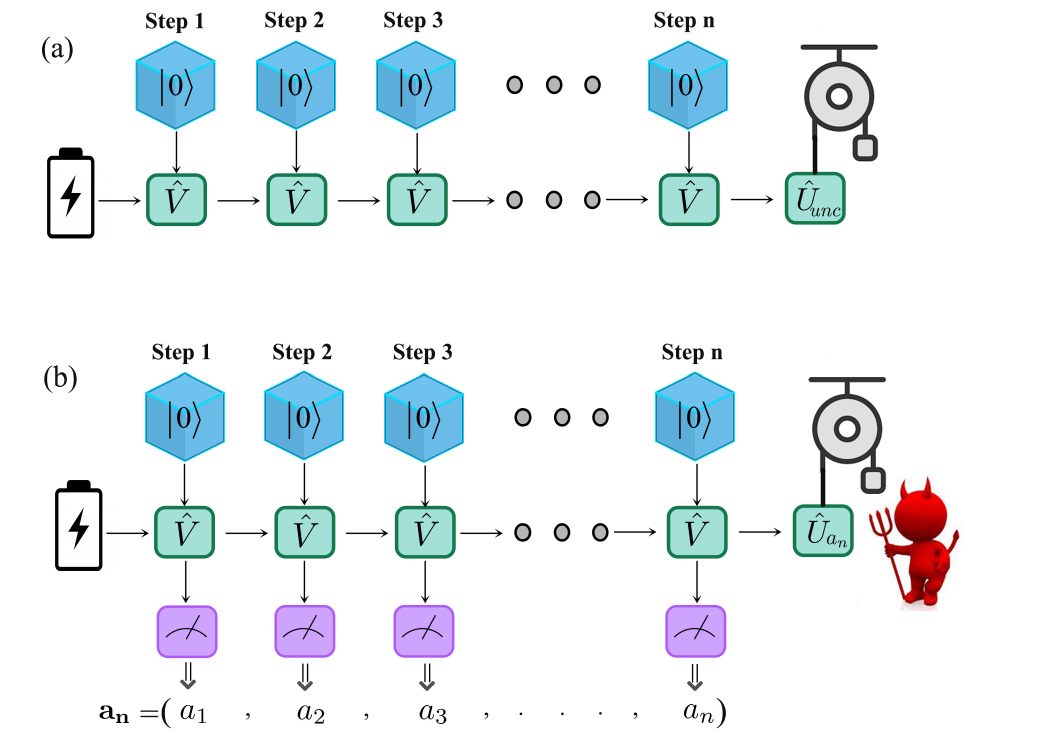}
\caption{
a) Pictorial description of a charging and work extraction protocol for an open QB via a CM: a two-level quantum battery interacts sequentially with auxiliary qubit systems; after $n$ steps a work extraction unitary is performed on the quantum state. b) Pictorial description of a daemonic charging and work extraction protocol for an open QB via a CMCM: in this case the auxiliary qubits are measured after the interaction. After $n$-steps, the measurement outcomes identify one of the $2^n$ quantum trajectories for the QB state and thus identify also the corresponding optimal work extraction optimal unitary.}
\label{f:QBmodel}
\end{figure}
The manuscript is structured as follows: in Sec. \ref{sec:QBs}, we revisit the definition of quantum batteries and the figures of merit we aim to investigate, including ergotropy and daemonic ergotropy. In Sec. \ref{sec:CMCM} we introduce the formalism of (continuously monitored) collisional models, with a particular emphasis on how to assess daemonic work extraction protocols in this framework. 
In Sec.~\ref{s:idealCMCM} we introduce the physically relevant CMCM that we have implemented on the IBM quantum platform, and we first present the experimental results of the daemonic work extraction protocol, by assuming a noiseless implementation of the corresponding quantum circuit. In Sec.~\ref{s:noisyCMCM} we discuss how one can improve the characterization and the performances of the work extraction protocol by properly modeling the unavoidable noise acting on the physical qubits, and we show the corresponding experimental results. Sec.~\ref{sec:conclusions} concludes the papers with some remarks and outlooks.
\section{Work extraction from quantum systems, ergotropy and daemonic ergotropy}
\label{sec:QBs}
Quantum batteries are devices governed by quantum mechanics that function as energy storage units. They operate in a similar manner to their classical counterparts, which store and retrieve energy as extractable work. Let us assume that the QB is prepared in a quantum state $\varrho$, and it is described by an internal Hamiltonian 
\begin{align}
\hat{H}_{0} = \sum_{i}\epsilon_{i}\ketbra{  \epsilon_{i}}{  \epsilon_{i}} \,,\, \textrm{with}\,\, \epsilon_{i}<\epsilon_{i+1} \, ,
\end{align}
and where, without loss of generality, we will always consider the energy of the ground state equal to zero, $E_0=0$.
We can then define the extracted work via a given unitary operation $\hat{U}$ as the difference between the energy before and after the unitary evolution, that is
\begin{align}
\mathcal{W}_{\hat{U}}(\varrho) = E(\varrho) - E(\hat{U}\varrho \hat{U}^\dag) \,. \label{eq:extractedwork}
\end{align}
where we have defined the average energy of a quantum state as $E(\varrho) = \hbox{Tr}[\varrho H_0]$.

The ergotropy for a state $\varrho$ is defined  as the extracted work maximized over all the possible unitary operations~\cite{allahverdyanMaximalWorkExtraction2004}, in formula
\begin{align}
\label{eq:ergotropy}
\mathcal{E}(\varrho) &= \max_{\hat{U}} \mathcal{W}_{\hat{U}}(\varrho)  \, ,\nonumber \\
&= E(\varrho) - \min_{\hat{U}} E(\hat{U}\varrho \hat{U}^\dag) \,.
\end{align}
Ergotropy is always greater or equal to zero; in particular one defines {\em passive states} $\varrho_p$ those states having zero ergotropy, $\mathcal{E}(\varrho_p)=0$. Passive states are diagonal in the eigenbasis of the Hamiltonian $\hat{H}_0$, $\varrho_p =  \sum_{i} p_{i}\ketbra{ \epsilon_{i}}{ \epsilon_{i}}$, while their eigenvalues do not admit energy extraction via energy inversion, i.e. $p_{i}>p_{i+1}$. 

The evaluation of the ergotropy thus corresponds to identifying the passive state $\varrho_p$ corresponding to the input quantum state $\varrho$, and the optimal unitary connecting these two quantum states. As one can always send an input pure state $\varrho = |\psi\rangle\langle\psi|$ to the ground state of the Hamiltonian $\varrho_0 = |\epsilon_0\rangle\langle\epsilon_0|$, it is straightforward to conclude that the ergotropy of pure states is in fact equal to their average energy, $\mathcal{E}(|\psi\rangle\langle\psi|) = E(|\psi\rangle\langle\psi|)$. 
In general, for finite dimensional systems, one can always identify the unitary sending an input (mixed) state $\varrho$ to its passive state, and one can derive an analytical formula for the ergotropy just in terms of eigenvalues and eigenstates of $\varrho$ and $\hat{H}_0$~\cite{allahverdyanMaximalWorkExtraction2004}. \\

Let us now consider a bipartite quantum system described by a quantum state $\varrho^{BE}$; with the aim of better describing the physical scenario we will be interested in the rest of this work, we will refer to system $B$ as the {\em battery}, and system $E$ as the {\em environment}. 
If one has no access to the environment, then the maximal amount of work extractable is simply equal to $\mathcal{E}(\varrho_B)$, that is to the erogotropy of the reduced state of subsystem $B$, $\varrho_B = \hbox{Tr}_E[\varrho_{BE}]$. 

Let us now assume that a measurement described by a set of POVM operators $\boldsymbol{\Pi} = \{\Pi_a\}$ is performed on subsystem $E$. For each possible outcome $a$ of the measurement we can identify a conditional state for system B
\begin{align}
\varrho_{a}^{B}= \frac{\hbox{Tr}_E[\varrho^{BE} (\mathbbm{1} \otimes \Pi_a)]}{p_a}
\end{align} 
where $p_a = \hbox{Tr}[\varrho^{BE} (\mathbbm{1} \otimes \Pi_a)]$ denotes the corresponding outcome probability. If we now assume that the work extraction unitary can be optimized for each conditional state, that is by exploiting the knowledge obtained from the measurement performed on $E$, then the maximum amount of work extractable is the average ergotropy of the conditional states, i.e.
\begin{align}
\overline{\mathcal{E}}_{\{\Pi_a\}} &= \sum_a p_a \mathcal{E}(\varrho_a^{B}) \,,\\
&= E(\varrho^{\tiny B}) - \sum_a p_a \min_{\hat{U}_a} E(\hat{U}_a \varrho_a^{\tiny B} \hat{U}_a^\dag)\, .
\label{eq:daemonicergo}
\end{align}

By exploiting the convexity property of the ergotropy, and by observing that one can always write $\varrho^B = \sum_a p_a \varrho_a^B$, it is then easy to show that $\overline{\mathcal{E}}_{\{\Pi_a^E\}} \geq \mathcal{E}(\varrho^B)$. We are indeed mathematically observing how a Maxwell demon can enhance the amount of work extracted from a quantum system $B$ by performing a measurement on a correlated system $E$, and thus optimizing the work extraction operation. 

For this reason $\overline{\mathcal{E}}_{\{\Pi_a^E\}}$ has been dubbed {\em daemonic ergotropy} and it has been introduced in~\cite{francicaDaemonicErgotropyEnhanced2017} to investigate its relationship with the quantum and classical correlations owned by the bipartite state $\varrho^{BE}$. We remark again that the daemonic ergotropy is the maximal amount of work that can be extracted by exploiting the information obtained by measuring $E$ via the POVM $\{\Pi_a\}$, that is the work extracted when the unitary operators are optimized for each conditional state.

If one considers a set of non-optimal work extraction unitaries $\{\hat{U}_a\}$ for each conditional state, then one can evaluate the {\em daemonic extracted work} as
\begin{align}
\overline{\mathcal{W}}_{\{\Pi_a,\hat{U}_a\}} = \sum_a p_a\, \mathcal{W}_{\hat{U}_a}(\varrho_a)\,. \label{eq:averagework}
\end{align}
This quantity is clearly upper bounded by the daemonic ergotropy, i.e. $\overline{\mathcal{W}}_{\{\Pi_a,\hat{U}_a\}} \leq \overline{\mathcal{E}}_{\{\Pi_a\}}$.\\

The daemonic ergotropy has been extended to the open quantum system scenario in~\cite{MorronePRApp}, by considering how to enhance the work extracted by exploiting the information obtained by continuously monitoring the environment. We refer to~\cite{MorronePRApp} for more details on this quantity, and for more remarks and examples concerning specific strategies based on continuous photo-detection and homodyne detection. In the next section we will consider the same scenario, but with the collisional model approach that we will adopt for the experimental implementation.
\section{Daemonic work extraction in a CMCM}
\label{sec:CMCM}
In this section we will first describe the formalism of CM for open quantum systems, extending it to CMCMs in order to discuss the daemonic work extraction in this scenario. 

CMs are a powerful and flexible tool to deal with open quantum systems, based on the discretization of both time and environment. While this approach allows us to treat systematically memory effects and in general non-Markovian dynamics, we will here stick to the simpler Markovian case. In this case, we will consider the initial state of the system (battery $B$ and environment $E$) in the following form, which is generally called Born approximation,
\begin{align}
\label{Initial State}
R_{0} = \varrho^{B}_{0}\otimes \varrho^{E_1}...\otimes\varrho^{E_n} \, .
\end{align}
Here,  $\varrho^{B}_{0}$ and $\varrho^{E_j}$ represent the initial states of the QB and of the $j^{th}$ auxiliary system (AS), respectively. We will also assume that the quantum states  of the ASs are identical and pure, $\varrho^{E_j} = |0\rangle_j {}_j \langle 0|$ (where the state $|0\rangle_j$ denotes a fixed pure reference state).

The dynamics of the QB in this scenario occur over subsequent steps, during which the QB interacts with each AS  via a unitary operator $\hat{V}_j$ acting only on the corresponding Hilbert spaces. 
After $n$ collisions, the joint $B+E$ quantum state can be written as
\begin{align}
\label{Iteration}
R_{n} =  \hat{V}_{n}... \hat{V}_{1} R_{0} \hat{V}_{1}^{\dag}... \hat{V}_{n}^{\dag}\, .
\end{align}
Now, by tracing out the environmental degrees of freedom, we can obtain the dynamics of the QB in terms of the number of collisions such that 
\begin{align}
\varrho^{B}_{n} = \hbox{Tr}_{E_n,\dots,E_1}[ R_{n}]\, .
\end{align}

It is easy to show how, under the assumptions above, it is possible to write explicitly the evolution for each collision in terms of a completely positive and trace-preserving map (CPTP), i.e.
\begin{align}
\varrho^{B}_n &= \mathcal{M}(\varrho^{B}_{n-1}) \,\\
&= \hbox{Tr}_{E_n}[ \hat{V}_n (\varrho^B_{n-1} \otimes \varrho^{E_n}) \hat{V}_n^\dag]\,,
\end{align}
highlighting how the evolution is in fact Markovian, that is, it is not influenced by what happened at the previous steps.

While typically the collision between the system and the environmental subsystems is described by a unitary interaction, in order to better describe the actual experimental implementations in the next sections, we will generalize the formalism described above by replacing it with a more general CPTP map $\mathcal{V}_n$ acting jointly on the battery system and the $n$-th AS, i.e.
\begin{align}
\hat{V}_n (\varrho^B_n \otimes \varrho^{E_n}) \hat{V}_n^\dag \, \rightarrow \, \mathcal{V}_n (\varrho^B_n \otimes \varrho^{E_n}) \,,
\end{align}
such that we can write the single-step evolution as 
\begin{align}
\varrho^{B}_n &= \mathcal{M}(\varrho^{B}_{n-1}) \,\\
&= \hbox{Tr}_{E_n}[ \mathcal{V}_n(\varrho^B_{n-1} \otimes \varrho^{E_n})]\,, \label{eq:CMevolution}
\end{align}

The maximum amount of work extractable from the battery system after $n$ steps is then equal to the {\em unconditional} ergotropy (the term {\em unconditional} will be more clear in the following)
\begin{align}
\mathcal{E}_{{\sf unc},n} = \mathcal{E}(\varrho^B_n) \,,
\end{align}
which is in turn upper bounded by the internal energy of the state $\mathcal{E}_{{\sf unc},n} \leq E(\varrho^B_n)$.

We will now assume that the auxiliary systems that play the role of the environment can be measured via a given POVM $\boldsymbol{\Pi} = \{\Pi_a\}$ after their interaction. In this case, we will obtain a conditional evolution for the quantum state of the battery dependent on all the measurement results ${\bf a}_n = (a_1,\dots,a_n)$ obtained at each step of the evolution, in the formula
\begin{align}
\varrho^{B}_{{\bf a}_n} = \frac{\hbox{Tr}_{E_n,\dots,E_1}\left[ R_{n} \left(\mathbbm{1}_B \otimes \Pi_{a_1} \otimes \dots \otimes \Pi_{a_n}\right)\right]}{p_{{\bf a}_n}}\,,
\end{align}
where $p_{{\bf a}_n}= \hbox{Tr}\left[ R_{n} \left(\mathbbm{1}_B \otimes \Pi_{a_1} \otimes \dots \otimes \Pi_{a_n}\right)\right]$ denotes the probability of measuring the outcomes ${\bf a}_n$ and thus of obtaining the conditional states $\varrho^B_{{\bf a}_n}$. We will often refer to the conditional quantum state $\varrho^{B}_{{\bf a}_n}$ as a {\em quantum trajectory}. Also in this case one can write the single-step conditional evolution of the quantum trajectory, obtaining
\begin{align}
\varrho^{B}_{{\bf a}_n} = \frac{\hbox{Tr}_{E_n}\left[ \mathcal{V} (\varrho^{B}_{{\bf a}_{n-1}} \otimes \varrho^{E_n}) (\mathbbm{1}\otimes \Pi_{a_n}) \right]}{p_{a_n}} \,, 
\label{eq:conditionalstate}
\end{align} 
where $p_{a_n} = \hbox{Tr}\left[ \mathcal{V}_n (\varrho^{B}_{{\bf a}_{n-1}} \otimes \varrho^{E_n}) (\mathbbm{1}\otimes \Pi_{a_n}) \right]$ is the probability of obtaining the outcome $a_n$ at the $n$-th step, conditioned on having obtained the state $\varrho^{B}_{{\bf a}_{n-1}}$ from the previous collisions. 

The formalism we have just presented correspond to a CMCM that has been introduced in~\cite{LandiPRXQuantum2022,BelenchiaPRA2022} to discuss entropy production in quantum trajectories. Here we will exploit it as the perfect playground to discuss daemonic work extraction in open quantum systems: by assuming that we will associate a certain work extraction unitary $\hat{U}_{{\bf a}_n}$ to each stream of measurement outcomes ${\bf a}_n$, and thus to the corresponding conditional quantum state, we can generalize the average work extracted in Eq.~(\ref{eq:averagework}) as 
\begin{align}
\overline{\mathcal{W}}_{\{\Pi_{{\bf a}_n},\hat{U}_{{\bf a}_n} \}} &= \sum_{{\bf a}_n} p_{{\bf a}_n}\, \mathcal{W}_{\hat{U}_{{\bf a}_n}}(\varrho^B_{{\bf a}_n})\,, \notag \\
&= E(\varrho^{\tiny B}_n) - \sum_{{\bf a}_n} p_{{\bf a}_n} E(\hat{U}_{{\bf a}_n} \varrho_{{\bf a}_n}^{\tiny B} \hat{U}_{{\bf a}_n}^\dag)\, ,
 \label{eq:qtaveragework}
\end{align}
and similarly its upper bound, that is the {\em unravelling} daemonic ergotropy~\cite{MorronePRApp},
\begin{align}
\overline{\mathcal{E}}_{\{\Pi_{{\bf a}_n}\}} &= \sum_{{\bf a}_n} p_{{\bf a}_n} \mathcal{E}(\varrho_{{\bf a}_n}^{B}) \,, \notag \\
&= E(\varrho^{\tiny B}_n) - \sum_{{\bf a}_n} p_{{\bf a}_n} \min_{\hat{U}_{{\bf a}_n}} E(\hat{U}_{{\bf a}_n} \varrho_{{\bf a}_n}^{\tiny B} \hat{U}_{{\bf a}_n}^\dag)\, .
\label{eq:qtdaemonicergo}
\end{align}

As we discussed in the previous section, one can prove that $\overline{\mathcal{E}}_{\{\Pi_{{\bf a}_n}\}} \geq \mathcal{E}_{{\sf unc},n} $, that is, by measuring the environment subsystems and thus by optimizing the extraction unitary protocol for each quantum trajectory, one can increase the amount of work extracted. 
\begin{figure*}[t]
\begin{center}
\includegraphics[width=1\textwidth]{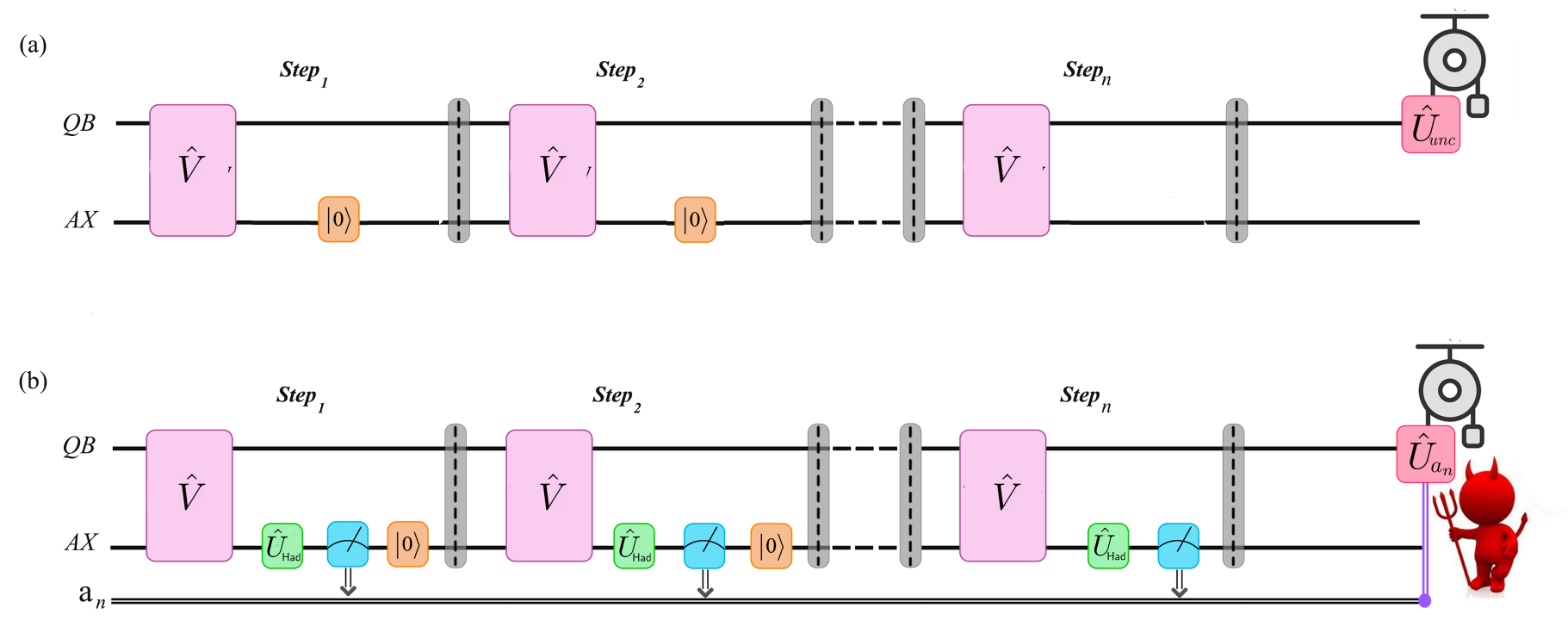}
\end{center}
\caption{a) pictorial representation of the implementation of the Markovian CM as a quantum circuit involving only two qubits: before each interaction, the auxiliary qubit is reset in the reference state $|0\rangle$; the work extraction is performed by a deterministic unitary gate, denoted as $\hat{U}_{\sf unc}$, which only depends on the parameters characterizing the interaction $\hat{V}$ and the number of steps of the CM. b) pictorial representation of the implementation of the Markovian CMCM as a quantum circuit involving only two qubits: after each interaction, the auxiliary qubit is measured in the $\hat{\sigma}_x$ basis (this is obtained by performing a Hadamard gate before the $\hat{\sigma}_z$ measurement) and then reset in the reference state $|0\rangle$; unlike the previous case, where the work extraction unitary applied was deterministic, here the gate implemented will depend on the outcomes ${\bf a}_n$ obtained from the auxiliary qubit's measurements.}
\label{f:idealmodel}
\end{figure*}
\begin{figure*}[bt]
\begin{center}
\includegraphics[width=0.45\textwidth]{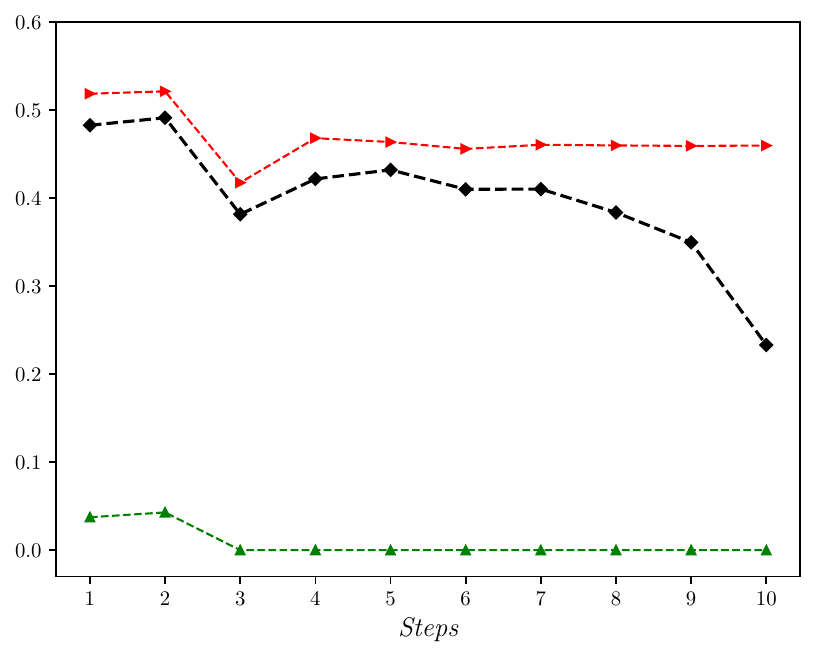} 
 \put(-115,185){\small (a)}
\includegraphics[width=0.45\textwidth]{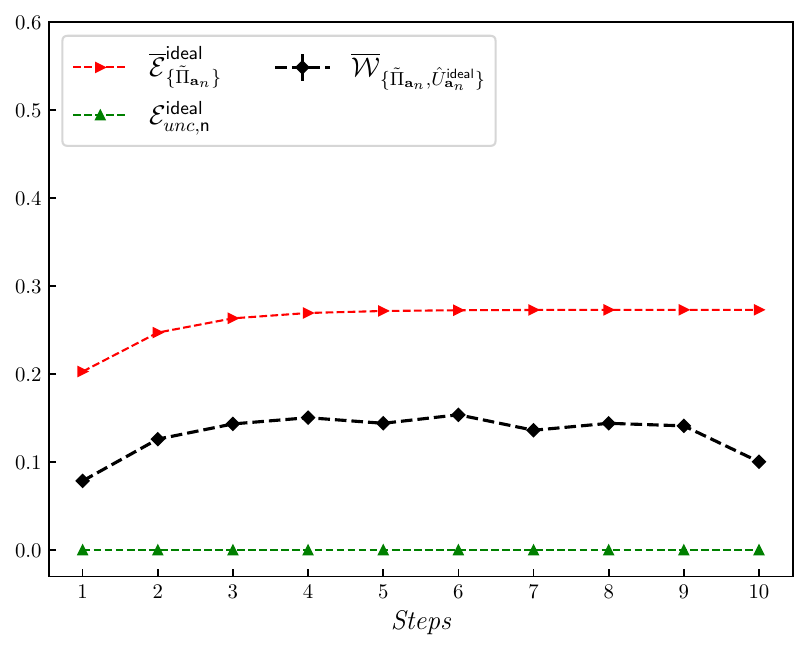}
 \put(-115,185){\small (b)}
\end{center}
\caption{Theoretical values of the unconditional ergotropy  $\mathcal{E}_{{\sf unc},n}^{\sf ideal}$ (green circles) and of the daemonic ergotropy $\overline{\mathcal{E}}_{\{\Pi_{{\bf a}_n}\}}^{\sf ideal}$ for the noiseless model (red triangles), along with the experimental daemonic extracted work $\overline{\mathcal{W}}_{\{\Pi_{{\bf a}_n},\hat{U}_{{\bf a}_n}^{\sf ideal} \} }$ obtained by employing the conditional unitaries $\hat{U}_{{\bf a}_n}^{\sf ideal}$ evaluated according to the noiseless ideal model. The experimental results have been obtained with $10^4$ executions for each experiment. Left panel: $\kappa=\alpha=1$; right panel: $\kappa=2\alpha=2$.
More details on the implementation on IBM quantum computers, along with the physical parameters values characterizing the experiment can be found in the Appendix \ref{a:IBMqdetails} (see Table.~\ref{table}).}
\label{f:results1} 
\end{figure*}
\section{Experimental simulation of daemonic work extraction in a CMCM}
\label{sec:results}
In this section we will describe a particular (physically motivated) CMCM and we will show how to implement it on a digital quantum computer, along with the conditional work extraction unitary, and how to evaluate the corresponding figures of merit in order to demonstrate the daemonic enhancement due to information obtained from the environment. We performed the experiments on IBM Quantum devices as they provided all the required instructions, including mid-circuit measurements, feed-forward and conditional instructions. We will first discuss the results for the ideal noiseless CMCM, and we will then show how, by properly modeling the noise acting on the quantum device, we can better characterize and optimize the work extraction protocol. 
\subsection{Ideal noiseless CMCM} \label{s:idealCMCM}
In what follows, both the QB system and the auxiliary systems in the collisional model will correspond to two-level quantum systems (qubits). In particular, we will also assume that the internal Hamiltonian of the QB that defines its energy is
\begin{align}
\hat{H}_0 = \frac{\omega_0}{2} \left(\hat{\mathbbm{1}}^B - \hat{\sigma}_z^B \right) \,,
\end{align}
where $\hat{\sigma}_z$ denotes the z-Pauli operator, and thus its eigenstates $|0\rangle_B$ and $|1\rangle_B$ respectively correspond to the ground and excited state. We will now assume that the QB and all the auxiliary systems will be initially prepared in the ground states, $|0\rangle_B$ and $|0\rangle_{E_j}$. The charging of the QB and the interaction with the environment will then be described in the collisional model at each step $j$ by the unitary $\hat{V}_j = e^{-i \hat{H}_j}$, where the interaction Hamiltonian reads
\begin{align}
\hat{H}_j = \alpha \left( \hat{\sigma}_x^B \otimes \hat{\mathbbm{1}}^{E_j}\right) + \kappa \left( \hat{\sigma}_+^B \otimes \hat{\sigma}_-^{E_j} + h.c. \right) \,,
\label{eq:hamiltonian}
\end{align}
with $\hat{\sigma}_x$ denoting the x-Pauli operator, while $\hat{\sigma}_+$ and $\hat{\sigma}_-$ correspond respectively to the raising and lowering operator. We are assuming to work in interaction picture respect to both the QB system Hamiltonian $\hat{H}_0$ and the environment Hamiltonian, such that, following for example the time discretization of the environment described in \cite{AlbarelliPLA2024}, the auxiliary systems play the role of input temporal modes interacting with the QB. This CM is indeed particularly relevant from a physical point of view as, by taking its correct infinitesimal time limit~\cite{CiccarelloPR2022} and by considering no measurements on the auxiliary systems of the environment, it describes the evolution of the following Markovian master equation
\begin{align}
\frac{d\varrho_{\sf unc}^B}{dt} = - i \tilde{\alpha} [\hat{\sigma}_x^B, \varrho_{\sf unc}^B] + \tilde\kappa \mathcal{D}[\hat{\sigma}_-^B]\varrho_{\sf unc}^B \,,
\end{align}
where $\mathcal{D}[\hat{O}]\varrho = \hat{O}\varrho \hat{O}^\dag - (\hat{O}^\dag \hat{O} \varrho + \varrho \hat{O}^\dag \hat{O})/2$, while $\tilde{\alpha}$ and $\tilde{\kappa}$ are properly renormalized constants. This master equation describes a two-level atom driven by a resonant classical field of intensity $\tilde\alpha$, and spontaneously emitting with rate $\tilde\kappa$, and it has been studied in~\cite{farinaChargermediatedEnergyTransfer2019} as the paradigmatic example of open QB, and in~\cite{MorronePRApp} to discuss the daemonic ergotropy under different continuous time unravellings (as pointed out before, one should remark that this physical interpretation is valid under the assumption that we are working in interaction picture respect to the battery Hamiltonian $\hat{H}_0$). 

The CM can be readily implemented by using only two qubits (a battery qubit and an auxiliary qubit) on a digital quantum computer as described in Fig.~\ref{f:idealmodel}(a): the initial QB battery qubit and the auxiliary qubit can be initialized in the corresponding ground states, while the interaction unitary $\hat{V}_j$ can be implemented by setting the chosen values of the parameters $\alpha$ and $\kappa$. After the interaction, the auxiliary qubit can be reinitialized in the ground state $\ket{0}$ using a \textsc{reset} gate, and the next steps of the CM can simply follow as before. After $n$ steps one will have obtained the corresponding {\em unconditional} battery quantum state $\varrho_n^B$, and, as explained before, the amount of work that can be unitarily extracted is upper bounded by the unconditional ergotropy.

We now assume that the auxiliary systems after their interaction with the QB are sequentially measured in the $\hat{\sigma}_x^{E_j}$ basis (one should notice that in the noiseless case, any ideal measurement would lead to the same values of daemonic ergotropy~\cite{MorronePRApp}; we eventually decided to measure in the $\hat{\sigma}_x$ basis as this corresponds to a nearly-optimal measurement for the amount and kind of noise present in the experimental platform that we will discuss later in the paper, yielding in particular slightly larger results of daemonic ergotropy respect to the $\hat{\sigma}_z$ measurement). The corresponding CMCM can be easily implemented by exploiting the {\em mid-circuit measurement} feature of IBM quantum computers: after the unitary interaction $\hat{V}_j$ the $\hat{\sigma}_x$ measurement can be performed by first acting with a Hadamard gate on the auxiliary qubit, and then by performing a native $\hat{\sigma}_z$ measurement. After the measurement, the auxiliary qubit is  reinitialized in the ground state $\ket{0}$ using a conditional \textsc{x} gate and one can repeat the operations, interaction, and measurement, described before at each step. 

A classical register is used to keep track of the measurement outcomes ${\bf a}_n = (a_1,\dots,a_n)$, where in this case $a_j =\{0,1\}$ denotes the two possible outcomes from the measurement on the auxiliary qubit. In this case the optimal extraction unitary will depend not only on the parameters $\alpha$ and $\kappa$, but also on all the measurement results ${\bf a}_n$ that define the corresponding quantum trajectory $\varrho_{{\bf a}_n}^B$; one then needs to perform a real-time feedback unitary operation $\hat{U}_{{\bf a}_n}$ on the battery qubit. 

This is indeed possible on IBM quantum computers, at the only cost of previously simulating all the possible $2^n$ trajectories and then storing in the memory the corresponding possible $2^n$ unitary operations. This has been accomplished by writing a separate Python code that exploits the QuTiP package~\cite{johanssonQuTiPOpensourcePython2012} to simulate the corresponding quantum circuit for each trajectory and generate a list of all the $2^n$ two-dimensional matrices that univocally describe the corresponding work extraction unitary $\hat{U}_{{\bf a}_n}$ on the qubit.
Notice that pre-computing the unitary operations for the $2^n$ trajectories is not a fundamental requirement, but rather a technical limitation of the current IBM systems, where the instructions with the conditional operations need to be provided at compilation time. Further technological advancements could allow one to compute the unitaries on the fly.

In order to experimentally evaluate the average daemonic work $\overline{\mathcal{W}}_{\{\Pi_{{\bf a}_n},\hat{U}_{{\bf a}_n} \} }$ we can evaluate the two terms in Eq.~(\ref{eq:qtaveragework}) separately: the energy of the unconditional state $E(\varrho_n^B)$ can be evaluated by running several instances of the circuit in Fig.~\ref{f:idealmodel}(a) without the final extraction unitary operator $\hat{U}_{\sf unc}$ on the battery and by performing a final measurement of the battery in the $\hat{\sigma}_z$ basis.  The second term in Eq.~(\ref{eq:qtaveragework}) can be in turn obtained by experimentally running several times the CMCM and then perform the energy $\hat{\sigma}_z$-measurement after the conditional unitaries $\hat{U}_{{\bf a}_n}$; this allows to jointly sample both on the trajectories probability distribution $p_{{\bf a}_n}$ and on the probability distribution of the Hamiltonian operator $\hat{H}_0$, needed to evaluate the quantum trajectories average energies $E(\hat{U}_{{\bf a}_n} \varrho_{{\bf a}_n}^B \hat{U}_{{\bf a}_n}^{\dag})$.\\

It is important to remark again how the optimal work extraction unitaries for the different quantum trajectories $\varrho^B_{{\bf a}_n}$ are predetermined by assuming that the circuit that has been experimentally implemented on the IBM quantum computer is noiseless and thus effectively implement the CMCM we have just described. As we described before, we have determined them by running a separate simulation of the circuit; we will denote these unitary operators as $\hat{U}_{{\bf a}_n}^{\sf ideal}$.  By exploiting them we have then experimentally evaluated the daemonic extracted work $\overline{\mathcal{W}}_{\{\Pi_{{\bf a}_n},\hat{U}_{{\bf a}_n}^{\sf ideal} \} }$ and compared it with its respective upper bound, that is the daemonic ergotropy $\overline{\mathcal{E}}_{\{\Pi_{{\bf a}_n}\}}^{\sf ideal}$, and, in order to experimentally verify an actual daemonic enhancement in the work extraction, with the unconditional ergotropy $\mathcal{E}_{{\sf unc},n}^{\sf ideal}$ (these quantities have been separately numerically evaluated). We also remind that, in the noiseless CMCM described above, all the quantum trajectories remain in a pure state during the evolution, $\varrho^B_{{\bf a}_n} = |\psi_{{\bf a}_n}\rangle\langle\psi_{{\bf a}_n}|$; as a consequence the daemonic ergotropy is in fact equal to its upper bound, that is the energy of the unconditional state $\overline{\mathcal{E}}_{\{\Pi_{{\bf a}_n}\}}^{\sf ideal} = E(\varrho_n^B)$, regardless on the measurement performed on the auxiliary systems. \\

More details about the implementation of these experiments on the IBM quantum computers can be found in Appendix~\ref{a:IBMqdetails}, while the results are reported in Fig.~\ref{f:results1} for two different choices of the parameters $\alpha$ and $\kappa$: we immediately observe how in general $\overline{\mathcal{W}}_{\{\Pi_{{\bf a}_n},\hat{U}_{{\bf a}_n}^{\sf ideal} \} } > \mathcal{E}_{{\sf unc},n}^{\sf ideal}$; this clearly proves a daemonic enhancement, as the work extracted by exploiting the information obtained from the measurement is larger than the unconditional ergotropy, that is the maximum amount of work that one can extract without measuring the environment. 

On the other hand we also observe how in both cases $\overline{\mathcal{W}}_{\{\Pi_{{\bf a}_n},\hat{U}_{{\bf a}_n}^{\sf ideal} \} }$ is definitely smaller than the daemonic ergotropy $\overline{\mathcal{E}}_{\{\Pi_{{\bf a}_n}\},{\sf ideal}}$. As we are employing the optimized unitary operators $\hat{U}_{{\bf a}_n}^{\sf ideal}$, we would expect that the two quantities should be equal. The gap is however clearly due to the fact that the (noiseless and ideal) model we have employed to evaluate both the daemonic ergotropy and the optimized unitaries is not describing accurately the experiment that we are actually performing on the IBM quantum computer, where in fact unwanted noise is present and is playing a role. In the next subsection, we will show how we can in fact characterize the noise affecting the system and also improve the work extraction by exploiting this information.
\subsection{Noise characterization of the CMCM}\label{s:noisyCMCM}
As all the current experimental quantum platforms aiming to run quantum algorithms or to be used as feasible testbeds for simulating quantum systems, IBM quantum computers are affected by noisy processes, and several attempts have been proposed in order to design error correction algorithms or error mitigation techniques to alleviate these issues \cite{cao2021nisq}. These errors can stem from various sources, including interactions with nearby defects \cite{gordon2022environmental} and fluctuations in charge noise \cite{riste2013millisecond}. Such fluctuations can induce variations in the qubit's energy levels and coupling strengths, thereby introducing errors during quantum operations. 
Additionally, errors can occur during the measurement process, leading to inaccuracies in determining the qubit state \cite{khezri2015qubit}.

The different types of error affecting quantum computations have different magnitudes.
On superconducting qubit devices like IBM's, readout errors are the largest and range between $1-5\%$, while two-qubit gates have a median error rate around $1\%$. Single-qubit gates have errors that are one to two orders of magnitude smaller than two-qubit gates and can typically be neglected. Moreover, physical qubits tend to undergo decoherence due to spontaneous relaxation to the ground state (amplitude damping) and dephasing. This decoherence time typically allows for the execution of several layers of gates, but in our specific simulation, we are performing repeated mid-circuit measurements, which take significant time compared to the total coherence time of the qubits. Finally, gates (mostly the two-qubit ones) and readouts tend to perturb the state of other qubits nearby their target qubits, in what is called a cross-talk effect. This effect can be relevant for larger circuits, but it is not a leading source of error in our two-qubit circuit and we can neglect it.

In this section, we improve the characterization of the CMCM and the work extraction protocol by creating a simple yet effective model of the leading sources of noise in the circuit of interest, which are the readout error and the qubit decoherence affecting the battery while the ancillary qubit is measured.

The IBM quantum platform allows one to estimate the parameters characterizing the different kinds of noise affecting the qubits on the device, in particular the spontaneous relaxation to the ground state (amplitude damping) and dephasing. These are quantified, respectively, by the coherence times $T_1$ and $T_2$, which are indeed available information for all the qubits in the IBM quantum backends. The other relevant available information is the readout time $T_r$, which quantifies the time required to perform a measurement on a given qubit. 

\begin{figure*}[t!]
\begin{center}
\includegraphics[width=1\textwidth]{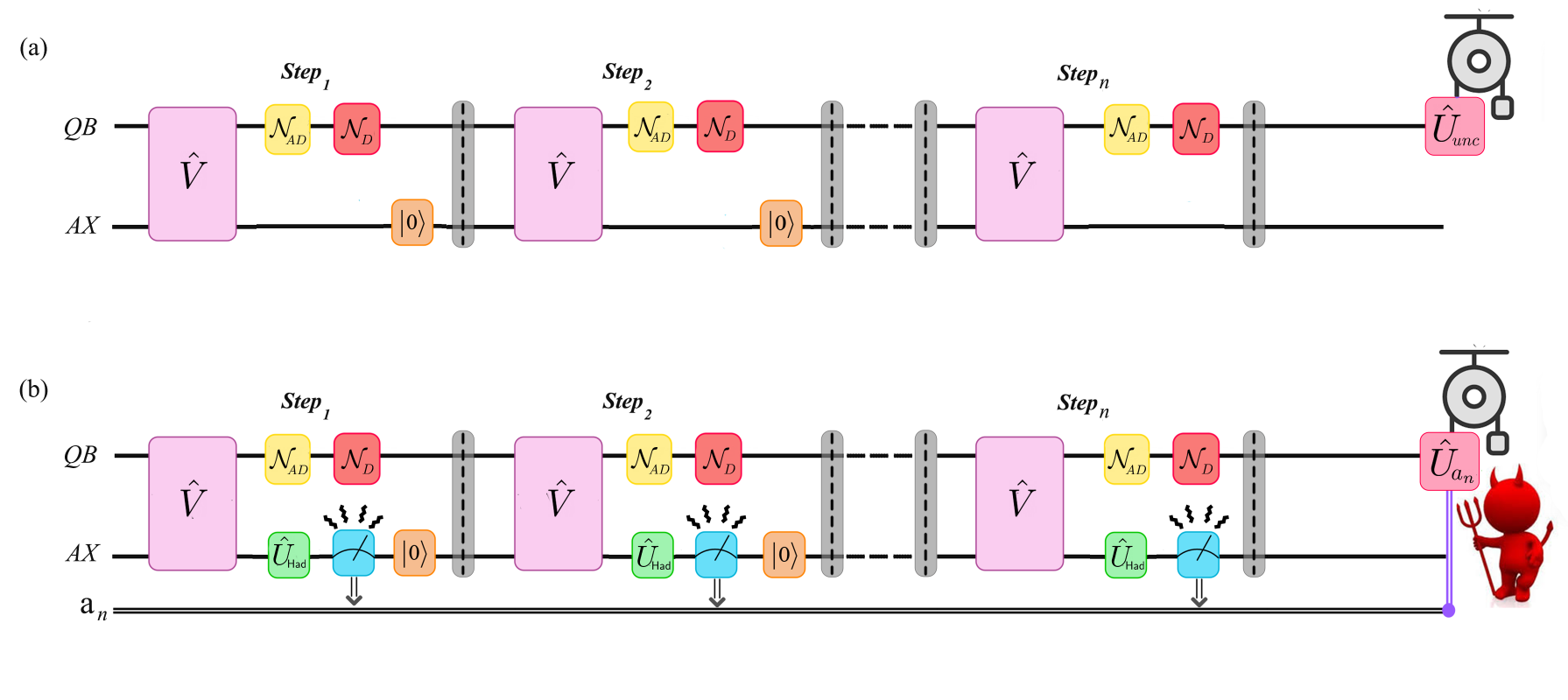}
\end{center}
\caption{A pictorial representation of the noise model applied to our circuit for both protocols to achieve unconditional ergotropy (a) and daemonic ergotropy (b). The dephasing channel, denoted by DP, and the Amplitude Damping channel, shown by AD, are applied at each step using parameters obtained from IBM devices. However, since measurement occurs only in the daemonic ergotropy case, we have applied the noisy measurement only on the ancilla to find the trajectories. }
\label{f:noisymodel}
\end{figure*}

We here assume that both dephasing and amplitude damping channels are acting on the battery qubit during the measurement, as depicted in the {\em noisy CM} in Fig.~\ref{f:noisymodel}(a). The qubit battery state at each step can be thus calculated as in Eq.~(\ref{eq:CMevolution}) 
\begin{align}
\varrho_n^B = \hbox{Tr}_{E_n}[\mathcal{V}_n (\varrho_{n-1}^B \otimes |0\rangle_n{}_n\langle n|)]
\end{align}
where now the map $\mathcal{V}_j$ reads
\begin{align}
\mathcal{V}_j = (\mathcal{N}_{AD} \otimes \mathcal{I}_{E_j}) \circ (\mathcal{N}_{D} \otimes \mathcal{I}_{E_j}) \circ \mathcal{U}_j
\end{align} 
with $\mathcal{U}_j(\bullet) = e^{-i \hat{H}_j} \bullet e^{i \hat{H}_j}$ denoting the unitary map generated by the Hamiltonian in Eq.~(\ref{eq:hamiltonian}), while the maps $\mathcal{N}_{AD}$ and $\mathcal{N}_{D}$ correspond respectively to the amplitude damping and dephasing channels acting on the battery qubit.
The amplitude damping channel can be represented as $\mathcal{N}_{AD}(\varrho)= \sum_{l=0,1} \hat{K}_l \varrho \hat{K}_l^\dag$ where the Kraus operators read
\begin{align}
\hat{K}_{0} = 
\begin{pmatrix}
1 & 0 \\
0 & \sqrt{1-P_{AD}}
\end{pmatrix}
,\,\,
\hat{K}_{1} = 
\begin{pmatrix}
0 & \sqrt{P_{AD}} \\
0 & 0
\end{pmatrix}
%\\
%\phi_{AD}(\varrho) = K_{0}\varrho K_{0}^{\dag} + K_{1}\varrho K_{1}^{\dag}
\end{align}
The probability $P_{AD}$ can be calculated based on the $T_1$ time for the battery qubit and on the readout time $T_r$ for the auxiliary qubit, via the formula $ P_{AD} = 1-e^{-T_{r}/T_1} $. Also the dephasing channel can be written in terms of just two Kraus operators $\mathcal{N}_{D}(\varrho)= \sum_{l=0,1} \hat{N}_l \varrho \hat{N}_l^\dag$, where
\begin{align}
\hat{N}_{0} = \sqrt{1-P_D} \hat{\mathbbm{1}} \,, \,\,\, 
\hat{N}_{1} = \sqrt{P_D}\hat{\sigma}_{z}  \,,
\end{align}
where the dephasing probability can be obtained again from the readout time of the auxiliary qubit $T_r$ and the $T_2$ time of the battery qubit as $P_D =1 - e^{-T_r/T_2}$. 

As we described above, we apply these channels only on the QB qubit, as it is more susceptible to these noisy processes while the auxiliary qubit is measured. On the other end, the second qubit is reset at every step, eliminating concerns about its relaxation and dephasing, while the most significant error in this respect regards the measurement process. The simplest way to model a noisy ${\sigma}_z$ measurement is to replace the ideal $\hat{\sigma}_z$ projectors $\{{\pi}_0 = |0\rangle\langle 0|, \hat{\pi}_1=|1\rangle\langle 1|\}$, with the following POVM operators
\begin{align}
{\tilde{\pi}}_{0} = P_{00}\ketbra{0}{0} + P_{01}\ketbra{1}{1}
\\
{\tilde{\pi}}_{1} = P_{10}\ketbra{0}{0} + P_{11}\ketbra{1}{1}
\end{align}
where the parameters $P_{ab}$ quantify the probability of obtaining the outcome $\{a\}$ when one has prepared the $\hat{\sigma}_z$ eigenstate $|b\rangle$ (in Appendix~\ref{a:IBMqdetails} we describe in more detail how we estimate the set of probabilities $\{P_{ab}\}$ along with other details on the experimental implementations on the IBM quantum platform). The noisy ${\sigma}_x$ measurement is then described by the POVM operators ${\tilde{\Pi}}_a = \hat{U}_{\sf Had}\, {\tilde{\pi}}_a\, \hat{U}_{\sf Had}$, obtained by acting with a Hadamard gate $\hat{U}_{\sf Had}$ before the $\hat{\sigma}_z$ measurement. The corresponding CMCM is then the one described in Fig.~\ref{f:noisymodel}(b), such that the corresponding conditional state can be evaluated as in Eq.~(\ref{eq:conditionalstate}).\\
\begin{figure*}[ht!]
\begin{center}
\includegraphics[width=0.45\textwidth]{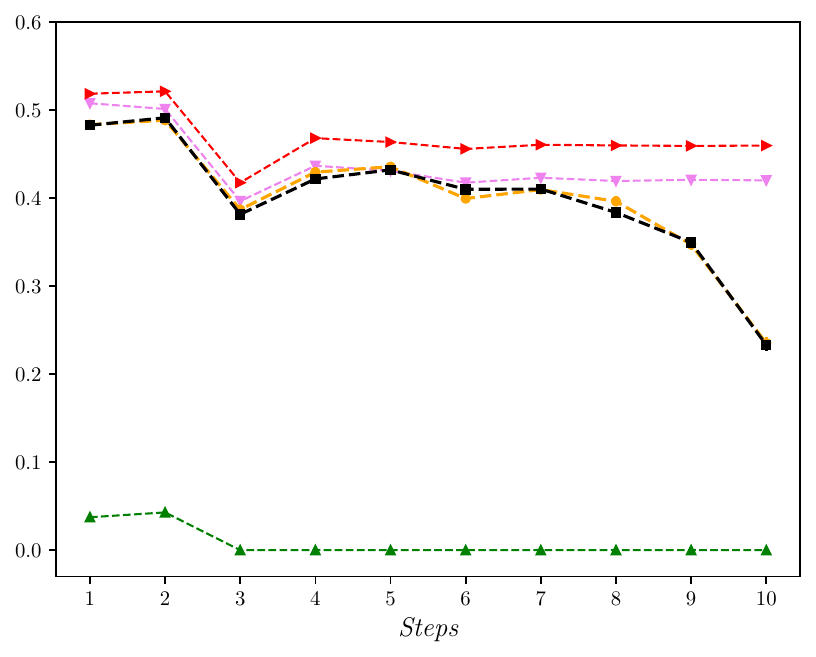}
 \put(-115,185){\small (a)}
\includegraphics[width=0.45\textwidth]{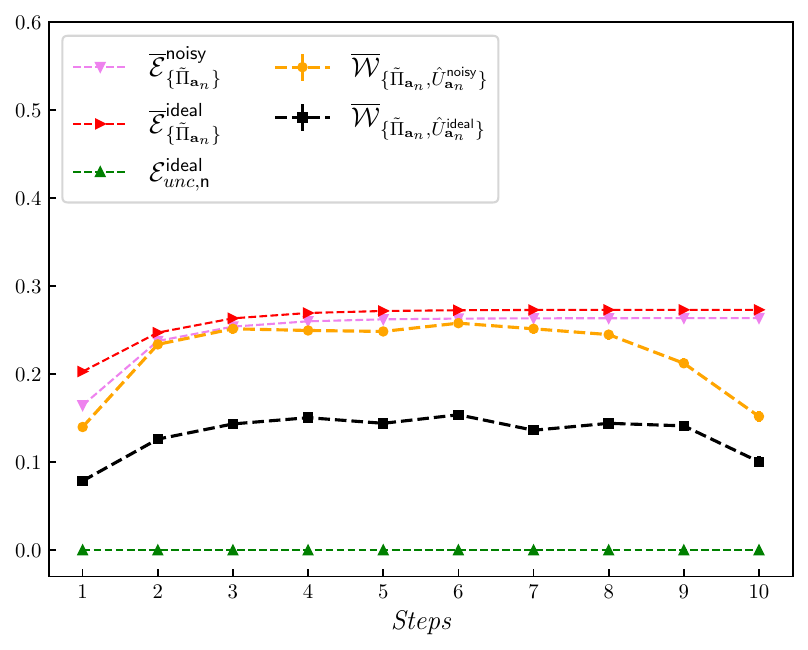}
 \put(-115,185){\small (b)}
\end{center}
\caption{Theoretical daemonic ergotropy $\overline{\mathcal{E}}_{{\{\tilde{\Pi}}_{{\bf a}_n}\}}^{\sf noisy}$ for the noisy model (pink triangles) and the experimental daemonic extracted work $\overline{\mathcal{W}}_{\{\tilde{\Pi}_{{\bf a}_n},\hat{U}_{{\bf a}_n}^{\sf noisy} \} }$ obtained by employing the conditional unitaries $\hat{U}_{{\bf a}_n}^{\sf noisy}$ evaluated according to the noisy model, along with the other quantities already present in Fig.~\ref{f:results1} (the unconditional ergotropy  ${\mathcal{E}}_{{\sf unc},n}^{\sf ideal}$ (green circles) and the daemonic ergotropy $\overline{\mathcal{E}}_{\{\Pi_{{\bf a}_n}\}}^{\sf ideal}$ (red triangles) for the noiseless model, the experimental daemonic extracted work $\overline{\mathcal{W}}_{\{\Pi_{{\bf a}_n},\hat{U}_{{\bf a}_n}^{\sf ideal} \} }$ obtained by employing the ideal conditional unitaries). The experimental results have been obtained with $10^4$ executions for each experiment. Left panel: $\kappa=\alpha=1$; right panel: $\kappa=2\alpha=2$. More details on the implementation on IBM quantum computers, along with the physical parameters values characterizing the experiment can be found in the Appendix \ref{a:IBMqdetails} (see Table~\ref{table}).}
\label{f:results2}
\end{figure*}
We have then considered this noisy CMCM to evaluate numerically the tigther upper bound on the extractable work, that is daemonic ergotropy $\overline{\mathcal{E}}_{\{\tilde{\Pi}_{{\bf a}_n}\}}^{\sf noisy}$, and to precalculate the proper conditional unitaries $\hat{U}_{{\bf a}_n}^{\sf noisy}$. This has been done with a separate numerical simulation of the CMCM; we have then run the protocol on IBM devices, with the gates shown in Fig.~\ref{f:idealmodel} unchanged, except for the conditional unitary extraction $\hat{U}_{{\bf a}_n}^{\sf noisy}$, optimized according to the noisy model. 

The corresponding experimental results are reported in Fig.~\ref{f:results2} which represents the main results of our work: the experimental daemonic extracted work with noisy-optimized conditional unitaries $\overline{\mathcal{W}}_{\{\tilde{\Pi}_{{\bf a}_n},\hat{U}_{{\bf a}_n}^{\sf noisy} \} }$ is added, together with the corresponding noisy daemonic ergotropy $\overline{\mathcal{E}}_{\{\tilde{\Pi}_{{\bf a}_n}\}}^{\sf noisy}$, to the experimental and theoretical data previously reported in Fig.~\ref{f:results1}. 

If we look at panel (a), that is for $\kappa=\alpha=1$, we readily observe that $\overline{\mathcal{W}}_{\{\tilde{\Pi}_{{\bf a}_n},\hat{U}_{{\bf a}_n}^{\sf noisy} \} }\approx \overline{\mathcal{W}}_{\{\Pi_{{\bf a}_n},\hat{U}_{{\bf a}_n}^{\sf ideal} \} }$, that is employing the conditional work extraction unitary optimized according to the noisy model is not really improving the work extraction protocol. 

However, we also observe that, at least for the first $n=6$ steps of the CM,  $\overline{\mathcal{W}}_{\{\tilde{\Pi}_{{\bf a}_n},\hat{U}_{{\bf a}_n}^{\sf noisy} \} }\approx \overline{\mathcal{E}}_{\{\tilde{\Pi}_{{\bf a}_n}\}}^{\sf noisy}$, that is the daemonic extracted work approaches its ultimate upper bound quantified by the daemonic ergotropy calculated by using the noisy model above. 
The fact that for a larger number of steps ($n>6$) the two quantities begin to differ is probably due to the fact that our simple noisy model is no longer able to capture all the noisy processes affecting the two qubits. 

The results shown in the panel (b) of Fig.~\ref{f:results2}, that is for $\kappa=2\alpha=2$, are even more interesting: in fact here we not only observe that for the first steps of the CM $\overline{\mathcal{W}}_{\{\tilde{\Pi}_{{\bf a}_n},\hat{U}_{{\bf a}_n}^{\sf noisy} \} }\approx \overline{\mathcal{E}}_{\{\tilde{\Pi}_{{\bf a}_n}\}}^{\sf noisy}$, that is the daemonic work approaches the daemonic ergotropy, we also find that $\overline{\mathcal{W}}_{\{\tilde{\Pi}_{{\bf a}_n},\hat{U}_{{\bf a}_n}^{\sf noisy} \} } > \overline{\mathcal{W}}_{\{\Pi_{{\bf a}_n},\hat{U}_{{\bf a}_n}^{\sf ideal} \} }$, that is by employing the conditional unitaries optimized according to the noise model one can sensibly increase the amount of work extracted respect to the ones that one can obtain with the noiseless ideal model and approach the ultimate limit set by the daemonic ergotropy. Also for these values of the parameters, we observe a sharp decrease of the extracted work for larger number of steps $n>6$, as our model is not able to properly describe the noisy processes that are affecting the qubits.

In general, our results clearly show how a proper characterization of the noise acting on the quantum device is fundamental, not only to properly assess the performance of the device, as in our case comparing the extracted work with its actual ultimate limits but also to improve the protocol by properly optimizing it.
\section{Conclusions}
\label{sec:conclusions}
The possibility of exploiting the information obtained from a measurement to increase the amount of work that can be extracted from a system has been thoroughly investigated both in the classical and in the quantum context. Daemonic work extraction from a quantum system and its relationship with quantum correlations has been recently demonstrated experimentally in~\cite{stahl2024demonstrationenergyextractiongain} in an experimental platform using trapped ions. 

In our work we have experimentally demonstrated the daemonic work extraction in a different physical platform, i.e. via the superconducting qubits employed in the IBM quantum devices. More importantly, we have remarkably proven the possibility of efficiently simulating an open quantum system and a feedback quantum control scenario via a CMCM. In particular, we have shown how a proper, yet simple, characterization of the noise acting on the quantum device allows not only to better characterize but also to better optimize the work extraction protocol. 

A more accurate characterization of the noise, including potential coherent and incoherent noise on the gates applied to implement the interaction unitary, inaccurate qubit \textsc{reset}s, as well as cross-talks, could further improve the characterization of the daemonic ergotropy and extracted work. Together with active error suppression methods like dynamical decoupling, this could allow to push the simulation to a larger number of steps.

We believe that our results will help in shedding light on the role of measurement in quantum thermodynamics protocol, and will also help in encouraging to use of available quantum devices, such as IBM quantum computers, as useful testbeds for simulating open quantum systems and quantum control protocols. As quantum devices become faster and less error-prone, and as error mitigation and suppression techniques improve, we foresee that more complex models and larger systems will be addressable.\\

\textbf{Acknowledgments}-- We acknowledge the use of IBM Quantum services for this work. MGG acknowledges support from Italian Ministry of Research and Next Generation EU via the PRIN 2022 project CONTRABASS (contract n.2022KB2JJM) and the project PE0000023-NQSTI-QMORE.
We would like to thank F. Albarelli, D. Morrone and A. Smirne for useful discussions.
S.N Elyasi would like to express his heartfelt gratitude to her late sister, Nashmin Elyasi, whose passion for research at TUM University and kind support have always been a source of inspiration in his life. Her memory continues to motivate him, and he dedicates this work to her.

%%%%%%%%%%%%%%%%%%%%%%%%%%%%%%%%%%%%%%%%%%%%%%
%%%%%%%%%%%%%%%%%%%%%%%%%%%%%%%%%%%%%%%%%%%%%%
%%%%%%%%%%%%%%%%%%%%%%%%%%%%%%%%%%%%%%%%%%%%%%
%%%%%%%%%%%%%%%%%%%%%%%%%%%%%%%%%%%%%%%%%%%%%%
%%%%%%%%%%%%%%%%%%%%%%%%%%%%%%%%%%%%%%%%%%%%%%
\textbf{Data Availability}-- 
The data and the code used to obtain the results presented in this manuscript are available from the authors upon reasonable request.
%
%
%

%%%%%%%%%%%%%%%%%%%%%%%%%%%%%%%%%%%%%%%%%%%%%%
%%%%%%%%%%%%%%%%%%%%%%%%%%%%%%%%%%%%%%%%%%%%%%
\bibliography{sample}

\begin{thebibliography}{113}%
\makeatletter
\providecommand \@ifxundefined [1]{%
 \@ifx{#1\undefined}
}%
\providecommand \@ifnum [1]{%
 \ifnum #1\expandafter \@firstoftwo
 \else \expandafter \@secondoftwo
 \fi
}%
\providecommand \@ifx [1]{%
 \ifx #1\expandafter \@firstoftwo
 \else \expandafter \@secondoftwo
 \fi
}%
\providecommand \natexlab [1]{#1}%
\providecommand \enquote  [1]{``#1''}%
\providecommand \bibnamefont  [1]{#1}%
\providecommand \bibfnamefont [1]{#1}%
\providecommand \citenamefont [1]{#1}%
\providecommand \href@noop [0]{\@secondoftwo}%
\providecommand \href [0]{\begingroup \@sanitize@url \@href}%
\providecommand \@href[1]{\@@startlink{#1}\@@href}%
\providecommand \@@href[1]{\endgroup#1\@@endlink}%
\providecommand \@sanitize@url [0]{\catcode `\\12\catcode `\$12\catcode
  `\&12\catcode `\#12\catcode `\^12\catcode `\_12\catcode `\%12\relax}%
\providecommand \@@startlink[1]{}%
\providecommand \@@endlink[0]{}%
\providecommand \url  [0]{\begingroup\@sanitize@url \@url }%
\providecommand \@url [1]{\endgroup\@href {#1}{\urlprefix }}%
\providecommand \urlprefix  [0]{URL }%
\providecommand \Eprint [0]{\href }%
\providecommand \doibase [0]{https://doi.org/}%
\providecommand \selectlanguage [0]{\@gobble}%
\providecommand \bibinfo  [0]{\@secondoftwo}%
\providecommand \bibfield  [0]{\@secondoftwo}%
\providecommand \translation [1]{[#1]}%
\providecommand \BibitemOpen [0]{}%
\providecommand \bibitemStop [0]{}%
\providecommand \bibitemNoStop [0]{.\EOS\space}%
\providecommand \EOS [0]{\spacefactor3000\relax}%
\providecommand \BibitemShut  [1]{\csname bibitem#1\endcsname}%
\let\auto@bib@innerbib\@empty
%</preamble>
\bibitem [{\citenamefont {Binder}\ \emph {et~al.}(2018)\citenamefont {Binder},
  \citenamefont {Correa}, \citenamefont {Gogolin}, \citenamefont {Anders},\
  and\ \citenamefont {Adesso}}]{binderThermodynamicsQuantumRegime2018}%
  \BibitemOpen
  \bibinfo {editor} {\bibfnamefont {F.}~\bibnamefont {Binder}}, \bibinfo
  {editor} {\bibfnamefont {L.~A.}\ \bibnamefont {Correa}}, \bibinfo {editor}
  {\bibfnamefont {C.}~\bibnamefont {Gogolin}}, \bibinfo {editor} {\bibfnamefont
  {J.}~\bibnamefont {Anders}},\ and\ \bibinfo {editor} {\bibfnamefont
  {G.}~\bibnamefont {Adesso}},\ eds.,\ \href@noop {} {\emph {\bibinfo {title}
  {Thermodynamics in the Quantum Regime}}},\ \bibinfo {series} {Fundamental
  {{Theories}} of {{Physics}}}, Vol.\ \bibinfo {volume} {195}\ (\bibinfo
  {publisher} {{Springer International Publishing}},\ \bibinfo {address}
  {{Cham}},\ \bibinfo {year} {2018})\BibitemShut {NoStop}%
\bibitem [{\citenamefont {Francica}\ \emph {et~al.}(2017)\citenamefont
  {Francica}, \citenamefont {Goold}, \citenamefont {Plastina},\ and\
  \citenamefont {Paternostro}}]{francicaDaemonicErgotropyEnhanced2017}%
  \BibitemOpen
  \bibfield  {author} {\bibinfo {author} {\bibfnamefont {G.}~\bibnamefont
  {Francica}}, \bibinfo {author} {\bibfnamefont {J.}~\bibnamefont {Goold}},
  \bibinfo {author} {\bibfnamefont {F.}~\bibnamefont {Plastina}},\ and\
  \bibinfo {author} {\bibfnamefont {M.}~\bibnamefont {Paternostro}},\
  }\bibfield  {title} {\bibinfo {title} {Daemonic ergotropy: Enhanced work
  extraction from quantum correlations},\ }\href
  {https://doi.org/10.1038/s41534-017-0012-8} {\bibfield  {journal} {\bibinfo
  {journal} {npj Quantum Information}\ }\textbf {\bibinfo {volume} {3}},\
  \bibinfo {pages} {1} (\bibinfo {year} {2017})}\BibitemShut {NoStop}%
\bibitem [{\citenamefont {Manzano}\ \emph {et~al.}(2018)\citenamefont
  {Manzano}, \citenamefont {Plastina},\ and\ \citenamefont
  {Zambrini}}]{manzanoOptimalWorkExtraction2018}%
  \BibitemOpen
  \bibfield  {author} {\bibinfo {author} {\bibfnamefont {G.}~\bibnamefont
  {Manzano}}, \bibinfo {author} {\bibfnamefont {F.}~\bibnamefont {Plastina}},\
  and\ \bibinfo {author} {\bibfnamefont {R.}~\bibnamefont {Zambrini}},\
  }\bibfield  {title} {\bibinfo {title} {Optimal {{Work Extraction}} and
  {{Thermodynamics}} of {{Quantum Measurements}} and {{Correlations}}},\ }\href
  {https://doi.org/10.1103/PhysRevLett.121.120602} {\bibfield  {journal}
  {\bibinfo  {journal} {Physical Review Letters}\ }\textbf {\bibinfo {volume}
  {121}},\ \bibinfo {pages} {120602} (\bibinfo {year} {2018})}\BibitemShut
  {NoStop}%
\bibitem [{\citenamefont {Morrone}\ \emph
  {et~al.}(2023{\natexlab{a}})\citenamefont {Morrone}, \citenamefont {Rossi},\
  and\ \citenamefont {Genoni}}]{MorronePRApp}%
  \BibitemOpen
  \bibfield  {author} {\bibinfo {author} {\bibfnamefont {D.}~\bibnamefont
  {Morrone}}, \bibinfo {author} {\bibfnamefont {M.~A.}\ \bibnamefont {Rossi}},\
  and\ \bibinfo {author} {\bibfnamefont {M.~G.}\ \bibnamefont {Genoni}},\
  }\bibfield  {title} {\bibinfo {title} {Daemonic ergotropy in continuously
  monitored open quantum batteries},\ }\href
  {https://doi.org/10.1103/PhysRevApplied.20.044073} {\bibfield  {journal}
  {\bibinfo  {journal} {Phys. Rev. Appl.}\ }\textbf {\bibinfo {volume} {20}},\
  \bibinfo {pages} {044073} (\bibinfo {year} {2023}{\natexlab{a}})}\BibitemShut
  {NoStop}%
\bibitem [{\citenamefont {Stahl}\ \emph {et~al.}(2024)\citenamefont {Stahl},
  \citenamefont {Kewming}, \citenamefont {Goold}, \citenamefont {Hilder},
  \citenamefont {Poschinger},\ and\ \citenamefont
  {Schmidt-Kaler}}]{stahl2024demonstrationenergyextractiongain}%
  \BibitemOpen
  \bibfield  {author} {\bibinfo {author} {\bibfnamefont {A.}~\bibnamefont
  {Stahl}}, \bibinfo {author} {\bibfnamefont {M.}~\bibnamefont {Kewming}},
  \bibinfo {author} {\bibfnamefont {J.}~\bibnamefont {Goold}}, \bibinfo
  {author} {\bibfnamefont {J.}~\bibnamefont {Hilder}}, \bibinfo {author}
  {\bibfnamefont {U.~G.}\ \bibnamefont {Poschinger}},\ and\ \bibinfo {author}
  {\bibfnamefont {F.}~\bibnamefont {Schmidt-Kaler}},\ }\href
  {https://arxiv.org/abs/2404.14838} {\bibinfo {title} {Demonstration of energy
  extraction gain from non-classical correlations}} (\bibinfo {year} {2024}),\
  \Eprint {https://arxiv.org/abs/2404.14838} {arXiv:2404.14838 [quant-ph]}
  \BibitemShut {NoStop}%
\bibitem [{\citenamefont {Campisi}\ \emph {et~al.}(2017)\citenamefont
  {Campisi}, \citenamefont {Pekola},\ and\ \citenamefont
  {Fazio}}]{Campisi_2017}%
  \BibitemOpen
  \bibfield  {author} {\bibinfo {author} {\bibfnamefont {M.}~\bibnamefont
  {Campisi}}, \bibinfo {author} {\bibfnamefont {J.}~\bibnamefont {Pekola}},\
  and\ \bibinfo {author} {\bibfnamefont {R.}~\bibnamefont {Fazio}},\ }\bibfield
   {title} {\bibinfo {title} {Feedback-controlled heat transport in quantum
  devices: theory and solid-state experimental proposal},\ }\href
  {https://doi.org/10.1088/1367-2630/aa6acb} {\bibfield  {journal} {\bibinfo
  {journal} {New Journal of Physics}\ }\textbf {\bibinfo {volume} {19}},\
  \bibinfo {pages} {053027} (\bibinfo {year} {2017})}\BibitemShut {NoStop}%
\bibitem [{\citenamefont {Elouard}\ \emph {et~al.}(2017)\citenamefont
  {Elouard}, \citenamefont {Herrera-Mart\'{\i}}, \citenamefont {Huard},\ and\
  \citenamefont {Auff\`eves}}]{ElouardPRL2017}%
  \BibitemOpen
  \bibfield  {author} {\bibinfo {author} {\bibfnamefont {C.}~\bibnamefont
  {Elouard}}, \bibinfo {author} {\bibfnamefont {D.}~\bibnamefont
  {Herrera-Mart\'{\i}}}, \bibinfo {author} {\bibfnamefont {B.}~\bibnamefont
  {Huard}},\ and\ \bibinfo {author} {\bibfnamefont {A.}~\bibnamefont
  {Auff\`eves}},\ }\bibfield  {title} {\bibinfo {title} {Extracting work from
  quantum measurement in maxwell's demon engines},\ }\href
  {https://doi.org/10.1103/PhysRevLett.118.260603} {\bibfield  {journal}
  {\bibinfo  {journal} {Phys. Rev. Lett.}\ }\textbf {\bibinfo {volume} {118}},\
  \bibinfo {pages} {260603} (\bibinfo {year} {2017})}\BibitemShut {NoStop}%
\bibitem [{\citenamefont {Yi}\ \emph {et~al.}(2017)\citenamefont {Yi},
  \citenamefont {Talkner},\ and\ \citenamefont {Kim}}]{YiPRE2017}%
  \BibitemOpen
  \bibfield  {author} {\bibinfo {author} {\bibfnamefont {J.}~\bibnamefont
  {Yi}}, \bibinfo {author} {\bibfnamefont {P.}~\bibnamefont {Talkner}},\ and\
  \bibinfo {author} {\bibfnamefont {Y.~W.}\ \bibnamefont {Kim}},\ }\bibfield
  {title} {\bibinfo {title} {Single-temperature quantum engine without feedback
  control},\ }\href {https://doi.org/10.1103/PhysRevE.96.022108} {\bibfield
  {journal} {\bibinfo  {journal} {Phys. Rev. E}\ }\textbf {\bibinfo {volume}
  {96}},\ \bibinfo {pages} {022108} (\bibinfo {year} {2017})}\BibitemShut
  {NoStop}%
\bibitem [{\citenamefont {Mohammady}\ and\ \citenamefont
  {Anders}(2017)}]{Mohammady_2017}%
  \BibitemOpen
  \bibfield  {author} {\bibinfo {author} {\bibfnamefont {M.~H.}\ \bibnamefont
  {Mohammady}}\ and\ \bibinfo {author} {\bibfnamefont {J.}~\bibnamefont
  {Anders}},\ }\bibfield  {title} {\bibinfo {title} {A quantum szilard engine
  without heat from a thermal reservoir},\ }\href
  {https://doi.org/10.1088/1367-2630/aa8ba1} {\bibfield  {journal} {\bibinfo
  {journal} {New Journal of Physics}\ }\textbf {\bibinfo {volume} {19}},\
  \bibinfo {pages} {113026} (\bibinfo {year} {2017})}\BibitemShut {NoStop}%
\bibitem [{\citenamefont {Buffoni}\ \emph {et~al.}(2019)\citenamefont
  {Buffoni}, \citenamefont {Solfanelli}, \citenamefont {Verrucchi},
  \citenamefont {Cuccoli},\ and\ \citenamefont {Campisi}}]{BuffoniPRL2019}%
  \BibitemOpen
  \bibfield  {author} {\bibinfo {author} {\bibfnamefont {L.}~\bibnamefont
  {Buffoni}}, \bibinfo {author} {\bibfnamefont {A.}~\bibnamefont {Solfanelli}},
  \bibinfo {author} {\bibfnamefont {P.}~\bibnamefont {Verrucchi}}, \bibinfo
  {author} {\bibfnamefont {A.}~\bibnamefont {Cuccoli}},\ and\ \bibinfo {author}
  {\bibfnamefont {M.}~\bibnamefont {Campisi}},\ }\bibfield  {title} {\bibinfo
  {title} {Quantum measurement cooling},\ }\href
  {https://doi.org/10.1103/PhysRevLett.122.070603} {\bibfield  {journal}
  {\bibinfo  {journal} {Phys. Rev. Lett.}\ }\textbf {\bibinfo {volume} {122}},\
  \bibinfo {pages} {070603} (\bibinfo {year} {2019})}\BibitemShut {NoStop}%
\bibitem [{\citenamefont {Stevens}\ \emph {et~al.}(2022)\citenamefont
  {Stevens}, \citenamefont {Szombati}, \citenamefont {Maffei}, \citenamefont
  {Elouard}, \citenamefont {Assouly}, \citenamefont {Cottet}, \citenamefont
  {Dassonneville}, \citenamefont {Ficheux}, \citenamefont {Zeppetzauer},
  \citenamefont {Bienfait}, \citenamefont {Jordan}, \citenamefont
  {Auff\`eves},\ and\ \citenamefont {Huard}}]{StevensPRL2022}%
  \BibitemOpen
  \bibfield  {author} {\bibinfo {author} {\bibfnamefont {J.}~\bibnamefont
  {Stevens}}, \bibinfo {author} {\bibfnamefont {D.}~\bibnamefont {Szombati}},
  \bibinfo {author} {\bibfnamefont {M.}~\bibnamefont {Maffei}}, \bibinfo
  {author} {\bibfnamefont {C.}~\bibnamefont {Elouard}}, \bibinfo {author}
  {\bibfnamefont {R.}~\bibnamefont {Assouly}}, \bibinfo {author} {\bibfnamefont
  {N.}~\bibnamefont {Cottet}}, \bibinfo {author} {\bibfnamefont
  {R.}~\bibnamefont {Dassonneville}}, \bibinfo {author} {\bibfnamefont
  {Q.}~\bibnamefont {Ficheux}}, \bibinfo {author} {\bibfnamefont
  {S.}~\bibnamefont {Zeppetzauer}}, \bibinfo {author} {\bibfnamefont
  {A.}~\bibnamefont {Bienfait}}, \bibinfo {author} {\bibfnamefont {A.~N.}\
  \bibnamefont {Jordan}}, \bibinfo {author} {\bibfnamefont {A.}~\bibnamefont
  {Auff\`eves}},\ and\ \bibinfo {author} {\bibfnamefont {B.}~\bibnamefont
  {Huard}},\ }\bibfield  {title} {\bibinfo {title} {Energetics of a single
  qubit gate},\ }\href {https://doi.org/10.1103/PhysRevLett.129.110601}
  {\bibfield  {journal} {\bibinfo  {journal} {Phys. Rev. Lett.}\ }\textbf
  {\bibinfo {volume} {129}},\ \bibinfo {pages} {110601} (\bibinfo {year}
  {2022})}\BibitemShut {NoStop}%
\bibitem [{\citenamefont {Yanik}\ \emph {et~al.}(2022)\citenamefont {Yanik},
  \citenamefont {Bhandari}, \citenamefont {Manikandan},\ and\ \citenamefont
  {Jordan}}]{yanikThermodynamicsQuantumMeasurement2022}%
  \BibitemOpen
  \bibfield  {author} {\bibinfo {author} {\bibfnamefont {K.}~\bibnamefont
  {Yanik}}, \bibinfo {author} {\bibfnamefont {B.}~\bibnamefont {Bhandari}},
  \bibinfo {author} {\bibfnamefont {S.~K.}\ \bibnamefont {Manikandan}},\ and\
  \bibinfo {author} {\bibfnamefont {A.~N.}\ \bibnamefont {Jordan}},\ }\bibfield
   {title} {\bibinfo {title} {Thermodynamics of quantum measurement and
  {{Maxwell}}'s demon's arrow of time},\ }\href
  {https://doi.org/10.1103/PhysRevA.106.042221} {\bibfield  {journal} {\bibinfo
   {journal} {Physical Review A}\ }\textbf {\bibinfo {volume} {106}},\ \bibinfo
  {pages} {042221} (\bibinfo {year} {2022})}\BibitemShut {NoStop}%
\bibitem [{\citenamefont {Jussiau}\ \emph {et~al.}(2023)\citenamefont
  {Jussiau}, \citenamefont {Bresque}, \citenamefont {Auff\`eves}, \citenamefont
  {Murch},\ and\ \citenamefont {Jordan}}]{JussiauPRR2023}%
  \BibitemOpen
  \bibfield  {author} {\bibinfo {author} {\bibfnamefont {E.}~\bibnamefont
  {Jussiau}}, \bibinfo {author} {\bibfnamefont {L.}~\bibnamefont {Bresque}},
  \bibinfo {author} {\bibfnamefont {A.}~\bibnamefont {Auff\`eves}}, \bibinfo
  {author} {\bibfnamefont {K.~W.}\ \bibnamefont {Murch}},\ and\ \bibinfo
  {author} {\bibfnamefont {A.~N.}\ \bibnamefont {Jordan}},\ }\bibfield  {title}
  {\bibinfo {title} {Many-body quantum vacuum fluctuation engines},\ }\href
  {https://doi.org/10.1103/PhysRevResearch.5.033122} {\bibfield  {journal}
  {\bibinfo  {journal} {Phys. Rev. Res.}\ }\textbf {\bibinfo {volume} {5}},\
  \bibinfo {pages} {033122} (\bibinfo {year} {2023})}\BibitemShut {NoStop}%
\bibitem [{\citenamefont {Linpeng}\ \emph {et~al.}(2024)\citenamefont
  {Linpeng}, \citenamefont {Piccione}, \citenamefont {Maffei}, \citenamefont
  {Bresque}, \citenamefont {Prasad}, \citenamefont {Jordan}, \citenamefont
  {Auff\`eves},\ and\ \citenamefont {Murch}}]{LinpengPRR2024}%
  \BibitemOpen
  \bibfield  {author} {\bibinfo {author} {\bibfnamefont {X.}~\bibnamefont
  {Linpeng}}, \bibinfo {author} {\bibfnamefont {N.}~\bibnamefont {Piccione}},
  \bibinfo {author} {\bibfnamefont {M.}~\bibnamefont {Maffei}}, \bibinfo
  {author} {\bibfnamefont {L.}~\bibnamefont {Bresque}}, \bibinfo {author}
  {\bibfnamefont {S.~P.}\ \bibnamefont {Prasad}}, \bibinfo {author}
  {\bibfnamefont {A.~N.}\ \bibnamefont {Jordan}}, \bibinfo {author}
  {\bibfnamefont {A.}~\bibnamefont {Auff\`eves}},\ and\ \bibinfo {author}
  {\bibfnamefont {K.~W.}\ \bibnamefont {Murch}},\ }\bibfield  {title} {\bibinfo
  {title} {Quantum energetics of a noncommuting measurement},\ }\href
  {https://doi.org/10.1103/PhysRevResearch.6.033045} {\bibfield  {journal}
  {\bibinfo  {journal} {Phys. Rev. Res.}\ }\textbf {\bibinfo {volume} {6}},\
  \bibinfo {pages} {033045} (\bibinfo {year} {2024})}\BibitemShut {NoStop}%
\bibitem [{\citenamefont {Allahverdyan}\ \emph {et~al.}(2004)\citenamefont
  {Allahverdyan}, \citenamefont {Balian},\ and\ \citenamefont
  {Nieuwenhuizen}}]{allahverdyanMaximalWorkExtraction2004}%
  \BibitemOpen
  \bibfield  {author} {\bibinfo {author} {\bibfnamefont {A.~E.}\ \bibnamefont
  {Allahverdyan}}, \bibinfo {author} {\bibfnamefont {R.}~\bibnamefont
  {Balian}},\ and\ \bibinfo {author} {\bibfnamefont {T.~M.}\ \bibnamefont
  {Nieuwenhuizen}},\ }\bibfield  {title} {\bibinfo {title} {Maximal work
  extraction from finite quantum systems},\ }\href
  {https://doi.org/10.1209/epl/i2004-10101-2} {\bibfield  {journal} {\bibinfo
  {journal} {Europhysics Letters}\ }\textbf {\bibinfo {volume} {67}},\ \bibinfo
  {pages} {565} (\bibinfo {year} {2004})}\BibitemShut {NoStop}%
\bibitem [{\citenamefont {Campaioli}\ \emph {et~al.}(2018)\citenamefont
  {Campaioli}, \citenamefont {Pollock},\ and\ \citenamefont
  {Vinjanampathy}}]{campaioliQuantumBatteriesReview2018}%
  \BibitemOpen
  \bibfield  {author} {\bibinfo {author} {\bibfnamefont {F.}~\bibnamefont
  {Campaioli}}, \bibinfo {author} {\bibfnamefont {F.~A.}\ \bibnamefont
  {Pollock}},\ and\ \bibinfo {author} {\bibfnamefont {S.}~\bibnamefont
  {Vinjanampathy}},\ }\href {https://doi.org/10.48550/arXiv.1805.05507}
  {\bibinfo {title} {Quantum {{Batteries}} - {{Review Chapter}}}} (\bibinfo
  {year} {2018})\BibitemShut {NoStop}%
\bibitem [{\citenamefont {Campaioli}\ \emph {et~al.}(2024)\citenamefont
  {Campaioli}, \citenamefont {Gherardini}, \citenamefont {Quach}, \citenamefont
  {Polini},\ and\ \citenamefont {Andolina}}]{CampaioliRMP2024}%
  \BibitemOpen
  \bibfield  {author} {\bibinfo {author} {\bibfnamefont {F.}~\bibnamefont
  {Campaioli}}, \bibinfo {author} {\bibfnamefont {S.}~\bibnamefont
  {Gherardini}}, \bibinfo {author} {\bibfnamefont {J.~Q.}\ \bibnamefont
  {Quach}}, \bibinfo {author} {\bibfnamefont {M.}~\bibnamefont {Polini}},\ and\
  \bibinfo {author} {\bibfnamefont {G.~M.}\ \bibnamefont {Andolina}},\
  }\bibfield  {title} {\bibinfo {title} {Colloquium: Quantum batteries},\
  }\href {https://doi.org/10.1103/RevModPhys.96.031001} {\bibfield  {journal}
  {\bibinfo  {journal} {Rev. Mod. Phys.}\ }\textbf {\bibinfo {volume} {96}},\
  \bibinfo {pages} {031001} (\bibinfo {year} {2024})}\BibitemShut {NoStop}%
\bibitem [{\citenamefont {Alicki}\ and\ \citenamefont
  {Fannes}(2013)}]{alickiEntanglementBoostExtractable2013}%
  \BibitemOpen
  \bibfield  {author} {\bibinfo {author} {\bibfnamefont {R.}~\bibnamefont
  {Alicki}}\ and\ \bibinfo {author} {\bibfnamefont {M.}~\bibnamefont
  {Fannes}},\ }\bibfield  {title} {\bibinfo {title} {Entanglement boost for
  extractable work from ensembles of quantum batteries},\ }\href
  {https://doi.org/10.1103/PhysRevE.87.042123} {\bibfield  {journal} {\bibinfo
  {journal} {Physical Review E}\ }\textbf {\bibinfo {volume} {87}},\ \bibinfo
  {pages} {042123} (\bibinfo {year} {2013})}\BibitemShut {NoStop}%
\bibitem [{\citenamefont {Binder}\ \emph {et~al.}(2015)\citenamefont {Binder},
  \citenamefont {Vinjanampathy}, \citenamefont {Modi},\ and\ \citenamefont
  {Goold}}]{binderQuantacellPowerfulCharging2015}%
  \BibitemOpen
  \bibfield  {author} {\bibinfo {author} {\bibfnamefont {F.~C.}\ \bibnamefont
  {Binder}}, \bibinfo {author} {\bibfnamefont {S.}~\bibnamefont
  {Vinjanampathy}}, \bibinfo {author} {\bibfnamefont {K.}~\bibnamefont
  {Modi}},\ and\ \bibinfo {author} {\bibfnamefont {J.}~\bibnamefont {Goold}},\
  }\bibfield  {title} {\bibinfo {title} {Quantacell: Powerful charging of
  quantum batteries},\ }\href {https://doi.org/10.1088/1367-2630/17/7/075015}
  {\bibfield  {journal} {\bibinfo  {journal} {New Journal of Physics}\ }\textbf
  {\bibinfo {volume} {17}},\ \bibinfo {pages} {075015} (\bibinfo {year}
  {2015})}\BibitemShut {NoStop}%
\bibitem [{\citenamefont {Campaioli}\ \emph {et~al.}(2017)\citenamefont
  {Campaioli}, \citenamefont {Pollock}, \citenamefont {Binder}, \citenamefont
  {C{\'e}leri}, \citenamefont {Goold}, \citenamefont {Vinjanampathy},\ and\
  \citenamefont {Modi}}]{campaioliEnhancingChargingPower2017}%
  \BibitemOpen
  \bibfield  {author} {\bibinfo {author} {\bibfnamefont {F.}~\bibnamefont
  {Campaioli}}, \bibinfo {author} {\bibfnamefont {F.~A.}\ \bibnamefont
  {Pollock}}, \bibinfo {author} {\bibfnamefont {F.~C.}\ \bibnamefont {Binder}},
  \bibinfo {author} {\bibfnamefont {L.}~\bibnamefont {C{\'e}leri}}, \bibinfo
  {author} {\bibfnamefont {J.}~\bibnamefont {Goold}}, \bibinfo {author}
  {\bibfnamefont {S.}~\bibnamefont {Vinjanampathy}},\ and\ \bibinfo {author}
  {\bibfnamefont {K.}~\bibnamefont {Modi}},\ }\bibfield  {title} {\bibinfo
  {title} {Enhancing the {{Charging Power}} of {{Quantum Batteries}}},\ }\href
  {https://doi.org/10.1103/PhysRevLett.118.150601} {\bibfield  {journal}
  {\bibinfo  {journal} {Physical Review Letters}\ }\textbf {\bibinfo {volume}
  {118}},\ \bibinfo {pages} {150601} (\bibinfo {year} {2017})}\BibitemShut
  {NoStop}%
\bibitem [{\citenamefont {Ferraro}\ \emph {et~al.}(2018)\citenamefont
  {Ferraro}, \citenamefont {Campisi}, \citenamefont {Andolina}, \citenamefont
  {Pellegrini},\ and\ \citenamefont
  {Polini}}]{ferraroHighPowerCollectiveCharging2018}%
  \BibitemOpen
  \bibfield  {author} {\bibinfo {author} {\bibfnamefont {D.}~\bibnamefont
  {Ferraro}}, \bibinfo {author} {\bibfnamefont {M.}~\bibnamefont {Campisi}},
  \bibinfo {author} {\bibfnamefont {G.~M.}\ \bibnamefont {Andolina}}, \bibinfo
  {author} {\bibfnamefont {V.}~\bibnamefont {Pellegrini}},\ and\ \bibinfo
  {author} {\bibfnamefont {M.}~\bibnamefont {Polini}},\ }\bibfield  {title}
  {\bibinfo {title} {High-{{Power Collective Charging}} of a {{Solid-State
  Quantum Battery}}},\ }\href {https://doi.org/10.1103/PhysRevLett.120.117702}
  {\bibfield  {journal} {\bibinfo  {journal} {Physical Review Letters}\
  }\textbf {\bibinfo {volume} {120}},\ \bibinfo {pages} {117702} (\bibinfo
  {year} {2018})}\BibitemShut {NoStop}%
\bibitem [{\citenamefont {Andolina}\ \emph {et~al.}(2019)\citenamefont
  {Andolina}, \citenamefont {Keck}, \citenamefont {Mari}, \citenamefont
  {Campisi}, \citenamefont {Giovannetti},\ and\ \citenamefont
  {Polini}}]{andolina_extractable_2019}%
  \BibitemOpen
  \bibfield  {author} {\bibinfo {author} {\bibfnamefont {G.~M.}\ \bibnamefont
  {Andolina}}, \bibinfo {author} {\bibfnamefont {M.}~\bibnamefont {Keck}},
  \bibinfo {author} {\bibfnamefont {A.}~\bibnamefont {Mari}}, \bibinfo {author}
  {\bibfnamefont {M.}~\bibnamefont {Campisi}}, \bibinfo {author} {\bibfnamefont
  {V.}~\bibnamefont {Giovannetti}},\ and\ \bibinfo {author} {\bibfnamefont
  {M.}~\bibnamefont {Polini}},\ }\bibfield  {title} {\bibinfo {title}
  {Extractable {Work}, the {Role} of {Correlations}, and {Asymptotic} {Freedom}
  in {Quantum} {Batteries}},\ }\href
  {https://doi.org/10.1103/PhysRevLett.122.047702} {\bibfield  {journal}
  {\bibinfo  {journal} {Physical Review Letters}\ }\textbf {\bibinfo {volume}
  {122}},\ \bibinfo {pages} {047702} (\bibinfo {year} {2019})}\BibitemShut
  {NoStop}%
\bibitem [{\citenamefont {Zhang}\ \emph {et~al.}(2019)\citenamefont {Zhang},
  \citenamefont {Yang}, \citenamefont {Fu},\ and\ \citenamefont
  {Wang}}]{ZhangPowerfulHarmonicCharging2019}%
  \BibitemOpen
  \bibfield  {author} {\bibinfo {author} {\bibfnamefont {Y.-Y.}\ \bibnamefont
  {Zhang}}, \bibinfo {author} {\bibfnamefont {T.-R.}\ \bibnamefont {Yang}},
  \bibinfo {author} {\bibfnamefont {L.}~\bibnamefont {Fu}},\ and\ \bibinfo
  {author} {\bibfnamefont {X.}~\bibnamefont {Wang}},\ }\bibfield  {title}
  {\bibinfo {title} {Powerful harmonic charging in a quantum battery},\ }\href
  {https://doi.org/10.1103/PhysRevE.99.052106} {\bibfield  {journal} {\bibinfo
  {journal} {Physical Review E}\ }\textbf {\bibinfo {volume} {99}},\ \bibinfo
  {pages} {052106} (\bibinfo {year} {2019})}\BibitemShut {NoStop}%
\bibitem [{\citenamefont {Crescente}\ \emph {et~al.}(2020)\citenamefont
  {Crescente}, \citenamefont {Carrega}, \citenamefont {Sassetti},\ and\
  \citenamefont {Ferraro}}]{CrescenteChargingEnergyFluctuations2020}%
  \BibitemOpen
  \bibfield  {author} {\bibinfo {author} {\bibfnamefont {A.}~\bibnamefont
  {Crescente}}, \bibinfo {author} {\bibfnamefont {M.}~\bibnamefont {Carrega}},
  \bibinfo {author} {\bibfnamefont {M.}~\bibnamefont {Sassetti}},\ and\
  \bibinfo {author} {\bibfnamefont {D.}~\bibnamefont {Ferraro}},\ }\bibfield
  {title} {\bibinfo {title} {Charging and energy fluctuations of a driven
  quantum battery},\ }\href {https://doi.org/10.1088/1367-2630/ab91fc}
  {\bibfield  {journal} {\bibinfo  {journal} {New Journal of Physics}\ }\textbf
  {\bibinfo {volume} {22}},\ \bibinfo {pages} {063057} (\bibinfo {year}
  {2020})}\BibitemShut {NoStop}%
\bibitem [{\citenamefont {Juli\`a-Farr\'e}\ \emph {et~al.}(2020)\citenamefont
  {Juli\`a-Farr\'e}, \citenamefont {Salamon}, \citenamefont {Riera},
  \citenamefont {Bera},\ and\ \citenamefont
  {Lewenstein}}]{JuliaFarreBounds2020}%
  \BibitemOpen
  \bibfield  {author} {\bibinfo {author} {\bibfnamefont {S.}~\bibnamefont
  {Juli\`a-Farr\'e}}, \bibinfo {author} {\bibfnamefont {T.}~\bibnamefont
  {Salamon}}, \bibinfo {author} {\bibfnamefont {A.}~\bibnamefont {Riera}},
  \bibinfo {author} {\bibfnamefont {M.~N.}\ \bibnamefont {Bera}},\ and\
  \bibinfo {author} {\bibfnamefont {M.}~\bibnamefont {Lewenstein}},\ }\bibfield
   {title} {\bibinfo {title} {Bounds on the capacity and power of quantum
  batteries},\ }\href {https://doi.org/10.1103/PhysRevResearch.2.023113}
  {\bibfield  {journal} {\bibinfo  {journal} {Physical Review Research}\
  }\textbf {\bibinfo {volume} {2}},\ \bibinfo {pages} {023113} (\bibinfo {year}
  {2020})}\BibitemShut {NoStop}%
\bibitem [{\citenamefont {Rossini}\ \emph {et~al.}(2020)\citenamefont
  {Rossini}, \citenamefont {Andolina}, \citenamefont {Rosa}, \citenamefont
  {Carrega},\ and\ \citenamefont
  {Polini}}]{rossiniQuantumAdvantageCharging2020}%
  \BibitemOpen
  \bibfield  {author} {\bibinfo {author} {\bibfnamefont {D.}~\bibnamefont
  {Rossini}}, \bibinfo {author} {\bibfnamefont {G.~M.}\ \bibnamefont
  {Andolina}}, \bibinfo {author} {\bibfnamefont {D.}~\bibnamefont {Rosa}},
  \bibinfo {author} {\bibfnamefont {M.}~\bibnamefont {Carrega}},\ and\ \bibinfo
  {author} {\bibfnamefont {M.}~\bibnamefont {Polini}},\ }\bibfield  {title}
  {\bibinfo {title} {Quantum {{Advantage}} in the {{Charging Process}} of
  {{Sachdev-Ye-Kitaev Batteries}}},\ }\href
  {https://doi.org/10.1103/PhysRevLett.125.236402} {\bibfield  {journal}
  {\bibinfo  {journal} {Physical Review Letters}\ }\textbf {\bibinfo {volume}
  {125}},\ \bibinfo {pages} {236402} (\bibinfo {year} {2020})}\BibitemShut
  {NoStop}%
\bibitem [{\citenamefont {Gyhm}\ \emph {et~al.}(2022)\citenamefont {Gyhm},
  \citenamefont {{\v S}afr{\'a}nek},\ and\ \citenamefont
  {Rosa}}]{gyhmQuantumChargingAdvantage2022}%
  \BibitemOpen
  \bibfield  {author} {\bibinfo {author} {\bibfnamefont {J.-Y.}\ \bibnamefont
  {Gyhm}}, \bibinfo {author} {\bibfnamefont {D.}~\bibnamefont {{\v
  S}afr{\'a}nek}},\ and\ \bibinfo {author} {\bibfnamefont {D.}~\bibnamefont
  {Rosa}},\ }\bibfield  {title} {\bibinfo {title} {Quantum {{Charging Advantage
  Cannot Be Extensive}} without {{Global Operations}}},\ }\href
  {https://doi.org/10.1103/PhysRevLett.128.140501} {\bibfield  {journal}
  {\bibinfo  {journal} {Physical Review Letters}\ }\textbf {\bibinfo {volume}
  {128}},\ \bibinfo {pages} {140501} (\bibinfo {year} {2022})}\BibitemShut
  {NoStop}%
\bibitem [{\citenamefont {Seah}\ \emph {et~al.}(2021)\citenamefont {Seah},
  \citenamefont {{Perarnau-Llobet}}, \citenamefont {Haack}, \citenamefont
  {Brunner},\ and\ \citenamefont
  {Nimmrichter}}]{seahQuantumSpeedupCollisional2021}%
  \BibitemOpen
  \bibfield  {author} {\bibinfo {author} {\bibfnamefont {S.}~\bibnamefont
  {Seah}}, \bibinfo {author} {\bibfnamefont {M.}~\bibnamefont
  {{Perarnau-Llobet}}}, \bibinfo {author} {\bibfnamefont {G.}~\bibnamefont
  {Haack}}, \bibinfo {author} {\bibfnamefont {N.}~\bibnamefont {Brunner}},\
  and\ \bibinfo {author} {\bibfnamefont {S.}~\bibnamefont {Nimmrichter}},\
  }\bibfield  {title} {\bibinfo {title} {Quantum {{Speed-Up}} in {{Collisional
  Battery Charging}}},\ }\href {https://doi.org/10.1103/PhysRevLett.127.100601}
  {\bibfield  {journal} {\bibinfo  {journal} {Physical Review Letters}\
  }\textbf {\bibinfo {volume} {127}},\ \bibinfo {pages} {100601} (\bibinfo
  {year} {2021})}\BibitemShut {NoStop}%
\bibitem [{\citenamefont {Salvia}\ \emph {et~al.}(2022)\citenamefont {Salvia},
  \citenamefont {{Perarnau-Llobet}}, \citenamefont {Haack}, \citenamefont
  {Brunner},\ and\ \citenamefont
  {Nimmrichter}}]{salviaQuantumAdvantageCharging2022}%
  \BibitemOpen
  \bibfield  {author} {\bibinfo {author} {\bibfnamefont {R.}~\bibnamefont
  {Salvia}}, \bibinfo {author} {\bibfnamefont {M.}~\bibnamefont
  {{Perarnau-Llobet}}}, \bibinfo {author} {\bibfnamefont {G.}~\bibnamefont
  {Haack}}, \bibinfo {author} {\bibfnamefont {N.}~\bibnamefont {Brunner}},\
  and\ \bibinfo {author} {\bibfnamefont {S.}~\bibnamefont {Nimmrichter}},\
  }\bibfield  {title} {\bibinfo {title} {Quantum advantage in charging cavity
  and spin batteries by repeated interactions},\ }\Eprint
  {https://arxiv.org/abs/2205.00026} {arxiv:2205.00026 [cond-mat,
  physics:quant-ph]}  (\bibinfo {year} {2022})\BibitemShut {NoStop}%
\bibitem [{\citenamefont {Landi}(2021)}]{landiBatteryChargingCollision2021}%
  \BibitemOpen
  \bibfield  {author} {\bibinfo {author} {\bibfnamefont {G.~T.}\ \bibnamefont
  {Landi}},\ }\bibfield  {title} {\bibinfo {title} {Battery {{Charging}} in
  {{Collision Models}} with {{Bayesian Risk Strategies}}},\ }\href
  {https://doi.org/10.3390/e23121627} {\bibfield  {journal} {\bibinfo
  {journal} {Entropy}\ }\textbf {\bibinfo {volume} {23}},\ \bibinfo {pages}
  {1627} (\bibinfo {year} {2021})}\BibitemShut {NoStop}%
\bibitem [{\citenamefont {Mayo}\ and\ \citenamefont
  {Roncaglia}(2022)}]{mayoCollectiveEffectsQuantum2022}%
  \BibitemOpen
  \bibfield  {author} {\bibinfo {author} {\bibfnamefont {F.}~\bibnamefont
  {Mayo}}\ and\ \bibinfo {author} {\bibfnamefont {A.~J.}\ \bibnamefont
  {Roncaglia}},\ }\bibfield  {title} {\bibinfo {title} {Collective effects and
  quantum coherence in dissipative charging of quantum batteries},\ }\href
  {https://doi.org/10.1103/PhysRevA.105.062203} {\bibfield  {journal} {\bibinfo
   {journal} {Physical Review A}\ }\textbf {\bibinfo {volume} {105}},\ \bibinfo
  {pages} {062203} (\bibinfo {year} {2022})}\BibitemShut {NoStop}%
\bibitem [{\citenamefont
  {Barra}(2022)}]{barraEfficiencyFluctuationsQuantum2022}%
  \BibitemOpen
  \bibfield  {author} {\bibinfo {author} {\bibfnamefont {F.}~\bibnamefont
  {Barra}},\ }\bibfield  {title} {\bibinfo {title} {Efficiency fluctuations in
  a quantum battery charged by a repeated interaction process},\ }\href
  {https://doi.org/10.3390/e24060820} {\bibfield  {journal} {\bibinfo
  {journal} {Entropy}\ }\textbf {\bibinfo {volume} {24}},\ \bibinfo {pages}
  {820} (\bibinfo {year} {2022})}\BibitemShut {NoStop}%
\bibitem [{\citenamefont {Rodriguez}\ \emph {et~al.}(2022)\citenamefont
  {Rodriguez}, \citenamefont {Ahmadi}, \citenamefont {Suarez}, \citenamefont
  {Mazurek}, \citenamefont {Barzanjeh},\ and\ \citenamefont
  {Horodecki}}]{rodriguezOptimalQuantumControl2022}%
  \BibitemOpen
  \bibfield  {author} {\bibinfo {author} {\bibfnamefont {R.~R.}\ \bibnamefont
  {Rodriguez}}, \bibinfo {author} {\bibfnamefont {B.}~\bibnamefont {Ahmadi}},
  \bibinfo {author} {\bibfnamefont {G.}~\bibnamefont {Suarez}}, \bibinfo
  {author} {\bibfnamefont {P.}~\bibnamefont {Mazurek}}, \bibinfo {author}
  {\bibfnamefont {S.}~\bibnamefont {Barzanjeh}},\ and\ \bibinfo {author}
  {\bibfnamefont {P.}~\bibnamefont {Horodecki}},\ }\bibfield  {title} {\bibinfo
  {title} {Optimal quantum control of charging quantum batteries},\ }\Eprint
  {https://arxiv.org/abs/2207.00094} {arxiv:2207.00094 [quant-ph]}  (\bibinfo
  {year} {2022})\BibitemShut {NoStop}%
\bibitem [{\citenamefont {Qi}\ and\ \citenamefont
  {Jing}(2021)}]{QiMagnonMediatedQuantum2021}%
  \BibitemOpen
  \bibfield  {author} {\bibinfo {author} {\bibfnamefont {S.-f.}\ \bibnamefont
  {Qi}}\ and\ \bibinfo {author} {\bibfnamefont {J.}~\bibnamefont {Jing}},\
  }\bibfield  {title} {\bibinfo {title} {Magnon-mediated quantum battery under
  systematic errors},\ }\href {https://doi.org/10.1103/PhysRevA.104.032606}
  {\bibfield  {journal} {\bibinfo  {journal} {Physical Review A}\ }\textbf
  {\bibinfo {volume} {104}},\ \bibinfo {pages} {032606} (\bibinfo {year}
  {2021})}\BibitemShut {NoStop}%
\bibitem [{\citenamefont {Crescente}\ \emph {et~al.}(2022)\citenamefont
  {Crescente}, \citenamefont {Ferraro}, \citenamefont {Carrega},\ and\
  \citenamefont {Sassetti}}]{CrescenteEnhancingCoherentEnergy2022}%
  \BibitemOpen
  \bibfield  {author} {\bibinfo {author} {\bibfnamefont {A.}~\bibnamefont
  {Crescente}}, \bibinfo {author} {\bibfnamefont {D.}~\bibnamefont {Ferraro}},
  \bibinfo {author} {\bibfnamefont {M.}~\bibnamefont {Carrega}},\ and\ \bibinfo
  {author} {\bibfnamefont {M.}~\bibnamefont {Sassetti}},\ }\bibfield  {title}
  {\bibinfo {title} {Enhancing coherent energy transfer between quantum devices
  via a mediator},\ }\href {https://doi.org/10.1103/PhysRevResearch.4.033216}
  {\bibfield  {journal} {\bibinfo  {journal} {Physical Review Research}\
  }\textbf {\bibinfo {volume} {4}},\ \bibinfo {pages} {033216} (\bibinfo {year}
  {2022})}\BibitemShut {NoStop}%
\bibitem [{\citenamefont {Mazzoncini}\ \emph {et~al.}(2023)\citenamefont
  {Mazzoncini}, \citenamefont {Cavina}, \citenamefont {Andolina}, \citenamefont
  {Erdman},\ and\ \citenamefont
  {Giovannetti}}]{mazzonciniOptimalControlMethods2023}%
  \BibitemOpen
  \bibfield  {author} {\bibinfo {author} {\bibfnamefont {F.}~\bibnamefont
  {Mazzoncini}}, \bibinfo {author} {\bibfnamefont {V.}~\bibnamefont {Cavina}},
  \bibinfo {author} {\bibfnamefont {G.~M.}\ \bibnamefont {Andolina}}, \bibinfo
  {author} {\bibfnamefont {P.~A.}\ \bibnamefont {Erdman}},\ and\ \bibinfo
  {author} {\bibfnamefont {V.}~\bibnamefont {Giovannetti}},\ }\bibfield
  {title} {\bibinfo {title} {Optimal {{Control Methods}} for {{Quantum
  Batteries}}},\ }\href {https://doi.org/10.1103/PhysRevA.107.032218}
  {\bibfield  {journal} {\bibinfo  {journal} {Physical Review A}\ }\textbf
  {\bibinfo {volume} {107}},\ \bibinfo {pages} {032218} (\bibinfo {year}
  {2023})}\BibitemShut {NoStop}%
\bibitem [{\citenamefont {Rodr{\'i}guez}\ \emph {et~al.}(2023)\citenamefont
  {Rodr{\'i}guez}, \citenamefont {Rosa},\ and\ \citenamefont
  {Olle}}]{rodriguezAIdiscoveryNewCharging2023}%
  \BibitemOpen
  \bibfield  {author} {\bibinfo {author} {\bibfnamefont {C.}~\bibnamefont
  {Rodr{\'i}guez}}, \bibinfo {author} {\bibfnamefont {D.}~\bibnamefont
  {Rosa}},\ and\ \bibinfo {author} {\bibfnamefont {J.}~\bibnamefont {Olle}},\
  }\bibfield  {title} {\bibinfo {title} {Ai-discovery of a new charging
  protocol in a micromaser quantum battery},\ }\Eprint
  {https://arxiv.org/abs/2301.09408} {arxiv:2301.09408 [quant-ph]}  (\bibinfo
  {year} {2023})\BibitemShut {NoStop}%
\bibitem [{\citenamefont {Gyhm}\ and\ \citenamefont
  {Fischer}(2024)}]{Gyhm2024}%
  \BibitemOpen
  \bibfield  {author} {\bibinfo {author} {\bibfnamefont {J.-Y.}\ \bibnamefont
  {Gyhm}}\ and\ \bibinfo {author} {\bibfnamefont {U.~R.}\ \bibnamefont
  {Fischer}},\ }\bibfield  {title} {\bibinfo {title} {{Beneficial and
  detrimental entanglement for quantum battery charging}},\ }\href
  {https://doi.org/10.1116/5.0184903} {\bibfield  {journal} {\bibinfo
  {journal} {AVS Quantum Science}\ }\textbf {\bibinfo {volume} {6}},\ \bibinfo
  {pages} {012001} (\bibinfo {year} {2024})},\ \Eprint
  {https://arxiv.org/abs/https://pubs.aip.org/avs/aqs/article-pdf/doi/10.1116/5.0184903/18703178/012001\_1\_5.0184903.pdf}
  {https://pubs.aip.org/avs/aqs/article-pdf/doi/10.1116/5.0184903/18703178/012001\_1\_5.0184903.pdf}
  \BibitemShut {NoStop}%
\bibitem [{\citenamefont {Razzoli}\ \emph {et~al.}(2024)\citenamefont
  {Razzoli}, \citenamefont {Gemme}, \citenamefont {Khomchenko}, \citenamefont
  {Sassetti}, \citenamefont {Ouerdane}, \citenamefont {Ferraro},\ and\
  \citenamefont {Benenti}}]{RazzoliQST2024}%
  \BibitemOpen
  \bibfield  {author} {\bibinfo {author} {\bibfnamefont {L.}~\bibnamefont
  {Razzoli}}, \bibinfo {author} {\bibfnamefont {G.}~\bibnamefont {Gemme}},
  \bibinfo {author} {\bibfnamefont {I.}~\bibnamefont {Khomchenko}}, \bibinfo
  {author} {\bibfnamefont {M.}~\bibnamefont {Sassetti}}, \bibinfo {author}
  {\bibfnamefont {H.}~\bibnamefont {Ouerdane}}, \bibinfo {author}
  {\bibfnamefont {D.}~\bibnamefont {Ferraro}},\ and\ \bibinfo {author}
  {\bibfnamefont {G.}~\bibnamefont {Benenti}},\ }\bibfield  {title} {\bibinfo
  {title} {Cyclic solid-state quantum battery: Thermodynamic characterization
  and quantum hardware simulation},\ }\href
  {http://iopscience.iop.org/article/10.1088/2058-9565/ad9ed4} {\bibfield
  {journal} {\bibinfo  {journal} {Quantum Science and Technology}\ } (\bibinfo
  {year} {2024})}\BibitemShut {NoStop}%
\bibitem [{\citenamefont {Hu}\ \emph {et~al.}(2022)\citenamefont {Hu},
  \citenamefont {Qiu}, \citenamefont {Souza}, \citenamefont {Yuan},
  \citenamefont {Zhou}, \citenamefont {Zhang}, \citenamefont {Chu},
  \citenamefont {Pan}, \citenamefont {Hu}, \citenamefont {Li}, \citenamefont
  {Xu}, \citenamefont {Zhong}, \citenamefont {Liu}, \citenamefont {Yan},
  \citenamefont {Tan}, \citenamefont {Bachelard}, \citenamefont {Villas-Boas},
  \citenamefont {Santos},\ and\ \citenamefont {Yu}}]{hu2022optimal}%
  \BibitemOpen
  \bibfield  {author} {\bibinfo {author} {\bibfnamefont {C.-K.}\ \bibnamefont
  {Hu}}, \bibinfo {author} {\bibfnamefont {J.}~\bibnamefont {Qiu}}, \bibinfo
  {author} {\bibfnamefont {P.~J.~P.}\ \bibnamefont {Souza}}, \bibinfo {author}
  {\bibfnamefont {J.}~\bibnamefont {Yuan}}, \bibinfo {author} {\bibfnamefont
  {Y.}~\bibnamefont {Zhou}}, \bibinfo {author} {\bibfnamefont {L.}~\bibnamefont
  {Zhang}}, \bibinfo {author} {\bibfnamefont {J.}~\bibnamefont {Chu}}, \bibinfo
  {author} {\bibfnamefont {X.}~\bibnamefont {Pan}}, \bibinfo {author}
  {\bibfnamefont {L.}~\bibnamefont {Hu}}, \bibinfo {author} {\bibfnamefont
  {J.}~\bibnamefont {Li}}, \bibinfo {author} {\bibfnamefont {Y.}~\bibnamefont
  {Xu}}, \bibinfo {author} {\bibfnamefont {Y.}~\bibnamefont {Zhong}}, \bibinfo
  {author} {\bibfnamefont {S.}~\bibnamefont {Liu}}, \bibinfo {author}
  {\bibfnamefont {F.}~\bibnamefont {Yan}}, \bibinfo {author} {\bibfnamefont
  {D.}~\bibnamefont {Tan}}, \bibinfo {author} {\bibfnamefont {R.}~\bibnamefont
  {Bachelard}}, \bibinfo {author} {\bibfnamefont {C.~J.}\ \bibnamefont
  {Villas-Boas}}, \bibinfo {author} {\bibfnamefont {A.~C.}\ \bibnamefont
  {Santos}},\ and\ \bibinfo {author} {\bibfnamefont {D.}~\bibnamefont {Yu}},\
  }\bibfield  {title} {\bibinfo {title} {Optimal charging of a superconducting
  quantum battery},\ }\href {https://doi.org/10.1088/2058-9565/ac8444}
  {\bibfield  {journal} {\bibinfo  {journal} {Quantum Science and Technology}\
  }\textbf {\bibinfo {volume} {7}},\ \bibinfo {pages} {045018} (\bibinfo {year}
  {2022})}\BibitemShut {NoStop}%
\bibitem [{\citenamefont {Quach}\ \emph {et~al.}(2022)\citenamefont {Quach},
  \citenamefont {McGhee}, \citenamefont {Ganzer}, \citenamefont {Rouse},
  \citenamefont {Lovett}, \citenamefont {Gauger}, \citenamefont {Keeling},
  \citenamefont {Cerullo}, \citenamefont {Lidzey},\ and\ \citenamefont
  {Virgili}}]{quach2022superabsorption}%
  \BibitemOpen
  \bibfield  {author} {\bibinfo {author} {\bibfnamefont {J.~Q.}\ \bibnamefont
  {Quach}}, \bibinfo {author} {\bibfnamefont {K.~E.}\ \bibnamefont {McGhee}},
  \bibinfo {author} {\bibfnamefont {L.}~\bibnamefont {Ganzer}}, \bibinfo
  {author} {\bibfnamefont {D.~M.}\ \bibnamefont {Rouse}}, \bibinfo {author}
  {\bibfnamefont {B.~W.}\ \bibnamefont {Lovett}}, \bibinfo {author}
  {\bibfnamefont {E.~M.}\ \bibnamefont {Gauger}}, \bibinfo {author}
  {\bibfnamefont {J.}~\bibnamefont {Keeling}}, \bibinfo {author} {\bibfnamefont
  {G.}~\bibnamefont {Cerullo}}, \bibinfo {author} {\bibfnamefont {D.~G.}\
  \bibnamefont {Lidzey}},\ and\ \bibinfo {author} {\bibfnamefont
  {T.}~\bibnamefont {Virgili}},\ }\bibfield  {title} {\bibinfo {title}
  {Superabsorption in an organic microcavity: {{Toward}} a quantum battery},\
  }\href {https://doi.org/10.1126/sciadv.abk3160} {\bibfield  {journal}
  {\bibinfo  {journal} {Science Advances}\ }\textbf {\bibinfo {volume} {8}},\
  \bibinfo {pages} {eabk3160} (\bibinfo {year} {2022})}\BibitemShut {NoStop}%
\bibitem [{\citenamefont {Farina}\ \emph {et~al.}(2019)\citenamefont {Farina},
  \citenamefont {Andolina}, \citenamefont {Mari}, \citenamefont {Polini},\ and\
  \citenamefont {Giovannetti}}]{farinaChargermediatedEnergyTransfer2019}%
  \BibitemOpen
  \bibfield  {author} {\bibinfo {author} {\bibfnamefont {D.}~\bibnamefont
  {Farina}}, \bibinfo {author} {\bibfnamefont {G.~M.}\ \bibnamefont
  {Andolina}}, \bibinfo {author} {\bibfnamefont {A.}~\bibnamefont {Mari}},
  \bibinfo {author} {\bibfnamefont {M.}~\bibnamefont {Polini}},\ and\ \bibinfo
  {author} {\bibfnamefont {V.}~\bibnamefont {Giovannetti}},\ }\bibfield
  {title} {\bibinfo {title} {Charger-mediated energy transfer for quantum
  batteries: An open system approach},\ }\href
  {https://doi.org/10.1103/PhysRevB.99.035421} {\bibfield  {journal} {\bibinfo
  {journal} {Physical Review B}\ }\textbf {\bibinfo {volume} {99}},\ \bibinfo
  {pages} {035421} (\bibinfo {year} {2019})}\BibitemShut {NoStop}%
\bibitem [{\citenamefont {Morrone}\ \emph
  {et~al.}(2023{\natexlab{b}})\citenamefont {Morrone}, \citenamefont {Rossi},
  \citenamefont {Smirne},\ and\ \citenamefont
  {Genoni}}]{morroneChargingQuantumBattery2023}%
  \BibitemOpen
  \bibfield  {author} {\bibinfo {author} {\bibfnamefont {D.}~\bibnamefont
  {Morrone}}, \bibinfo {author} {\bibfnamefont {M.~A.~C.}\ \bibnamefont
  {Rossi}}, \bibinfo {author} {\bibfnamefont {A.}~\bibnamefont {Smirne}},\ and\
  \bibinfo {author} {\bibfnamefont {M.~G.}\ \bibnamefont {Genoni}},\ }\bibfield
   {title} {\bibinfo {title} {Charging a quantum battery in a non-{{Markovian}}
  environment: A collisional model approach},\ }\href
  {https://doi.org/10.1088/2058-9565/accca4} {\bibfield  {journal} {\bibinfo
  {journal} {Quantum Science and Technology}\ }\textbf {\bibinfo {volume}
  {8}},\ \bibinfo {pages} {035007} (\bibinfo {year}
  {2023}{\natexlab{b}})}\BibitemShut {NoStop}%
\bibitem [{\citenamefont {Santos}\ \emph {et~al.}(2019)\citenamefont {Santos},
  \citenamefont {{\c C}akmak}, \citenamefont {Campbell},\ and\ \citenamefont
  {Zinner}}]{santosStableAdiabaticQuantum2019}%
  \BibitemOpen
  \bibfield  {author} {\bibinfo {author} {\bibfnamefont {A.~C.}\ \bibnamefont
  {Santos}}, \bibinfo {author} {\bibfnamefont {B.}~\bibnamefont {{\c C}akmak}},
  \bibinfo {author} {\bibfnamefont {S.}~\bibnamefont {Campbell}},\ and\
  \bibinfo {author} {\bibfnamefont {N.~T.}\ \bibnamefont {Zinner}},\ }\bibfield
   {title} {\bibinfo {title} {Stable adiabatic quantum batteries},\ }\href
  {https://doi.org/10.1103/PhysRevE.100.032107} {\bibfield  {journal} {\bibinfo
   {journal} {Physical Review E}\ }\textbf {\bibinfo {volume} {100}},\ \bibinfo
  {pages} {032107} (\bibinfo {year} {2019})}\BibitemShut {NoStop}%
\bibitem [{\citenamefont {Rodríguez}\ \emph {et~al.}(2024)\citenamefont
  {Rodríguez}, \citenamefont {Ahmadi}, \citenamefont {Suárez}, \citenamefont
  {Mazurek}, \citenamefont {Barzanjeh},\ and\ \citenamefont
  {Horodecki}}]{rodriguez2022optimal}%
  \BibitemOpen
  \bibfield  {author} {\bibinfo {author} {\bibfnamefont {R.~R.}\ \bibnamefont
  {Rodríguez}}, \bibinfo {author} {\bibfnamefont {B.}~\bibnamefont {Ahmadi}},
  \bibinfo {author} {\bibfnamefont {G.}~\bibnamefont {Suárez}}, \bibinfo
  {author} {\bibfnamefont {P.}~\bibnamefont {Mazurek}}, \bibinfo {author}
  {\bibfnamefont {S.}~\bibnamefont {Barzanjeh}},\ and\ \bibinfo {author}
  {\bibfnamefont {P.}~\bibnamefont {Horodecki}},\ }\bibfield  {title} {\bibinfo
  {title} {Optimal quantum control of charging quantum batteries},\ }\href
  {https://doi.org/10.1088/1367-2630/ad3843} {\bibfield  {journal} {\bibinfo
  {journal} {New Journal of Physics}\ }\textbf {\bibinfo {volume} {26}},\
  \bibinfo {pages} {043004} (\bibinfo {year} {2024})}\BibitemShut {NoStop}%
\bibitem [{\citenamefont {Gherardini}\ \emph {et~al.}(2020)\citenamefont
  {Gherardini}, \citenamefont {Campaioli}, \citenamefont {Caruso},\ and\
  \citenamefont {Binder}}]{gherardini2020stabilizing}%
  \BibitemOpen
  \bibfield  {author} {\bibinfo {author} {\bibfnamefont {S.}~\bibnamefont
  {Gherardini}}, \bibinfo {author} {\bibfnamefont {F.}~\bibnamefont
  {Campaioli}}, \bibinfo {author} {\bibfnamefont {F.}~\bibnamefont {Caruso}},\
  and\ \bibinfo {author} {\bibfnamefont {F.~C.}\ \bibnamefont {Binder}},\
  }\bibfield  {title} {\bibinfo {title} {Stabilizing open quantum batteries by
  sequential measurements},\ }\href
  {https://doi.org/10.1103/PhysRevResearch.2.013095} {\bibfield  {journal}
  {\bibinfo  {journal} {Phys. Rev. Res.}\ }\textbf {\bibinfo {volume} {2}},\
  \bibinfo {pages} {013095} (\bibinfo {year} {2020})}\BibitemShut {NoStop}%
\bibitem [{\citenamefont {Mitchison}\ \emph {et~al.}(2021)\citenamefont
  {Mitchison}, \citenamefont {Goold},\ and\ \citenamefont
  {Prior}}]{mitchisonChargingQuantumBattery2021}%
  \BibitemOpen
  \bibfield  {author} {\bibinfo {author} {\bibfnamefont {M.~T.}\ \bibnamefont
  {Mitchison}}, \bibinfo {author} {\bibfnamefont {J.}~\bibnamefont {Goold}},\
  and\ \bibinfo {author} {\bibfnamefont {J.}~\bibnamefont {Prior}},\ }\bibfield
   {title} {\bibinfo {title} {Charging a quantum battery with linear feedback
  control},\ }\href {https://doi.org/10.22331/q-2021-07-13-500} {\bibfield
  {journal} {\bibinfo  {journal} {Quantum}\ }\textbf {\bibinfo {volume} {5}},\
  \bibinfo {pages} {500} (\bibinfo {year} {2021})}\BibitemShut {NoStop}%
\bibitem [{\citenamefont {Yao}\ and\ \citenamefont
  {Shao}(2022)}]{yaoOptimalChargingOpen2022a}%
  \BibitemOpen
  \bibfield  {author} {\bibinfo {author} {\bibfnamefont {Y.}~\bibnamefont
  {Yao}}\ and\ \bibinfo {author} {\bibfnamefont {X.~Q.}\ \bibnamefont {Shao}},\
  }\bibfield  {title} {\bibinfo {title} {Optimal charging of open spin-chain
  quantum batteries via homodyne-based feedback control},\ }\href
  {https://doi.org/10.1103/PhysRevE.106.014138} {\bibfield  {journal} {\bibinfo
   {journal} {Physical Review E}\ }\textbf {\bibinfo {volume} {106}},\ \bibinfo
  {pages} {014138} (\bibinfo {year} {2022})}\BibitemShut {NoStop}%
\bibitem [{\citenamefont {Ahmadi}\ \emph {et~al.}(2024)\citenamefont {Ahmadi},
  \citenamefont {Mazurek}, \citenamefont {Horodecki},\ and\ \citenamefont
  {Barzanjeh}}]{Ahmadi2024}%
  \BibitemOpen
  \bibfield  {author} {\bibinfo {author} {\bibfnamefont {B.}~\bibnamefont
  {Ahmadi}}, \bibinfo {author} {\bibfnamefont {P.}~\bibnamefont {Mazurek}},
  \bibinfo {author} {\bibfnamefont {P.}~\bibnamefont {Horodecki}},\ and\
  \bibinfo {author} {\bibfnamefont {S.}~\bibnamefont {Barzanjeh}},\ }\bibfield
  {title} {\bibinfo {title} {Nonreciprocal quantum batteries},\ }\href
  {https://doi.org/10.1103/PhysRevLett.132.210402} {\bibfield  {journal}
  {\bibinfo  {journal} {Phys. Rev. Lett.}\ }\textbf {\bibinfo {volume} {132}},\
  \bibinfo {pages} {210402} (\bibinfo {year} {2024})}\BibitemShut {NoStop}%
\bibitem [{\citenamefont {Lu}\ \emph {et~al.}(2024)\citenamefont {Lu},
  \citenamefont {Tian}, \citenamefont {Lü},\ and\ \citenamefont
  {Shang}}]{Lu2024}%
  \BibitemOpen
  \bibfield  {author} {\bibinfo {author} {\bibfnamefont {Z.-G.}\ \bibnamefont
  {Lu}}, \bibinfo {author} {\bibfnamefont {G.}~\bibnamefont {Tian}}, \bibinfo
  {author} {\bibfnamefont {X.-Y.}\ \bibnamefont {Lü}},\ and\ \bibinfo {author}
  {\bibfnamefont {C.}~\bibnamefont {Shang}},\ }\href
  {https://arxiv.org/abs/2405.03675} {\bibinfo {title} {Topological quantum
  batteries}} (\bibinfo {year} {2024}),\ \Eprint
  {https://arxiv.org/abs/2405.03675} {arXiv:2405.03675 [quant-ph]} \BibitemShut
  {NoStop}%
\bibitem [{\citenamefont {Wiseman}\ and\ \citenamefont
  {Milburn}(2009)}]{wisemanQuantumMeasurementControl2009}%
  \BibitemOpen
  \bibfield  {author} {\bibinfo {author} {\bibfnamefont {H.~M.}\ \bibnamefont
  {Wiseman}}\ and\ \bibinfo {author} {\bibfnamefont {G.~J.}\ \bibnamefont
  {Milburn}},\ }\href {https://doi.org/10.1017/CBO9780511813948} {\emph
  {\bibinfo {title} {Quantum Measurement and Control}}}\ (\bibinfo  {publisher}
  {{Cambridge University Press}},\ \bibinfo {address} {{Cambridge}},\ \bibinfo
  {year} {2009})\BibitemShut {NoStop}%
\bibitem [{\citenamefont {Albarelli}\ and\ \citenamefont
  {Genoni}(2024)}]{AlbarelliPLA2024}%
  \BibitemOpen
  \bibfield  {author} {\bibinfo {author} {\bibfnamefont {F.}~\bibnamefont
  {Albarelli}}\ and\ \bibinfo {author} {\bibfnamefont {M.~G.}\ \bibnamefont
  {Genoni}},\ }\bibfield  {title} {\bibinfo {title} {A pedagogical introduction
  to continuously monitored quantum systems and measurement-based feedback},\
  }\href {https://doi.org/https://doi.org/10.1016/j.physleta.2023.129260}
  {\bibfield  {journal} {\bibinfo  {journal} {Physics Letters A}\ }\textbf
  {\bibinfo {volume} {494}},\ \bibinfo {pages} {129260} (\bibinfo {year}
  {2024})}\BibitemShut {NoStop}%
\bibitem [{\citenamefont {Wiseman}\ and\ \citenamefont
  {Milburn}(1993)}]{wisemanQuantumTheoryOptical1993}%
  \BibitemOpen
  \bibfield  {author} {\bibinfo {author} {\bibfnamefont {H.~M.}\ \bibnamefont
  {Wiseman}}\ and\ \bibinfo {author} {\bibfnamefont {G.~J.}\ \bibnamefont
  {Milburn}},\ }\bibfield  {title} {\bibinfo {title} {Quantum theory of optical
  feedback via homodyne detection},\ }\href
  {https://doi.org/10.1103/PhysRevLett.70.548} {\bibfield  {journal} {\bibinfo
  {journal} {Physical Review Letters}\ }\textbf {\bibinfo {volume} {70}},\
  \bibinfo {pages} {548} (\bibinfo {year} {1993})}\BibitemShut {NoStop}%
\bibitem [{\citenamefont {Doherty}\ and\ \citenamefont
  {Jacobs}(1999)}]{dohertyFeedbackControlQuantum1999}%
  \BibitemOpen
  \bibfield  {author} {\bibinfo {author} {\bibfnamefont {A.~C.}\ \bibnamefont
  {Doherty}}\ and\ \bibinfo {author} {\bibfnamefont {K.}~\bibnamefont
  {Jacobs}},\ }\bibfield  {title} {\bibinfo {title} {Feedback control of
  quantum systems using continuous state estimation},\ }\href
  {https://doi.org/10.1103/PhysRevA.60.2700} {\bibfield  {journal} {\bibinfo
  {journal} {Physical Review A}\ }\textbf {\bibinfo {volume} {60}},\ \bibinfo
  {pages} {2700} (\bibinfo {year} {1999})}\BibitemShut {NoStop}%
\bibitem [{\citenamefont {Thomsen}\ \emph {et~al.}(2002)\citenamefont
  {Thomsen}, \citenamefont {Mancini},\ and\ \citenamefont
  {Wiseman}}]{thomsenSpinSqueezingQuantum2002}%
  \BibitemOpen
  \bibfield  {author} {\bibinfo {author} {\bibfnamefont {L.~K.}\ \bibnamefont
  {Thomsen}}, \bibinfo {author} {\bibfnamefont {S.}~\bibnamefont {Mancini}},\
  and\ \bibinfo {author} {\bibfnamefont {H.~M.}\ \bibnamefont {Wiseman}},\
  }\bibfield  {title} {\bibinfo {title} {Spin squeezing via quantum feedback},\
  }\href {https://doi.org/10.1103/PhysRevA.65.061801} {\bibfield  {journal}
  {\bibinfo  {journal} {Physical Review A}\ }\textbf {\bibinfo {volume} {65}},\
  \bibinfo {pages} {061801} (\bibinfo {year} {2002})}\BibitemShut {NoStop}%
\bibitem [{\citenamefont {Serafini}\ and\ \citenamefont
  {Mancini}(2010)}]{serafiniDeterminationMaximalGaussian2010}%
  \BibitemOpen
  \bibfield  {author} {\bibinfo {author} {\bibfnamefont {A.}~\bibnamefont
  {Serafini}}\ and\ \bibinfo {author} {\bibfnamefont {S.}~\bibnamefont
  {Mancini}},\ }\bibfield  {title} {\bibinfo {title} {Determination of
  {{Maximal Gaussian Entanglement Achievable}} by {{Feedback-Controlled
  Dynamics}}},\ }\href {https://doi.org/10.1103/PhysRevLett.104.220501}
  {\bibfield  {journal} {\bibinfo  {journal} {Physical Review Letters}\
  }\textbf {\bibinfo {volume} {104}},\ \bibinfo {pages} {220501} (\bibinfo
  {year} {2010})}\BibitemShut {NoStop}%
\bibitem [{\citenamefont {Genoni}\ \emph {et~al.}(2013)\citenamefont {Genoni},
  \citenamefont {Mancini},\ and\ \citenamefont
  {Serafini}}]{genoniOptimalFeedbackControl2013}%
  \BibitemOpen
  \bibfield  {author} {\bibinfo {author} {\bibfnamefont {M.~G.}\ \bibnamefont
  {Genoni}}, \bibinfo {author} {\bibfnamefont {S.}~\bibnamefont {Mancini}},\
  and\ \bibinfo {author} {\bibfnamefont {A.}~\bibnamefont {Serafini}},\
  }\bibfield  {title} {\bibinfo {title} {Optimal feedback control of linear
  quantum systems in the presence of thermal noise},\ }\href
  {https://doi.org/10.1103/PhysRevA.87.042333} {\bibfield  {journal} {\bibinfo
  {journal} {Physical Review A}\ }\textbf {\bibinfo {volume} {87}},\ \bibinfo
  {pages} {042333} (\bibinfo {year} {2013})}\BibitemShut {NoStop}%
\bibitem [{\citenamefont {Genoni}\ \emph {et~al.}(2015)\citenamefont {Genoni},
  \citenamefont {Zhang}, \citenamefont {Millen}, \citenamefont {Barker},\ and\
  \citenamefont {Serafini}}]{genoniQuantumCoolingSqueezing2015}%
  \BibitemOpen
  \bibfield  {author} {\bibinfo {author} {\bibfnamefont {M.~G.}\ \bibnamefont
  {Genoni}}, \bibinfo {author} {\bibfnamefont {J.}~\bibnamefont {Zhang}},
  \bibinfo {author} {\bibfnamefont {J.}~\bibnamefont {Millen}}, \bibinfo
  {author} {\bibfnamefont {P.~F.}\ \bibnamefont {Barker}},\ and\ \bibinfo
  {author} {\bibfnamefont {A.}~\bibnamefont {Serafini}},\ }\bibfield  {title}
  {\bibinfo {title} {Quantum cooling and squeezing of a levitating nanosphere
  via time-continuous measurements},\ }\href
  {https://doi.org/10.1088/1367-2630/17/7/073019} {\bibfield  {journal}
  {\bibinfo  {journal} {New Journal of Physics}\ }\textbf {\bibinfo {volume}
  {17}},\ \bibinfo {pages} {073019} (\bibinfo {year} {2015})}\BibitemShut
  {NoStop}%
\bibitem [{\citenamefont {Brunelli}\ \emph {et~al.}(2019)\citenamefont
  {Brunelli}, \citenamefont {Malz},\ and\ \citenamefont
  {Nunnenkamp}}]{brunelliConditionalDynamicsOptomechanical2019}%
  \BibitemOpen
  \bibfield  {author} {\bibinfo {author} {\bibfnamefont {M.}~\bibnamefont
  {Brunelli}}, \bibinfo {author} {\bibfnamefont {D.}~\bibnamefont {Malz}},\
  and\ \bibinfo {author} {\bibfnamefont {A.}~\bibnamefont {Nunnenkamp}},\
  }\bibfield  {title} {\bibinfo {title} {Conditional {{Dynamics}} of
  {{Optomechanical Two-Tone Backaction-Evading Measurements}}},\ }\href
  {https://doi.org/10.1103/PhysRevLett.123.093602} {\bibfield  {journal}
  {\bibinfo  {journal} {Physical Review Letters}\ }\textbf {\bibinfo {volume}
  {123}},\ \bibinfo {pages} {093602} (\bibinfo {year} {2019})}\BibitemShut
  {NoStop}%
\bibitem [{\citenamefont {Di~Giovanni}\ \emph {et~al.}(2021)\citenamefont
  {Di~Giovanni}, \citenamefont {Brunelli},\ and\ \citenamefont
  {Genoni}}]{digiovanniUnconditionalMechanicalSqueezing2021}%
  \BibitemOpen
  \bibfield  {author} {\bibinfo {author} {\bibfnamefont {A.}~\bibnamefont
  {Di~Giovanni}}, \bibinfo {author} {\bibfnamefont {M.}~\bibnamefont
  {Brunelli}},\ and\ \bibinfo {author} {\bibfnamefont {M.~G.}\ \bibnamefont
  {Genoni}},\ }\bibfield  {title} {\bibinfo {title} {Unconditional mechanical
  squeezing via backaction-evading measurements and nonoptimal feedback
  control},\ }\href {https://doi.org/10.1103/PhysRevA.103.022614} {\bibfield
  {journal} {\bibinfo  {journal} {Physical Review A}\ }\textbf {\bibinfo
  {volume} {103}},\ \bibinfo {pages} {022614} (\bibinfo {year}
  {2021})}\BibitemShut {NoStop}%
\bibitem [{\citenamefont {Candeloro}\ \emph {et~al.}(2023)\citenamefont
  {Candeloro}, \citenamefont {Benedetti}, \citenamefont {Genoni},\ and\
  \citenamefont {Paris}}]{candeloroFeedbackAssistedQuantumSearch2023}%
  \BibitemOpen
  \bibfield  {author} {\bibinfo {author} {\bibfnamefont {A.}~\bibnamefont
  {Candeloro}}, \bibinfo {author} {\bibfnamefont {C.}~\bibnamefont
  {Benedetti}}, \bibinfo {author} {\bibfnamefont {M.~G.}\ \bibnamefont
  {Genoni}},\ and\ \bibinfo {author} {\bibfnamefont {M.~G.~A.}\ \bibnamefont
  {Paris}},\ }\bibfield  {title} {\bibinfo {title} {Feedback-{{Assisted Quantum
  Search}} by {{Continuous-Time Quantum Walks}}},\ }\href
  {https://doi.org/10.1002/qute.202200093} {\bibfield  {journal} {\bibinfo
  {journal} {Advanced Quantum Technologies}\ }\textbf {\bibinfo {volume} {6}},\
  \bibinfo {pages} {2200093} (\bibinfo {year} {2023})}\BibitemShut {NoStop}%
\bibitem [{\citenamefont {Isaksen}\ and\ \citenamefont
  {Andersen}(2023)}]{isaksenMechanicalCoolingSqueezing2023}%
  \BibitemOpen
  \bibfield  {author} {\bibinfo {author} {\bibfnamefont {F.~W.}\ \bibnamefont
  {Isaksen}}\ and\ \bibinfo {author} {\bibfnamefont {U.~L.}\ \bibnamefont
  {Andersen}},\ }\bibfield  {title} {\bibinfo {title} {Mechanical cooling and
  squeezing using optimal control},\ }\href
  {https://doi.org/10.1103/PhysRevA.107.023512} {\bibfield  {journal} {\bibinfo
   {journal} {Physical Review A}\ }\textbf {\bibinfo {volume} {107}},\ \bibinfo
  {pages} {023512} (\bibinfo {year} {2023})}\BibitemShut {NoStop}%
\bibitem [{\citenamefont {Gu{\c t}{\u a}}\ \emph {et~al.}(2008)\citenamefont
  {Gu{\c t}{\u a}}, \citenamefont {Janssens},\ and\ \citenamefont
  {Kahn}}]{gutaOptimalEstimationQubit2008}%
  \BibitemOpen
  \bibfield  {author} {\bibinfo {author} {\bibfnamefont {M.}~\bibnamefont
  {Gu{\c t}{\u a}}}, \bibinfo {author} {\bibfnamefont {B.}~\bibnamefont
  {Janssens}},\ and\ \bibinfo {author} {\bibfnamefont {J.}~\bibnamefont
  {Kahn}},\ }\bibfield  {title} {\bibinfo {title} {Optimal {{Estimation}} of
  {{Qubit States}} with {{Continuous Time Measurements}}},\ }\href
  {https://doi.org/10.1007/s00220-007-0357-5} {\bibfield  {journal} {\bibinfo
  {journal} {Communications in Mathematical Physics}\ }\textbf {\bibinfo
  {volume} {277}},\ \bibinfo {pages} {127} (\bibinfo {year}
  {2008})}\BibitemShut {NoStop}%
\bibitem [{\citenamefont {Tsang}(2010)}]{tsangOptimalWaveformEstimation2010}%
  \BibitemOpen
  \bibfield  {author} {\bibinfo {author} {\bibfnamefont {M.}~\bibnamefont
  {Tsang}},\ }\bibfield  {title} {\bibinfo {title} {Optimal waveform estimation
  for classical and quantum systems via time-symmetric smoothing. ii.
  applications to atomic magnetometry and hardy's paradox},\ }\href
  {https://doi.org/10.1103/PhysRevA.81.013824} {\bibfield  {journal} {\bibinfo
  {journal} {Phys. Rev. A}\ }\textbf {\bibinfo {volume} {81}},\ \bibinfo
  {pages} {013824} (\bibinfo {year} {2010})}\BibitemShut {NoStop}%
\bibitem [{\citenamefont {Tsang}(2013)}]{tsangQuantumMetrologyOpen2013}%
  \BibitemOpen
  \bibfield  {author} {\bibinfo {author} {\bibfnamefont {M.}~\bibnamefont
  {Tsang}},\ }\bibfield  {title} {\bibinfo {title} {Quantum metrology with open
  dynamical systems},\ }\href {https://doi.org/10.1088/1367-2630/15/7/073005}
  {\bibfield  {journal} {\bibinfo  {journal} {New Journal of Physics}\ }\textbf
  {\bibinfo {volume} {15}},\ \bibinfo {pages} {073005} (\bibinfo {year}
  {2013})}\BibitemShut {NoStop}%
\bibitem [{\citenamefont {Gammelmark}\ and\ \citenamefont
  {M{\o}lmer}(2013)}]{gammelmarkBayesianParameterInference2013}%
  \BibitemOpen
  \bibfield  {author} {\bibinfo {author} {\bibfnamefont {S.}~\bibnamefont
  {Gammelmark}}\ and\ \bibinfo {author} {\bibfnamefont {K.}~\bibnamefont
  {M{\o}lmer}},\ }\bibfield  {title} {\bibinfo {title} {Bayesian parameter
  inference from continuously monitored quantum systems},\ }\href
  {https://doi.org/10.1103/PhysRevA.87.032115} {\bibfield  {journal} {\bibinfo
  {journal} {Physical Review A}\ }\textbf {\bibinfo {volume} {87}},\ \bibinfo
  {pages} {032115} (\bibinfo {year} {2013})}\BibitemShut {NoStop}%
\bibitem [{\citenamefont {Gammelmark}\ and\ \citenamefont
  {M{\o}lmer}(2014)}]{gammelmarkFisherInformationQuantum2014}%
  \BibitemOpen
  \bibfield  {author} {\bibinfo {author} {\bibfnamefont {S.}~\bibnamefont
  {Gammelmark}}\ and\ \bibinfo {author} {\bibfnamefont {K.}~\bibnamefont
  {M{\o}lmer}},\ }\bibfield  {title} {\bibinfo {title} {Fisher {{Information}}
  and the {{Quantum Cram}}\textbackslash 'er-{{Rao Sensitivity Limit}} of
  {{Continuous Measurements}}},\ }\href
  {https://doi.org/10.1103/PhysRevLett.112.170401} {\bibfield  {journal}
  {\bibinfo  {journal} {Physical Review Letters}\ }\textbf {\bibinfo {volume}
  {112}},\ \bibinfo {pages} {170401} (\bibinfo {year} {2014})}\BibitemShut
  {NoStop}%
\bibitem [{\citenamefont {Ralph}\ \emph {et~al.}(2017)\citenamefont {Ralph},
  \citenamefont {Maskell},\ and\ \citenamefont
  {Jacobs}}]{ralphMultiparameterEstimationQuantum2017}%
  \BibitemOpen
  \bibfield  {author} {\bibinfo {author} {\bibfnamefont {J.~F.}\ \bibnamefont
  {Ralph}}, \bibinfo {author} {\bibfnamefont {S.}~\bibnamefont {Maskell}},\
  and\ \bibinfo {author} {\bibfnamefont {K.}~\bibnamefont {Jacobs}},\
  }\bibfield  {title} {\bibinfo {title} {Multiparameter estimation along
  quantum trajectories with sequential {{Monte Carlo}} methods},\ }\href
  {https://doi.org/10.1103/PhysRevA.96.052306} {\bibfield  {journal} {\bibinfo
  {journal} {Physical Review A}\ }\textbf {\bibinfo {volume} {96}},\ \bibinfo
  {pages} {052306} (\bibinfo {year} {2017})}\BibitemShut {NoStop}%
\bibitem [{\citenamefont {Genoni}(2017)}]{genoniCramErRaoBound2017}%
  \BibitemOpen
  \bibfield  {author} {\bibinfo {author} {\bibfnamefont {M.~G.}\ \bibnamefont
  {Genoni}},\ }\bibfield  {title} {\bibinfo {title} {Cram\textbackslash
  'er-{{Rao}} bound for time-continuous measurements in linear {{Gaussian}}
  quantum systems},\ }\href {https://doi.org/10.1103/PhysRevA.95.012116}
  {\bibfield  {journal} {\bibinfo  {journal} {Physical Review A}\ }\textbf
  {\bibinfo {volume} {95}},\ \bibinfo {pages} {012116} (\bibinfo {year}
  {2017})}\BibitemShut {NoStop}%
\bibitem [{\citenamefont {Albarelli}\ \emph {et~al.}(2017)\citenamefont
  {Albarelli}, \citenamefont {Rossi}, \citenamefont {Paris},\ and\
  \citenamefont {Genoni}}]{albarelliUltimateLimitsQuantum2017}%
  \BibitemOpen
  \bibfield  {author} {\bibinfo {author} {\bibfnamefont {F.}~\bibnamefont
  {Albarelli}}, \bibinfo {author} {\bibfnamefont {M.~A.~C.}\ \bibnamefont
  {Rossi}}, \bibinfo {author} {\bibfnamefont {M.~G.~A.}\ \bibnamefont
  {Paris}},\ and\ \bibinfo {author} {\bibfnamefont {M.~G.}\ \bibnamefont
  {Genoni}},\ }\bibfield  {title} {\bibinfo {title} {Ultimate limits for
  quantum magnetometry via time-continuous measurements},\ }\href
  {https://doi.org/10.1088/1367-2630/aa9840} {\bibfield  {journal} {\bibinfo
  {journal} {New Journal of Physics}\ }\textbf {\bibinfo {volume} {19}},\
  \bibinfo {pages} {123011} (\bibinfo {year} {2017})}\BibitemShut {NoStop}%
\bibitem [{\citenamefont {Rossi}\ \emph
  {et~al.}(2020{\natexlab{a}})\citenamefont {Rossi}, \citenamefont {Albarelli},
  \citenamefont {Tamascelli},\ and\ \citenamefont {Genoni}}]{rossiPRL2020}%
  \BibitemOpen
  \bibfield  {author} {\bibinfo {author} {\bibfnamefont {M.~A.~C.}\
  \bibnamefont {Rossi}}, \bibinfo {author} {\bibfnamefont {F.}~\bibnamefont
  {Albarelli}}, \bibinfo {author} {\bibfnamefont {D.}~\bibnamefont
  {Tamascelli}},\ and\ \bibinfo {author} {\bibfnamefont {M.~G.}\ \bibnamefont
  {Genoni}},\ }\bibfield  {title} {\bibinfo {title} {Noisy quantum metrology
  enhanced by continuous nondemolition measurement},\ }\href
  {https://doi.org/10.1103/PhysRevLett.125.200505} {\bibfield  {journal}
  {\bibinfo  {journal} {Phys. Rev. Lett.}\ }\textbf {\bibinfo {volume} {125}},\
  \bibinfo {pages} {200505} (\bibinfo {year} {2020}{\natexlab{a}})}\BibitemShut
  {NoStop}%
\bibitem [{\citenamefont {Amorós-Binefa}\ and\ \citenamefont
  {Kołodyński}(2021)}]{AmorosBinefa2021}%
  \BibitemOpen
  \bibfield  {author} {\bibinfo {author} {\bibfnamefont {J.}~\bibnamefont
  {Amorós-Binefa}}\ and\ \bibinfo {author} {\bibfnamefont {J.}~\bibnamefont
  {Kołodyński}},\ }\bibfield  {title} {\bibinfo {title} {Noisy atomic
  magnetometry in real time},\ }\href
  {https://doi.org/10.1088/1367-2630/ac3b71} {\bibfield  {journal} {\bibinfo
  {journal} {New Journal of Physics}\ }\textbf {\bibinfo {volume} {23}},\
  \bibinfo {pages} {123030} (\bibinfo {year} {2021})}\BibitemShut {NoStop}%
\bibitem [{\citenamefont {Ilias}\ \emph {et~al.}(2022)\citenamefont {Ilias},
  \citenamefont {Yang}, \citenamefont {Huelga},\ and\ \citenamefont
  {Plenio}}]{iliasCriticalityEnhancedQuantumSensing2022}%
  \BibitemOpen
  \bibfield  {author} {\bibinfo {author} {\bibfnamefont {T.}~\bibnamefont
  {Ilias}}, \bibinfo {author} {\bibfnamefont {D.}~\bibnamefont {Yang}},
  \bibinfo {author} {\bibfnamefont {S.~F.}\ \bibnamefont {Huelga}},\ and\
  \bibinfo {author} {\bibfnamefont {M.~B.}\ \bibnamefont {Plenio}},\ }\bibfield
   {title} {\bibinfo {title} {Criticality-{{Enhanced Quantum Sensing}} via
  {{Continuous Measurement}}},\ }\href
  {https://doi.org/10.1103/PRXQuantum.3.010354} {\bibfield  {journal} {\bibinfo
   {journal} {PRX Quantum}\ }\textbf {\bibinfo {volume} {3}},\ \bibinfo {pages}
  {010354} (\bibinfo {year} {2022})}\BibitemShut {NoStop}%
\bibitem [{\citenamefont {Yang}\ \emph {et~al.}(2022)\citenamefont {Yang},
  \citenamefont {Huelga},\ and\ \citenamefont
  {Plenio}}]{yangEfficientInformationRetrieval2022}%
  \BibitemOpen
  \bibfield  {author} {\bibinfo {author} {\bibfnamefont {D.}~\bibnamefont
  {Yang}}, \bibinfo {author} {\bibfnamefont {S.~F.}\ \bibnamefont {Huelga}},\
  and\ \bibinfo {author} {\bibfnamefont {M.~B.}\ \bibnamefont {Plenio}},\
  }\bibfield  {title} {\bibinfo {title} {Efficient information retrieval for
  sensing via continuous measurement},\ }\Eprint
  {https://arxiv.org/abs/2209.08777} {arxiv:2209.08777 [quant-ph]}  (\bibinfo
  {year} {2022})\BibitemShut {NoStop}%
\bibitem [{\citenamefont {Fallani}\ \emph {et~al.}(2022)\citenamefont
  {Fallani}, \citenamefont {Rossi}, \citenamefont {Tamascelli},\ and\
  \citenamefont {Genoni}}]{FallaniPRXQuantum2022}%
  \BibitemOpen
  \bibfield  {author} {\bibinfo {author} {\bibfnamefont {A.}~\bibnamefont
  {Fallani}}, \bibinfo {author} {\bibfnamefont {M.~A.~C.}\ \bibnamefont
  {Rossi}}, \bibinfo {author} {\bibfnamefont {D.}~\bibnamefont {Tamascelli}},\
  and\ \bibinfo {author} {\bibfnamefont {M.~G.}\ \bibnamefont {Genoni}},\
  }\bibfield  {title} {\bibinfo {title} {Learning feedback control strategies
  for quantum metrology},\ }\href {https://doi.org/10.1103/PRXQuantum.3.020310}
  {\bibfield  {journal} {\bibinfo  {journal} {PRX Quantum}\ }\textbf {\bibinfo
  {volume} {3}},\ \bibinfo {pages} {020310} (\bibinfo {year}
  {2022})}\BibitemShut {NoStop}%
\bibitem [{\citenamefont {Amoros-Binefa}\ and\ \citenamefont
  {Kolodynski}(2024)}]{amorosbinefa2024}%
  \BibitemOpen
  \bibfield  {author} {\bibinfo {author} {\bibfnamefont {J.}~\bibnamefont
  {Amoros-Binefa}}\ and\ \bibinfo {author} {\bibfnamefont {J.}~\bibnamefont
  {Kolodynski}},\ }\href {https://arxiv.org/abs/2403.14764} {\bibinfo {title}
  {Noisy atomic magnetometry with kalman filtering and measurement-based
  feedback}} (\bibinfo {year} {2024}),\ \Eprint
  {https://arxiv.org/abs/2403.14764} {arXiv:2403.14764 [quant-ph]} \BibitemShut
  {NoStop}%
\bibitem [{\citenamefont {Elouard}\ and\ \citenamefont
  {Mohammady}(2018)}]{elouardWorkHeatEntropy2018}%
  \BibitemOpen
  \bibfield  {author} {\bibinfo {author} {\bibfnamefont {C.}~\bibnamefont
  {Elouard}}\ and\ \bibinfo {author} {\bibfnamefont {M.~H.}\ \bibnamefont
  {Mohammady}},\ }\bibfield  {title} {\bibinfo {title} {Work, {{Heat}} and
  {{Entropy Production Along Quantum Trajectories}}},\ }in\ \href
  {https://doi.org/10.1007/978-3-319-99046-0\_15} {\emph {\bibinfo {booktitle}
  {Thermodynamics in the {{Quantum Regime}}: {{Fundamental Aspects}} and {{New
  Directions}}}}},\ \bibinfo {series and number} {Fundamental {{Theories}} of
  {{Physics}}},\ \bibinfo {editor} {edited by\ \bibinfo {editor} {\bibfnamefont
  {F.}~\bibnamefont {Binder}}, \bibinfo {editor} {\bibfnamefont {L.~A.}\
  \bibnamefont {Correa}}, \bibinfo {editor} {\bibfnamefont {C.}~\bibnamefont
  {Gogolin}}, \bibinfo {editor} {\bibfnamefont {J.}~\bibnamefont {Anders}},\
  and\ \bibinfo {editor} {\bibfnamefont {G.}~\bibnamefont {Adesso}}}\ (\bibinfo
   {publisher} {{Springer International Publishing}},\ \bibinfo {address}
  {{Cham}},\ \bibinfo {year} {2018})\ pp.\ \bibinfo {pages}
  {363--393}\BibitemShut {NoStop}%
\bibitem [{\citenamefont {Manzano}\ and\ \citenamefont
  {Zambrini}(2022)}]{manzanoQuantumThermodynamicsContinuous2022}%
  \BibitemOpen
  \bibfield  {author} {\bibinfo {author} {\bibfnamefont {G.}~\bibnamefont
  {Manzano}}\ and\ \bibinfo {author} {\bibfnamefont {R.}~\bibnamefont
  {Zambrini}},\ }\bibfield  {title} {\bibinfo {title} {Quantum thermodynamics
  under continuous monitoring: {{A}} general framework},\ }\href
  {https://doi.org/10.1116/5.0079886} {\bibfield  {journal} {\bibinfo
  {journal} {AVS Quantum Science}\ }\textbf {\bibinfo {volume} {4}},\ \bibinfo
  {pages} {025302} (\bibinfo {year} {2022})}\BibitemShut {NoStop}%
\bibitem [{\citenamefont {Garrahan}\ and\ \citenamefont
  {Lesanovsky}(2010)}]{garrahanThermodynamicsQuantumJump2010}%
  \BibitemOpen
  \bibfield  {author} {\bibinfo {author} {\bibfnamefont {J.~P.}\ \bibnamefont
  {Garrahan}}\ and\ \bibinfo {author} {\bibfnamefont {I.}~\bibnamefont
  {Lesanovsky}},\ }\bibfield  {title} {\bibinfo {title} {Thermodynamics of
  {{Quantum Jump Trajectories}}},\ }\href
  {https://doi.org/10.1103/PhysRevLett.104.160601} {\bibfield  {journal}
  {\bibinfo  {journal} {Physical Review Letters}\ }\textbf {\bibinfo {volume}
  {104}},\ \bibinfo {pages} {160601} (\bibinfo {year} {2010})}\BibitemShut
  {NoStop}%
\bibitem [{\citenamefont {Garrahan}\ \emph {et~al.}(2011)\citenamefont
  {Garrahan}, \citenamefont {Armour},\ and\ \citenamefont
  {Lesanovsky}}]{garrahanQuantumTrajectoryPhase2011}%
  \BibitemOpen
  \bibfield  {author} {\bibinfo {author} {\bibfnamefont {J.~P.}\ \bibnamefont
  {Garrahan}}, \bibinfo {author} {\bibfnamefont {A.~D.}\ \bibnamefont
  {Armour}},\ and\ \bibinfo {author} {\bibfnamefont {I.}~\bibnamefont
  {Lesanovsky}},\ }\bibfield  {title} {\bibinfo {title} {Quantum trajectory
  phase transitions in the micromaser},\ }\href
  {https://doi.org/10.1103/PhysRevE.84.021115} {\bibfield  {journal} {\bibinfo
  {journal} {Physical Review E}\ }\textbf {\bibinfo {volume} {84}},\ \bibinfo
  {pages} {021115} (\bibinfo {year} {2011})}\BibitemShut {NoStop}%
\bibitem [{\citenamefont {Cilluffo}\ \emph {et~al.}(2019)\citenamefont
  {Cilluffo}, \citenamefont {Lorenzo}, \citenamefont {Palma},\ and\
  \citenamefont {Ciccarello}}]{cilluffoQuantumJumpStatistics2019}%
  \BibitemOpen
  \bibfield  {author} {\bibinfo {author} {\bibfnamefont {D.}~\bibnamefont
  {Cilluffo}}, \bibinfo {author} {\bibfnamefont {S.}~\bibnamefont {Lorenzo}},
  \bibinfo {author} {\bibfnamefont {G.~M.}\ \bibnamefont {Palma}},\ and\
  \bibinfo {author} {\bibfnamefont {F.}~\bibnamefont {Ciccarello}},\ }\bibfield
   {title} {\bibinfo {title} {Quantum jump statistics with a shifted jump
  operator in a chiral waveguide},\ }\href
  {https://doi.org/10.1088/1742-5468/ab371c} {\bibfield  {journal} {\bibinfo
  {journal} {Journal of Statistical Mechanics: Theory and Experiment}\ }\textbf
  {\bibinfo {volume} {2019}},\ \bibinfo {pages} {104004} (\bibinfo {year}
  {2019})}\BibitemShut {NoStop}%
\bibitem [{\citenamefont {Belenchia}\ \emph {et~al.}(2020)\citenamefont
  {Belenchia}, \citenamefont {Mancino}, \citenamefont {Landi},\ and\
  \citenamefont {Paternostro}}]{belenchiaEntropyProductionContinuously2020}%
  \BibitemOpen
  \bibfield  {author} {\bibinfo {author} {\bibfnamefont {A.}~\bibnamefont
  {Belenchia}}, \bibinfo {author} {\bibfnamefont {L.}~\bibnamefont {Mancino}},
  \bibinfo {author} {\bibfnamefont {G.~T.}\ \bibnamefont {Landi}},\ and\
  \bibinfo {author} {\bibfnamefont {M.}~\bibnamefont {Paternostro}},\
  }\bibfield  {title} {\bibinfo {title} {Entropy production in continuously
  measured {{Gaussian}} quantum systems},\ }\href
  {https://doi.org/10.1038/s41534-020-00334-6} {\bibfield  {journal} {\bibinfo
  {journal} {npj Quantum Information}\ }\textbf {\bibinfo {volume} {6}},\
  \bibinfo {pages} {1} (\bibinfo {year} {2020})}\BibitemShut {NoStop}%
\bibitem [{\citenamefont {Rossi}\ \emph
  {et~al.}(2020{\natexlab{b}})\citenamefont {Rossi}, \citenamefont {Mancino},
  \citenamefont {Landi}, \citenamefont {Paternostro}, \citenamefont
  {Schliesser},\ and\ \citenamefont
  {Belenchia}}]{rossiExperimentalAssessmentEntropy2020}%
  \BibitemOpen
  \bibfield  {author} {\bibinfo {author} {\bibfnamefont {M.}~\bibnamefont
  {Rossi}}, \bibinfo {author} {\bibfnamefont {L.}~\bibnamefont {Mancino}},
  \bibinfo {author} {\bibfnamefont {G.~T.}\ \bibnamefont {Landi}}, \bibinfo
  {author} {\bibfnamefont {M.}~\bibnamefont {Paternostro}}, \bibinfo {author}
  {\bibfnamefont {A.}~\bibnamefont {Schliesser}},\ and\ \bibinfo {author}
  {\bibfnamefont {A.}~\bibnamefont {Belenchia}},\ }\bibfield  {title} {\bibinfo
  {title} {Experimental {{Assessment}} of {{Entropy Production}} in a
  {{Continuously Measured Mechanical Resonator}}},\ }\href
  {https://doi.org/10.1103/PhysRevLett.125.080601} {\bibfield  {journal}
  {\bibinfo  {journal} {Physical Review Letters}\ }\textbf {\bibinfo {volume}
  {125}},\ \bibinfo {pages} {080601} (\bibinfo {year}
  {2020}{\natexlab{b}})}\BibitemShut {NoStop}%
\bibitem [{\citenamefont {Landi}\ and\ \citenamefont
  {Paternostro}(2021)}]{landiIrreversibleEntropyProduction2021}%
  \BibitemOpen
  \bibfield  {author} {\bibinfo {author} {\bibfnamefont {G.~T.}\ \bibnamefont
  {Landi}}\ and\ \bibinfo {author} {\bibfnamefont {M.}~\bibnamefont
  {Paternostro}},\ }\bibfield  {title} {\bibinfo {title} {Irreversible entropy
  production: {{From}} classical to quantum},\ }\href
  {https://doi.org/10.1103/RevModPhys.93.035008} {\bibfield  {journal}
  {\bibinfo  {journal} {Reviews of Modern Physics}\ }\textbf {\bibinfo {volume}
  {93}},\ \bibinfo {pages} {035008} (\bibinfo {year} {2021})}\BibitemShut
  {NoStop}%
\bibitem [{\citenamefont {Landi}\ \emph {et~al.}(2022)\citenamefont {Landi},
  \citenamefont {Paternostro},\ and\ \citenamefont
  {Belenchia}}]{LandiPRXQuantum2022}%
  \BibitemOpen
  \bibfield  {author} {\bibinfo {author} {\bibfnamefont {G.~T.}\ \bibnamefont
  {Landi}}, \bibinfo {author} {\bibfnamefont {M.}~\bibnamefont {Paternostro}},\
  and\ \bibinfo {author} {\bibfnamefont {A.}~\bibnamefont {Belenchia}},\
  }\bibfield  {title} {\bibinfo {title} {Informational steady states and
  conditional entropy production in continuously monitored systems},\ }\href
  {https://doi.org/10.1103/PRXQuantum.3.010303} {\bibfield  {journal} {\bibinfo
   {journal} {PRX Quantum}\ }\textbf {\bibinfo {volume} {3}},\ \bibinfo {pages}
  {010303} (\bibinfo {year} {2022})}\BibitemShut {NoStop}%
\bibitem [{\citenamefont {Bhandari}\ and\ \citenamefont
  {Jordan}(2022)}]{bhandariContinuousMeasurementBoosted2022}%
  \BibitemOpen
  \bibfield  {author} {\bibinfo {author} {\bibfnamefont {B.}~\bibnamefont
  {Bhandari}}\ and\ \bibinfo {author} {\bibfnamefont {A.~N.}\ \bibnamefont
  {Jordan}},\ }\bibfield  {title} {\bibinfo {title} {Continuous measurement
  boosted adiabatic quantum thermal machines},\ }\href
  {https://doi.org/10.1103/PhysRevResearch.4.033103} {\bibfield  {journal}
  {\bibinfo  {journal} {Physical Review Research}\ }\textbf {\bibinfo {volume}
  {4}},\ \bibinfo {pages} {033103} (\bibinfo {year} {2022})}\BibitemShut
  {NoStop}%
\bibitem [{\citenamefont {Bhandari}\ \emph {et~al.}(2023)\citenamefont
  {Bhandari}, \citenamefont {Czupryniak}, \citenamefont {Erdman},\ and\
  \citenamefont {Jordan}}]{bhandariMeasurementBasedQuantumThermal2023}%
  \BibitemOpen
  \bibfield  {author} {\bibinfo {author} {\bibfnamefont {B.}~\bibnamefont
  {Bhandari}}, \bibinfo {author} {\bibfnamefont {R.}~\bibnamefont
  {Czupryniak}}, \bibinfo {author} {\bibfnamefont {P.~A.}\ \bibnamefont
  {Erdman}},\ and\ \bibinfo {author} {\bibfnamefont {A.~N.}\ \bibnamefont
  {Jordan}},\ }\bibfield  {title} {\bibinfo {title} {Measurement-{{Based
  Quantum Thermal Machines}} with {{Feedback Control}}},\ }\href
  {https://doi.org/10.3390/e25020204} {\bibfield  {journal} {\bibinfo
  {journal} {Entropy}\ }\textbf {\bibinfo {volume} {25}},\ \bibinfo {pages}
  {204} (\bibinfo {year} {2023})}\BibitemShut {NoStop}%
\bibitem [{\citenamefont {Murch}\ \emph {et~al.}(2013)\citenamefont {Murch},
  \citenamefont {Weber}, \citenamefont {Macklin},\ and\ \citenamefont
  {Siddiqi}}]{Murch2013}%
  \BibitemOpen
  \bibfield  {author} {\bibinfo {author} {\bibfnamefont {K.~W.}\ \bibnamefont
  {Murch}}, \bibinfo {author} {\bibfnamefont {S.~J.}\ \bibnamefont {Weber}},
  \bibinfo {author} {\bibfnamefont {C.}~\bibnamefont {Macklin}},\ and\ \bibinfo
  {author} {\bibfnamefont {I.}~\bibnamefont {Siddiqi}},\ }\bibfield  {title}
  {\bibinfo {title} {Observing single quantum trajectories of a superconducting
  quantum bit},\ }\href {https://doi.org/10.1038/nature12539} {\bibfield
  {journal} {\bibinfo  {journal} {Nature}\ }\textbf {\bibinfo {volume} {502}},\
  \bibinfo {pages} {211} (\bibinfo {year} {2013})}\BibitemShut {NoStop}%
\bibitem [{\citenamefont {{Campagne-Ibarcq}}\ \emph {et~al.}(2016)\citenamefont
  {{Campagne-Ibarcq}}, \citenamefont {Six}, \citenamefont {Bretheau},
  \citenamefont {Sarlette}, \citenamefont {Mirrahimi}, \citenamefont
  {Rouchon},\ and\ \citenamefont {Huard}}]{Campagne-Ibarcq2016}%
  \BibitemOpen
  \bibfield  {author} {\bibinfo {author} {\bibfnamefont {P.}~\bibnamefont
  {{Campagne-Ibarcq}}}, \bibinfo {author} {\bibfnamefont {P.}~\bibnamefont
  {Six}}, \bibinfo {author} {\bibfnamefont {L.}~\bibnamefont {Bretheau}},
  \bibinfo {author} {\bibfnamefont {A.}~\bibnamefont {Sarlette}}, \bibinfo
  {author} {\bibfnamefont {M.}~\bibnamefont {Mirrahimi}}, \bibinfo {author}
  {\bibfnamefont {P.}~\bibnamefont {Rouchon}},\ and\ \bibinfo {author}
  {\bibfnamefont {B.}~\bibnamefont {Huard}},\ }\bibfield  {title} {\bibinfo
  {title} {Observing {{Quantum State Diffusion}} by {{Heterodyne Detection}} of
  {{Fluorescence}}},\ }\href {https://doi.org/10.1103/PhysRevX.6.011002}
  {\bibfield  {journal} {\bibinfo  {journal} {Phys. Rev. X}\ }\textbf {\bibinfo
  {volume} {6}},\ \bibinfo {pages} {011002} (\bibinfo {year}
  {2016})}\BibitemShut {NoStop}%
\bibitem [{\citenamefont {Ficheux}\ \emph {et~al.}(2018)\citenamefont
  {Ficheux}, \citenamefont {Jezouin}, \citenamefont {Leghtas},\ and\
  \citenamefont {Huard}}]{Ficheux2017}%
  \BibitemOpen
  \bibfield  {author} {\bibinfo {author} {\bibfnamefont {Q.}~\bibnamefont
  {Ficheux}}, \bibinfo {author} {\bibfnamefont {S.}~\bibnamefont {Jezouin}},
  \bibinfo {author} {\bibfnamefont {Z.}~\bibnamefont {Leghtas}},\ and\ \bibinfo
  {author} {\bibfnamefont {B.}~\bibnamefont {Huard}},\ }\bibfield  {title}
  {\bibinfo {title} {Dynamics of a qubit while simultaneously monitoring its
  relaxation and dephasing},\ }\href
  {https://doi.org/10.1038/s41467-018-04372-9} {\bibfield  {journal} {\bibinfo
  {journal} {Nat. Commun.}\ }\textbf {\bibinfo {volume} {9}},\ \bibinfo {pages}
  {1926} (\bibinfo {year} {2018})},\ \Eprint {https://arxiv.org/abs/1711.01208}
  {arXiv:1711.01208} \BibitemShut {NoStop}%
\bibitem [{\citenamefont {Minev}\ \emph {et~al.}(2019)\citenamefont {Minev},
  \citenamefont {Mundhada}, \citenamefont {Shankar}, \citenamefont {Reinhold},
  \citenamefont {{Guti{\'e}rrez-J{\'a}uregui}}, \citenamefont {Schoelkopf},
  \citenamefont {Mirrahimi}, \citenamefont {Carmichael},\ and\ \citenamefont
  {Devoret}}]{Minev2019}%
  \BibitemOpen
  \bibfield  {author} {\bibinfo {author} {\bibfnamefont {Z.~K.}\ \bibnamefont
  {Minev}}, \bibinfo {author} {\bibfnamefont {S.~O.}\ \bibnamefont {Mundhada}},
  \bibinfo {author} {\bibfnamefont {S.}~\bibnamefont {Shankar}}, \bibinfo
  {author} {\bibfnamefont {P.}~\bibnamefont {Reinhold}}, \bibinfo {author}
  {\bibfnamefont {R.}~\bibnamefont {{Guti{\'e}rrez-J{\'a}uregui}}}, \bibinfo
  {author} {\bibfnamefont {R.~J.}\ \bibnamefont {Schoelkopf}}, \bibinfo
  {author} {\bibfnamefont {M.}~\bibnamefont {Mirrahimi}}, \bibinfo {author}
  {\bibfnamefont {H.~J.}\ \bibnamefont {Carmichael}},\ and\ \bibinfo {author}
  {\bibfnamefont {M.~H.}\ \bibnamefont {Devoret}},\ }\bibfield  {title}
  {\bibinfo {title} {To catch and reverse a quantum jump mid-flight},\ }\href
  {https://doi.org/10.1038/s41586-019-1287-z} {\bibfield  {journal} {\bibinfo
  {journal} {Nature}\ }\textbf {\bibinfo {volume} {570}},\ \bibinfo {pages}
  {200} (\bibinfo {year} {2019})},\ \Eprint {https://arxiv.org/abs/1803.00545}
  {arXiv:1803.00545} \BibitemShut {NoStop}%
\bibitem [{\citenamefont {Wieczorek}\ \emph {et~al.}(2015)\citenamefont
  {Wieczorek}, \citenamefont {Hofer}, \citenamefont {{Hoelscher-Obermaier}},
  \citenamefont {Riedinger}, \citenamefont {Hammerer},\ and\ \citenamefont
  {Aspelmeyer}}]{Wieczorek2015}%
  \BibitemOpen
  \bibfield  {author} {\bibinfo {author} {\bibfnamefont {W.}~\bibnamefont
  {Wieczorek}}, \bibinfo {author} {\bibfnamefont {S.~G.}\ \bibnamefont
  {Hofer}}, \bibinfo {author} {\bibfnamefont {J.}~\bibnamefont
  {{Hoelscher-Obermaier}}}, \bibinfo {author} {\bibfnamefont {R.}~\bibnamefont
  {Riedinger}}, \bibinfo {author} {\bibfnamefont {K.}~\bibnamefont
  {Hammerer}},\ and\ \bibinfo {author} {\bibfnamefont {M.}~\bibnamefont
  {Aspelmeyer}},\ }\bibfield  {title} {\bibinfo {title} {Optimal {{State
  Estimation}} for {{Cavity Optomechanical Systems}}},\ }\href
  {https://doi.org/10.1103/PhysRevLett.114.223601} {\bibfield  {journal}
  {\bibinfo  {journal} {Phys. Rev. Lett.}\ }\textbf {\bibinfo {volume} {114}},\
  \bibinfo {pages} {223601} (\bibinfo {year} {2015})}\BibitemShut {NoStop}%
\bibitem [{\citenamefont {Rossi}\ \emph {et~al.}(2019)\citenamefont {Rossi},
  \citenamefont {Mason}, \citenamefont {Chen},\ and\ \citenamefont
  {Schliesser}}]{Rossi2018}%
  \BibitemOpen
  \bibfield  {author} {\bibinfo {author} {\bibfnamefont {M.}~\bibnamefont
  {Rossi}}, \bibinfo {author} {\bibfnamefont {D.}~\bibnamefont {Mason}},
  \bibinfo {author} {\bibfnamefont {J.}~\bibnamefont {Chen}},\ and\ \bibinfo
  {author} {\bibfnamefont {A.}~\bibnamefont {Schliesser}},\ }\bibfield  {title}
  {\bibinfo {title} {Observing and {{Verifying}} the {{Quantum Trajectory}} of
  a {{Mechanical Resonator}}},\ }\href
  {https://doi.org/10.1103/PhysRevLett.123.163601} {\bibfield  {journal}
  {\bibinfo  {journal} {Phys. Rev. Lett.}\ }\textbf {\bibinfo {volume} {123}},\
  \bibinfo {pages} {163601} (\bibinfo {year} {2019})},\ \Eprint
  {https://arxiv.org/abs/1812.00928} {arXiv:1812.00928} \BibitemShut {NoStop}%
\bibitem [{\citenamefont {Rossi}\ \emph {et~al.}(2018)\citenamefont {Rossi},
  \citenamefont {Mason}, \citenamefont {Chen}, \citenamefont {Tsaturyan},\ and\
  \citenamefont {Schliesser}}]{rossi2018control}%
  \BibitemOpen
  \bibfield  {author} {\bibinfo {author} {\bibfnamefont {M.}~\bibnamefont
  {Rossi}}, \bibinfo {author} {\bibfnamefont {D.}~\bibnamefont {Mason}},
  \bibinfo {author} {\bibfnamefont {J.}~\bibnamefont {Chen}}, \bibinfo {author}
  {\bibfnamefont {Y.}~\bibnamefont {Tsaturyan}},\ and\ \bibinfo {author}
  {\bibfnamefont {A.}~\bibnamefont {Schliesser}},\ }\bibfield  {title}
  {\bibinfo {title} {Measurement-based quantum control of mechanical motion},\
  }\href {https://doi.org/10.1038/s41586-018-0643-8} {\bibfield  {journal}
  {\bibinfo  {journal} {Nature}\ }\textbf {\bibinfo {volume} {563}},\ \bibinfo
  {pages} {53} (\bibinfo {year} {2018})}\BibitemShut {NoStop}%
\bibitem [{\citenamefont {Magrini}\ \emph {et~al.}(2021)\citenamefont
  {Magrini}, \citenamefont {Rosenzweig}, \citenamefont {Bach}, \citenamefont
  {{Deutschmann-Olek}}, \citenamefont {Hofer}, \citenamefont {Hong},
  \citenamefont {Kiesel}, \citenamefont {Kugi},\ and\ \citenamefont
  {Aspelmeyer}}]{Magrini2021}%
  \BibitemOpen
  \bibfield  {author} {\bibinfo {author} {\bibfnamefont {L.}~\bibnamefont
  {Magrini}}, \bibinfo {author} {\bibfnamefont {P.}~\bibnamefont {Rosenzweig}},
  \bibinfo {author} {\bibfnamefont {C.}~\bibnamefont {Bach}}, \bibinfo {author}
  {\bibfnamefont {A.}~\bibnamefont {{Deutschmann-Olek}}}, \bibinfo {author}
  {\bibfnamefont {S.~G.}\ \bibnamefont {Hofer}}, \bibinfo {author}
  {\bibfnamefont {S.}~\bibnamefont {Hong}}, \bibinfo {author} {\bibfnamefont
  {N.}~\bibnamefont {Kiesel}}, \bibinfo {author} {\bibfnamefont
  {A.}~\bibnamefont {Kugi}},\ and\ \bibinfo {author} {\bibfnamefont
  {M.}~\bibnamefont {Aspelmeyer}},\ }\bibfield  {title} {\bibinfo {title}
  {Real-time optimal quantum control of mechanical motion at room
  temperature},\ }\href {https://doi.org/10.1038/s41586-021-03602-3} {\bibfield
   {journal} {\bibinfo  {journal} {Nature}\ }\textbf {\bibinfo {volume}
  {595}},\ \bibinfo {pages} {373} (\bibinfo {year} {2021})}\BibitemShut
  {NoStop}%
\bibitem [{\citenamefont {Tebbenjohanns}\ \emph {et~al.}(2021)\citenamefont
  {Tebbenjohanns}, \citenamefont {Mattana}, \citenamefont {Rossi},
  \citenamefont {Frimmer},\ and\ \citenamefont {Novotny}}]{Tebbenjohanns2021}%
  \BibitemOpen
  \bibfield  {author} {\bibinfo {author} {\bibfnamefont {F.}~\bibnamefont
  {Tebbenjohanns}}, \bibinfo {author} {\bibfnamefont {M.~L.}\ \bibnamefont
  {Mattana}}, \bibinfo {author} {\bibfnamefont {M.}~\bibnamefont {Rossi}},
  \bibinfo {author} {\bibfnamefont {M.}~\bibnamefont {Frimmer}},\ and\ \bibinfo
  {author} {\bibfnamefont {L.}~\bibnamefont {Novotny}},\ }\bibfield  {title}
  {\bibinfo {title} {Quantum control of a nanoparticle optically levitated in
  cryogenic free space},\ }\href {https://doi.org/10.1038/s41586-021-03617-w}
  {\bibfield  {journal} {\bibinfo  {journal} {Nature}\ }\textbf {\bibinfo
  {volume} {595}},\ \bibinfo {pages} {378} (\bibinfo {year}
  {2021})}\BibitemShut {NoStop}%
\bibitem [{\citenamefont {Ciccarello}\ \emph {et~al.}(2013)\citenamefont
  {Ciccarello}, \citenamefont {Palma},\ and\ \citenamefont
  {Giovannetti}}]{CiccarelloPra2013}%
  \BibitemOpen
  \bibfield  {author} {\bibinfo {author} {\bibfnamefont {F.}~\bibnamefont
  {Ciccarello}}, \bibinfo {author} {\bibfnamefont {G.~M.}\ \bibnamefont
  {Palma}},\ and\ \bibinfo {author} {\bibfnamefont {V.}~\bibnamefont
  {Giovannetti}},\ }\bibfield  {title} {\bibinfo {title} {Collision-model-based
  approach to non-markovian quantum dynamics},\ }\href
  {https://doi.org/10.1103/PhysRevA.87.040103} {\bibfield  {journal} {\bibinfo
  {journal} {Phys. Rev. A}\ }\textbf {\bibinfo {volume} {87}},\ \bibinfo
  {pages} {040103} (\bibinfo {year} {2013})}\BibitemShut {NoStop}%
\bibitem [{\citenamefont {Ciccarello}\ \emph {et~al.}(2022)\citenamefont
  {Ciccarello}, \citenamefont {Lorenzo}, \citenamefont {Giovannetti},\ and\
  \citenamefont {Palma}}]{CiccarelloPR2022}%
  \BibitemOpen
  \bibfield  {author} {\bibinfo {author} {\bibfnamefont {F.}~\bibnamefont
  {Ciccarello}}, \bibinfo {author} {\bibfnamefont {S.}~\bibnamefont {Lorenzo}},
  \bibinfo {author} {\bibfnamefont {V.}~\bibnamefont {Giovannetti}},\ and\
  \bibinfo {author} {\bibfnamefont {G.~M.}\ \bibnamefont {Palma}},\ }\bibfield
  {title} {\bibinfo {title} {Quantum collision models: Open system dynamics
  from repeated interactions},\ }\href
  {https://doi.org/https://doi.org/10.1016/j.physrep.2022.01.001} {\bibfield
  {journal} {\bibinfo  {journal} {Physics Reports}\ }\textbf {\bibinfo {volume}
  {954}},\ \bibinfo {pages} {1} (\bibinfo {year} {2022})},\ \bibinfo {note}
  {quantum collision models: Open system dynamics from repeated
  interactions}\BibitemShut {NoStop}%
\bibitem [{\citenamefont {Campbell}\ and\ \citenamefont
  {Vacchini}(2021)}]{Campbell2021_CM}%
  \BibitemOpen
  \bibfield  {author} {\bibinfo {author} {\bibfnamefont {S.}~\bibnamefont
  {Campbell}}\ and\ \bibinfo {author} {\bibfnamefont {B.}~\bibnamefont
  {Vacchini}},\ }\bibfield  {title} {\bibinfo {title} {Collision models in open
  system dynamics: A versatile tool for deeper insights?},\ }\href
  {https://doi.org/10.1209/0295-5075/133/60001} {\bibfield  {journal} {\bibinfo
   {journal} {Europhysics Letters}\ }\textbf {\bibinfo {volume} {133}},\
  \bibinfo {pages} {60001} (\bibinfo {year} {2021})}\BibitemShut {NoStop}%
\bibitem [{\citenamefont {Cattaneo}\ \emph {et~al.}(2022)\citenamefont
  {Cattaneo}, \citenamefont {Giorgi}, \citenamefont {Zambrini},\ and\
  \citenamefont {Maniscalco}}]{cattaneoBriefJourneyCollision2022a}%
  \BibitemOpen
  \bibfield  {author} {\bibinfo {author} {\bibfnamefont {M.}~\bibnamefont
  {Cattaneo}}, \bibinfo {author} {\bibfnamefont {G.~L.}\ \bibnamefont
  {Giorgi}}, \bibinfo {author} {\bibfnamefont {R.}~\bibnamefont {Zambrini}},\
  and\ \bibinfo {author} {\bibfnamefont {S.}~\bibnamefont {Maniscalco}},\
  }\bibfield  {title} {\bibinfo {title} {A brief journey through collision
  models for multipartite open quantum dynamics},\ }\href
  {https://doi.org/10.1142/S1230161222500159} {\bibfield  {journal} {\bibinfo
  {journal} {Open Systems \& Information Dynamics}\ }\textbf {\bibinfo {volume}
  {29}},\ \bibinfo {pages} {2250015} (\bibinfo {year} {2022})}\BibitemShut
  {NoStop}%
\bibitem [{\citenamefont {Belenchia}\ \emph {et~al.}(2022)\citenamefont
  {Belenchia}, \citenamefont {Paternostro},\ and\ \citenamefont
  {Landi}}]{BelenchiaPRA2022}%
  \BibitemOpen
  \bibfield  {author} {\bibinfo {author} {\bibfnamefont {A.}~\bibnamefont
  {Belenchia}}, \bibinfo {author} {\bibfnamefont {M.}~\bibnamefont
  {Paternostro}},\ and\ \bibinfo {author} {\bibfnamefont {G.~T.}\ \bibnamefont
  {Landi}},\ }\bibfield  {title} {\bibinfo {title} {Informational steady states
  and conditional entropy production in continuously monitored systems: The
  case of gaussian systems},\ }\href
  {https://doi.org/10.1103/PhysRevA.105.022213} {\bibfield  {journal} {\bibinfo
   {journal} {Phys. Rev. A}\ }\textbf {\bibinfo {volume} {105}},\ \bibinfo
  {pages} {022213} (\bibinfo {year} {2022})}\BibitemShut {NoStop}%
\bibitem [{\citenamefont {Burnett}\ \emph {et~al.}(2019)\citenamefont
  {Burnett}, \citenamefont {Bengtsson}, \citenamefont {Scigliuzzo},
  \citenamefont {Niepce}, \citenamefont {Kudra}, \citenamefont {Delsing},\ and\
  \citenamefont {Bylander}}]{burnett2019decoherence}%
  \BibitemOpen
  \bibfield  {author} {\bibinfo {author} {\bibfnamefont {J.~J.}\ \bibnamefont
  {Burnett}}, \bibinfo {author} {\bibfnamefont {A.}~\bibnamefont {Bengtsson}},
  \bibinfo {author} {\bibfnamefont {M.}~\bibnamefont {Scigliuzzo}}, \bibinfo
  {author} {\bibfnamefont {D.}~\bibnamefont {Niepce}}, \bibinfo {author}
  {\bibfnamefont {M.}~\bibnamefont {Kudra}}, \bibinfo {author} {\bibfnamefont
  {P.}~\bibnamefont {Delsing}},\ and\ \bibinfo {author} {\bibfnamefont
  {J.}~\bibnamefont {Bylander}},\ }\bibfield  {title} {\bibinfo {title}
  {Decoherence benchmarking of superconducting qubits},\ }\href
  {https://doi.org/10.1038/s41534-019-0168-5} {\bibfield  {journal} {\bibinfo
  {journal} {npj Quantum Information}\ }\textbf {\bibinfo {volume} {5}},\
  \bibinfo {pages} {54} (\bibinfo {year} {2019})}\BibitemShut {NoStop}%
\bibitem [{\citenamefont {Gemme}\ \emph {et~al.}(2022)\citenamefont {Gemme},
  \citenamefont {Grossi}, \citenamefont {Ferraro}, \citenamefont {Vallecorsa},\
  and\ \citenamefont {Sassetti}}]{gemmeIBMQuantumPlatforms2022a}%
  \BibitemOpen
  \bibfield  {author} {\bibinfo {author} {\bibfnamefont {G.}~\bibnamefont
  {Gemme}}, \bibinfo {author} {\bibfnamefont {M.}~\bibnamefont {Grossi}},
  \bibinfo {author} {\bibfnamefont {D.}~\bibnamefont {Ferraro}}, \bibinfo
  {author} {\bibfnamefont {S.}~\bibnamefont {Vallecorsa}},\ and\ \bibinfo
  {author} {\bibfnamefont {M.}~\bibnamefont {Sassetti}},\ }\bibfield  {title}
  {\bibinfo {title} {{{IBM}} quantum platforms: A quantum battery
  perspective},\ }\href {https://doi.org/10.3390/batteries8050043} {\bibfield
  {journal} {\bibinfo  {journal} {Batteries}\ }\textbf {\bibinfo {volume}
  {8}},\ \bibinfo {pages} {43} (\bibinfo {year} {2022})}\BibitemShut {NoStop}%
\bibitem [{\citenamefont {{Garc{\'i}a-P{\'e}rez}}\ \emph
  {et~al.}(2020)\citenamefont {{Garc{\'i}a-P{\'e}rez}}, \citenamefont {Rossi},\
  and\ \citenamefont {Maniscalco}}]{garcia-perezIBMExperienceVersatile2020}%
  \BibitemOpen
  \bibfield  {author} {\bibinfo {author} {\bibfnamefont {G.}~\bibnamefont
  {{Garc{\'i}a-P{\'e}rez}}}, \bibinfo {author} {\bibfnamefont {M.~A.~C.}\
  \bibnamefont {Rossi}},\ and\ \bibinfo {author} {\bibfnamefont
  {S.}~\bibnamefont {Maniscalco}},\ }\bibfield  {title} {\bibinfo {title}
  {{{IBM Q Experience}} as a versatile experimental testbed for simulating open
  quantum systems},\ }\href {https://doi.org/10.1038/s41534-019-0235-y}
  {\bibfield  {journal} {\bibinfo  {journal} {npj Quantum Information}\
  }\textbf {\bibinfo {volume} {6}},\ \bibinfo {pages} {1} (\bibinfo {year}
  {2020})}\BibitemShut {NoStop}%
\bibitem [{\citenamefont {Cech}\ \emph {et~al.}(2023)\citenamefont {Cech},
  \citenamefont {Lesanovsky},\ and\ \citenamefont
  {Carollo}}]{cech2023thermodynamics}%
  \BibitemOpen
  \bibfield  {author} {\bibinfo {author} {\bibfnamefont {M.}~\bibnamefont
  {Cech}}, \bibinfo {author} {\bibfnamefont {I.}~\bibnamefont {Lesanovsky}},\
  and\ \bibinfo {author} {\bibfnamefont {F.}~\bibnamefont {Carollo}},\
  }\bibfield  {title} {\bibinfo {title} {Thermodynamics of quantum trajectories
  on a quantum computer},\ }\href
  {https://doi.org/10.1103/PhysRevLett.131.120401} {\bibfield  {journal}
  {\bibinfo  {journal} {Phys. Rev. Lett.}\ }\textbf {\bibinfo {volume} {131}},\
  \bibinfo {pages} {120401} (\bibinfo {year} {2023})}\BibitemShut {NoStop}%
\bibitem [{\citenamefont {Cattaneo}\ \emph {et~al.}(2023)\citenamefont
  {Cattaneo}, \citenamefont {Rossi}, \citenamefont {{Garc{\'i}a-P{\'e}rez}},
  \citenamefont {Zambrini},\ and\ \citenamefont
  {Maniscalco}}]{cattaneoQuantumSimulationDissipative2023}%
  \BibitemOpen
  \bibfield  {author} {\bibinfo {author} {\bibfnamefont {M.}~\bibnamefont
  {Cattaneo}}, \bibinfo {author} {\bibfnamefont {M.~A.~C.}\ \bibnamefont
  {Rossi}}, \bibinfo {author} {\bibfnamefont {G.}~\bibnamefont
  {{Garc{\'i}a-P{\'e}rez}}}, \bibinfo {author} {\bibfnamefont {R.}~\bibnamefont
  {Zambrini}},\ and\ \bibinfo {author} {\bibfnamefont {S.}~\bibnamefont
  {Maniscalco}},\ }\bibfield  {title} {\bibinfo {title} {Quantum simulation of
  dissipative collective effects on noisy quantum computers},\ }\href
  {https://doi.org/10.1103/PRXQuantum.4.010324} {\bibfield  {journal} {\bibinfo
   {journal} {PRX Quantum}\ }\textbf {\bibinfo {volume} {4}},\ \bibinfo {pages}
  {010324} (\bibinfo {year} {2023})}\BibitemShut {NoStop}%
\bibitem [{\citenamefont {Javadi-Abhari}\ \emph {et~al.}(2024)\citenamefont
  {Javadi-Abhari}, \citenamefont {Treinish}, \citenamefont {Krsulich},
  \citenamefont {Wood}, \citenamefont {Lishman}, \citenamefont {Gacon},
  \citenamefont {Martiel}, \citenamefont {Nation}, \citenamefont {Bishop},
  \citenamefont {Cross}, \citenamefont {Johnson},\ and\ \citenamefont
  {Gambetta}}]{javadiabhari2024quantumcomputingqiskit}%
  \BibitemOpen
  \bibfield  {author} {\bibinfo {author} {\bibfnamefont {A.}~\bibnamefont
  {Javadi-Abhari}}, \bibinfo {author} {\bibfnamefont {M.}~\bibnamefont
  {Treinish}}, \bibinfo {author} {\bibfnamefont {K.}~\bibnamefont {Krsulich}},
  \bibinfo {author} {\bibfnamefont {C.~J.}\ \bibnamefont {Wood}}, \bibinfo
  {author} {\bibfnamefont {J.}~\bibnamefont {Lishman}}, \bibinfo {author}
  {\bibfnamefont {J.}~\bibnamefont {Gacon}}, \bibinfo {author} {\bibfnamefont
  {S.}~\bibnamefont {Martiel}}, \bibinfo {author} {\bibfnamefont {P.~D.}\
  \bibnamefont {Nation}}, \bibinfo {author} {\bibfnamefont {L.~S.}\
  \bibnamefont {Bishop}}, \bibinfo {author} {\bibfnamefont {A.~W.}\
  \bibnamefont {Cross}}, \bibinfo {author} {\bibfnamefont {B.~R.}\ \bibnamefont
  {Johnson}},\ and\ \bibinfo {author} {\bibfnamefont {J.~M.}\ \bibnamefont
  {Gambetta}},\ }\href {https://arxiv.org/abs/2405.08810} {\bibinfo {title}
  {Quantum computing with qiskit}} (\bibinfo {year} {2024}),\ \Eprint
  {https://arxiv.org/abs/2405.08810} {arXiv:2405.08810 [quant-ph]} \BibitemShut
  {NoStop}%
\bibitem [{\citenamefont {Johansson}\ \emph {et~al.}(2012)\citenamefont
  {Johansson}, \citenamefont {Nation},\ and\ \citenamefont
  {Nori}}]{johanssonQuTiPOpensourcePython2012}%
  \BibitemOpen
  \bibfield  {author} {\bibinfo {author} {\bibfnamefont {J.~R.}\ \bibnamefont
  {Johansson}}, \bibinfo {author} {\bibfnamefont {P.~D.}\ \bibnamefont
  {Nation}},\ and\ \bibinfo {author} {\bibfnamefont {F.}~\bibnamefont {Nori}},\
  }\bibfield  {title} {\bibinfo {title} {{{QuTiP}}: {{An}} open-source
  {{Python}} framework for the dynamics of open quantum systems},\ }\href
  {https://doi.org/10.1016/j.cpc.2012.02.021} {\bibfield  {journal} {\bibinfo
  {journal} {Computer Physics Communications}\ }\textbf {\bibinfo {volume}
  {183}},\ \bibinfo {pages} {1760} (\bibinfo {year} {2012})}\BibitemShut
  {NoStop}%
\bibitem [{\citenamefont {Cao}\ \emph {et~al.}(2022)\citenamefont {Cao},
  \citenamefont {Lin}, \citenamefont {Kribs}, \citenamefont {Poon},
  \citenamefont {Zeng},\ and\ \citenamefont {Laflamme}}]{cao2021nisq}%
  \BibitemOpen
  \bibfield  {author} {\bibinfo {author} {\bibfnamefont {N.}~\bibnamefont
  {Cao}}, \bibinfo {author} {\bibfnamefont {J.}~\bibnamefont {Lin}}, \bibinfo
  {author} {\bibfnamefont {D.}~\bibnamefont {Kribs}}, \bibinfo {author}
  {\bibfnamefont {Y.-T.}\ \bibnamefont {Poon}}, \bibinfo {author}
  {\bibfnamefont {B.}~\bibnamefont {Zeng}},\ and\ \bibinfo {author}
  {\bibfnamefont {R.}~\bibnamefont {Laflamme}},\ }\href
  {https://arxiv.org/abs/2111.02345} {\bibinfo {title} {Nisq: Error correction,
  mitigation, and noise simulation}} (\bibinfo {year} {2022}),\ \Eprint
  {https://arxiv.org/abs/2111.02345} {arXiv:2111.02345 [quant-ph]} \BibitemShut
  {NoStop}%
\bibitem [{\citenamefont {Gordon}\ \emph {et~al.}(2022)\citenamefont {Gordon},
  \citenamefont {Murray}, \citenamefont {Kurter}, \citenamefont {Sandberg},
  \citenamefont {Hall}, \citenamefont {Balakrishnan}, \citenamefont {Shelby},
  \citenamefont {Wacaser}, \citenamefont {Stabile}, \citenamefont {Sleight},
  \citenamefont {Brink}, \citenamefont {Rothwell}, \citenamefont {Rodbell},
  \citenamefont {Dial},\ and\ \citenamefont
  {Steffen}}]{gordon2022environmental}%
  \BibitemOpen
  \bibfield  {author} {\bibinfo {author} {\bibfnamefont {R.~T.}\ \bibnamefont
  {Gordon}}, \bibinfo {author} {\bibfnamefont {C.~E.}\ \bibnamefont {Murray}},
  \bibinfo {author} {\bibfnamefont {C.}~\bibnamefont {Kurter}}, \bibinfo
  {author} {\bibfnamefont {M.}~\bibnamefont {Sandberg}}, \bibinfo {author}
  {\bibfnamefont {S.~A.}\ \bibnamefont {Hall}}, \bibinfo {author}
  {\bibfnamefont {K.}~\bibnamefont {Balakrishnan}}, \bibinfo {author}
  {\bibfnamefont {R.}~\bibnamefont {Shelby}}, \bibinfo {author} {\bibfnamefont
  {B.}~\bibnamefont {Wacaser}}, \bibinfo {author} {\bibfnamefont {A.~A.}\
  \bibnamefont {Stabile}}, \bibinfo {author} {\bibfnamefont {J.~W.}\
  \bibnamefont {Sleight}}, \bibinfo {author} {\bibfnamefont {M.}~\bibnamefont
  {Brink}}, \bibinfo {author} {\bibfnamefont {M.~B.}\ \bibnamefont {Rothwell}},
  \bibinfo {author} {\bibfnamefont {K.~P.}\ \bibnamefont {Rodbell}}, \bibinfo
  {author} {\bibfnamefont {O.}~\bibnamefont {Dial}},\ and\ \bibinfo {author}
  {\bibfnamefont {M.}~\bibnamefont {Steffen}},\ }\bibfield  {title} {\bibinfo
  {title} {{Environmental radiation impact on lifetimes and quasiparticle
  tunneling rates of fixed-frequency transmon qubits}},\ }\href
  {https://doi.org/10.1063/5.0078785} {\bibfield  {journal} {\bibinfo
  {journal} {Applied Physics Letters}\ }\textbf {\bibinfo {volume} {120}},\
  \bibinfo {pages} {074002} (\bibinfo {year} {2022})}\BibitemShut {NoStop}%
\bibitem [{\citenamefont {Rist{\`e}}\ \emph {et~al.}(2013)\citenamefont
  {Rist{\`e}}, \citenamefont {Bultink}, \citenamefont {Tiggelman},
  \citenamefont {Schouten}, \citenamefont {Lehnert},\ and\ \citenamefont
  {DiCarlo}}]{riste2013millisecond}%
  \BibitemOpen
  \bibfield  {author} {\bibinfo {author} {\bibfnamefont {D.}~\bibnamefont
  {Rist{\`e}}}, \bibinfo {author} {\bibfnamefont {C.~C.}\ \bibnamefont
  {Bultink}}, \bibinfo {author} {\bibfnamefont {M.~J.}\ \bibnamefont
  {Tiggelman}}, \bibinfo {author} {\bibfnamefont {R.~N.}\ \bibnamefont
  {Schouten}}, \bibinfo {author} {\bibfnamefont {K.~W.}\ \bibnamefont
  {Lehnert}},\ and\ \bibinfo {author} {\bibfnamefont {L.}~\bibnamefont
  {DiCarlo}},\ }\bibfield  {title} {\bibinfo {title} {Millisecond charge-parity
  fluctuations and induced decoherence in a superconducting transmon qubit},\
  }\href {https://doi.org/10.1038/ncomms2936} {\bibfield  {journal} {\bibinfo
  {journal} {Nature Communications}\ }\textbf {\bibinfo {volume} {4}},\
  \bibinfo {pages} {1913} (\bibinfo {year} {2013})}\BibitemShut {NoStop}%
\bibitem [{\citenamefont {Khezri}\ \emph {et~al.}(2015)\citenamefont {Khezri},
  \citenamefont {Dressel},\ and\ \citenamefont {Korotkov}}]{khezri2015qubit}%
  \BibitemOpen
  \bibfield  {author} {\bibinfo {author} {\bibfnamefont {M.}~\bibnamefont
  {Khezri}}, \bibinfo {author} {\bibfnamefont {J.}~\bibnamefont {Dressel}},\
  and\ \bibinfo {author} {\bibfnamefont {A.~N.}\ \bibnamefont {Korotkov}},\
  }\bibfield  {title} {\bibinfo {title} {Qubit measurement error from coupling
  with a detuned neighbor in circuit qed},\ }\href
  {https://doi.org/10.1103/PhysRevA.92.052306} {\bibfield  {journal} {\bibinfo
  {journal} {Phys. Rev. A}\ }\textbf {\bibinfo {volume} {92}},\ \bibinfo
  {pages} {052306} (\bibinfo {year} {2015})}\BibitemShut {NoStop}%
\bibitem [{\citenamefont {Nation}\ and\ \citenamefont
  {Treinish}(2023)}]{PRXQuantum.4.010327}%
  \BibitemOpen
  \bibfield  {author} {\bibinfo {author} {\bibfnamefont {P.~D.}\ \bibnamefont
  {Nation}}\ and\ \bibinfo {author} {\bibfnamefont {M.}~\bibnamefont
  {Treinish}},\ }\bibfield  {title} {\bibinfo {title} {Suppressing quantum
  circuit errors due to system variability},\ }\href
  {https://doi.org/10.1103/PRXQuantum.4.010327} {\bibfield  {journal} {\bibinfo
   {journal} {PRX Quantum}\ }\textbf {\bibinfo {volume} {4}},\ \bibinfo {pages}
  {010327} (\bibinfo {year} {2023})}\BibitemShut {NoStop}%
\end{thebibliography}%
%%%%%%%%%%%%%%%%%%%%%%%%%%%%%%%%%%%%%%%%%%%%%%
%%%%%%%%%%%%%%%%%%%%%%%%%%%%%%%%%%%%%%%%%%%%%%
%%%%%%%%%%%%%%%%%%%%%%%%%%%%%%%%%%%%%%%%%%%%%%
%%%%%%%%%%%%%%%%%%%%%%%%%%%%%%%%%%%%%%%%%%%%%%
%%%%%%%%%%%%%%%%%%%%%%%%%%%%%%%%%%%%%%%%%%%%%%
\clearpage

\appendix

\begin{center}
	\textbf{Appendix}
\end{center} 
\begin{figure*}[b]
    \begin{centering}     
\includegraphics[width=0.95\textwidth]{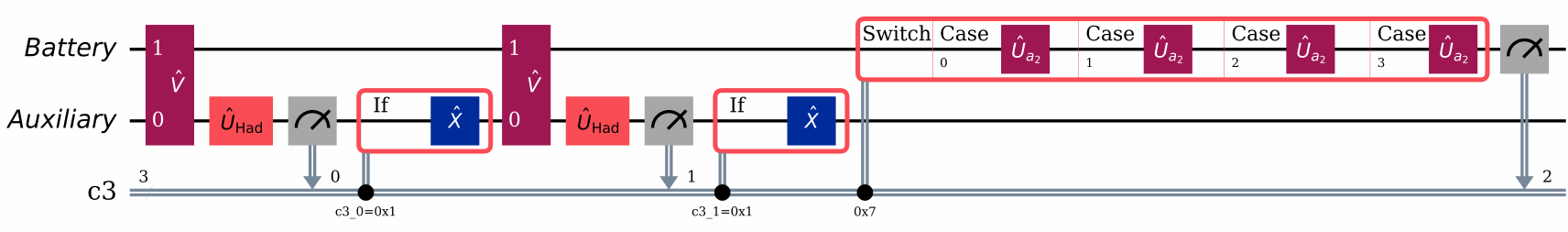}
\put(-500,60){\small (a)}  
\\
\includegraphics[width=0.95\textwidth]{qiskit_drawing_transpiled_1step.pdf}    
\put(-500,60){\small (b)}
    \end{centering}
    \caption{(a) A \texttt{qiskit} drawing representing the circuit that is effectively run on the IBM quantum computers to simulate a $n=2$ steps CMCM and the work extraction feedback protocol. (b) A \texttt{Transpiled} version of the former circuit for $n = 1$ in basis gates executable on IBM devices.}
    \label{f:qiskitdrawing}
\end{figure*}
This appendix is organized as follows: in Appendix~\ref{a:IBMqdetails} we present some further details on the implementation of the CMCM on the IBM quantum computers, along with their noise characterization, while in Appendix~\ref{a:additionalresults} we show some additional plots results corresponding to other experimental runs of the same CMCM.
\section{Details on the implementation of the CMCM on the IBM quantum computer}
\label{a:IBMqdetails}
To execute our protocol on real hardware, we utilized four 127-qubit backends with free access, which are based on Eagle processors, \texttt{ibm\_kyoto} and \texttt{ibm\_brisbane}. As we described in the main text, the simulation of the CMCM requires only two qubits, one designated as the QB qubit, while the other serves as the AS qubit, which can be reset after each step of the CM. In this regard, while the Qiskit platform offers a reset functionality, we avoid its use due to its time-consuming nature; instead, we employ a feed-forward operation that depends on the result obtained from measurement on the AS qubit: a $X$-gate is performed any time one obtains the result $1$, while no operation is performed when the outcome is $0$. In fact the reset functionality that can be implemented by Qiskit would actually perform the same kind of protocol, by performing a measurement on the qubit and a conditional operation depending on the outcome, as described above. A drawing of the \texttt{qiskit} circuit that is effectively run on the device for simulating $n=2$ steps and the corresponding feedback operation is pictured in Fig.~\ref{f:qiskitdrawing}, along with the corresponding transpiled version for $n=1$ steps.

The selection of the two qubits necessary to simulate the CMCM is optimized for each experiment by using the Mapomatic Python package\cite{PRXQuantum.4.010327} in order to reduce the noise acting on the quantum system.
Nevertheless, as we discussed in the main text, noise is still playing a non-negligible role in the circuit designed to simulate the work extraction protocol. For this reason in Sec.~\ref{s:noisyCMCM} we have introduced a noisy model able to better describe the CMCM implemented on the IBM quantum computers; in particular we have shown how the values of decoherence times $T_1$ and $T_2$ characterizing the QB qubit, along with the readout time $T_r$ of the AS qubit can be used to retrieve the probabilities
\begin{align}
P_{AD} &= 1 - e^{-T_r/T_1}  \, \nonumber \\
P_D &= 1 - e^{-T_r/T_2} \,, \nonumber
\end{align}
that identify respectively the amplitude damping channel $\mathcal{N}_{AD}$ and the dephasing channel $\mathcal{N}_{D}$ that at each step acts on QB qubit.

In order to model the noise acting also on the measurement apparatus implementing the $\sigma_z$ measurement on the AS qubit, we have also introduced the noisy POVM operators
\begin{align}
\tilde{\pi}_{0} = P_{00}\ketbra{0}{0} + P_{01}\ketbra{1}{1}
\\
\tilde{\pi}_{1} = P_{10}\ketbra{0}{0} + P_{11}\ketbra{1}{1}
\end{align}
where the parameters $P_{ab}$ quantify the probability of obtaining the outcome $\{a\}$ when one has prepared the $\hat{\sigma}_z$ eigenstate $|b\rangle$. These values are available within the IBM calibration data, which is updated every 30 minutes. However, to make our implementations more precise, we experimentally estimated these values using simple ad-hoc calibration circuits, involving only the AS qubit, before running the work extraction protocol. Because of the normalization condition $P_{0a} + P_{1a}=1$, we actually need to evaluate just two of these probabilities: $P_{10}$ can be evaluated as the ratio $N_1/N_{\sf tot}$ between the number of times one obtains the outcome $1$ and the total number of measurements $N_{\sf tot}$, when the AS qubit has been prepared in the state $|0\rangle$; similarly $P_{01}$ has been evaluated as the ratio $N_0/N_{\sf tot}$, where in this case the AS qubit has been initially prepared in the state $|1\rangle$ by acting with a $X$-gate on the reinialized qubit, before performing the measurement. Both probabilities have been obtained by repeating these two simple calibration circuits $N_{\sf tot} = 10^{4}$ times.

In Table~\ref{table}, we resume all the details of the experiments that correspond to the figures we have presented in this paper (both main text and next appendix). 
\begin{widetext}
\begin{table*}[h!]
\centering
\begin{tabular}{c|cccc|ccc|cccc}
%\hline
{\bf Figure} & {\bf IBM backend} & {\bf QB index} & {\bf AS index} & $N_{\sf exp}$ & $T_1$ & $T_2$ & $T_r$ & $P_{AD}$ & $P_D$ & $P_{01}$ & $P_{10}$ \\
\hline\hline
Fig.~\ref{f:results2}(a) and Fig.~\ref{f:additionalresults1}(a) & \texttt{ibm\_kyoto} & 13 & 12 & $10^4$ & $342.13 \mu s$ & $326.55 \mu s$ & $1440 \,ns$ & $0.0042$ & $0.0044$ & $0.0061$ & $0.0070$ \\
Fig.~\ref{f:results2}(b) and Fig.~\ref{f:additionalresults2}(a) & \texttt{ibm\_kyoto} & 66 & 73 & $10^4$ &  $334.16 \mu s$ & $202.09 \mu s$ & $1440\,ns$ & $0.0043$ & $0.0071$ & $0.0062$ & $0.0066$ \\
Fig.~\ref{f:additionalresults1}(b) & \texttt{ibm\_brisbane} & 12 & 13 & $10^4$ &  $303.33\mu s$ & $182.54\mu s$ & $1300\,ns$ & $0.0042$ & $0.0070$ & $0.0606$ & $0.0120$ \\
Fig.~\ref{f:additionalresults1}(c) & \texttt{ibm\_brisbane} & 12 & 13 & $10^4$ &  $303.33 \mu s$ & $182.54 \mu s$ & $1300\,ns$ & $0.0042$ & $0.0070$ & $0.0616$ & $0.0132$ \\
Fig.~\ref{f:additionalresults1}(d) & \texttt{ibm\_brisbane} & 12 & 13 & $10^4$ &  $303.33\,\mu s$ & $182.54\,\mu s$ & $1300\,ns$ & $0.0042$ & $0.0070$ & $0.0272$ & $0.0108$ \\
Fig.~\ref{f:additionalresults2}(b) & \texttt{ibm\_brisbane} & 12 & 13 & $10^4$ &  $295.36 \mu s$ & $164.04 \mu s$ & $1300\,ns$ & $0.0043$ & $0.0078$ & $0.0256$ & $0.0100$ \\
Fig.~\ref{f:additionalresults2}(c) & \texttt{ibm\_brisbane} & 12 & 13 & $10^4$ &  $303.33 \mu s$ & $182.54 \mu s$ & $1300\,ns$ & $0.0042$ & $0.0070$ & $0.0204$ & $0.0102$ \\
Fig.~\ref{f:additionalresults2}(d) & \texttt{ibm\_brisbane} & 12 & 13 & $10^4$ &  $295.36 \mu s$ & $164.04 \mu s$ & $1300\,ns$ & $0.0043$ & $0.0078$ & $0.0210$ & $0.0118$ \\

%\multicolumn{1}{c}{}& \multicolumn{4}{c}{Qubits Characteristics} & %\multicolumn{4}{c}{\textbf{Noise Factors}} \\ 
%\hline
%\textbf{Methods} & {Index} &{T1}&{T2}&{Readout Time}&{\( P_{AD} \)} & {\( P_{DP} \)} & {\( P_{01} \)} & {\( P_{10} \)} \\
\hline
%Physical Unitary & 0.0044& 0.9641&0.0080& 0.0051\\
%Ry+CNOT & 0.0085& 0.9452 & 0.0111 & 0.0043  \\
\hline
\end{tabular}
\caption{We resume the details regarding the experimental implementation of the daemonic work extraction protocols whose results are reported in the main text and in the Appendix. In particular for each figure we report on which IBM backend the experiment was implemented, the QB qubit and AS qubit indices, the number of experimental runs, the decoherence times $T_1$ and $T_2$ for the QB qubit and the readout time $T_r$ for the AS qubit, along with the corresponding values of the amplitude damping and dephasing probabilities $P_{AD}$ and $P_D$, and the probabilities $P_{01}$ and $P{10}$ characterizing the noisy measurement. 
\label{table}}
\end{table*}
\end{widetext}

\section{Additional results}
\label{a:additionalresults}
\begin{figure*}[ht]
\begin{center}
\includegraphics[width=0.45\textwidth]{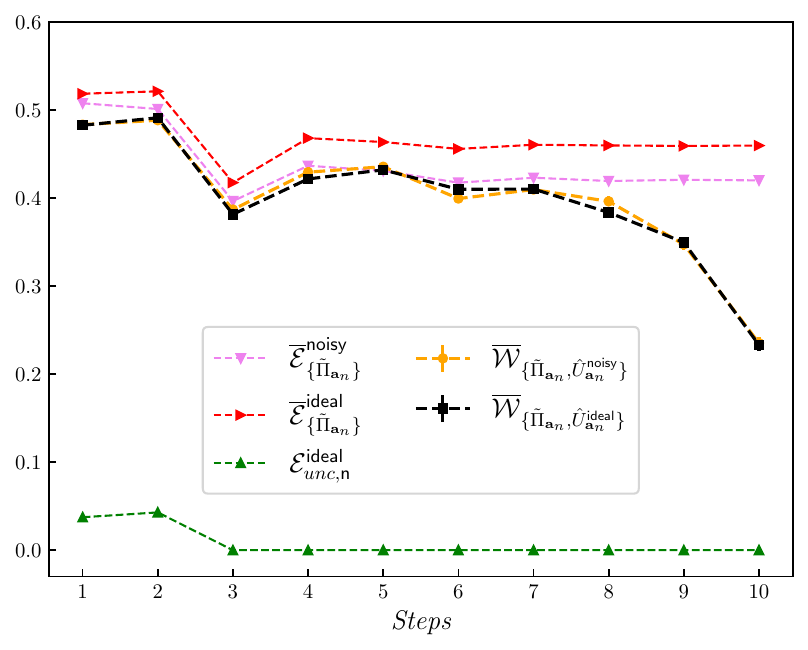}
 \put(-115,185){\small (a)}
\includegraphics[width=0.45\textwidth]{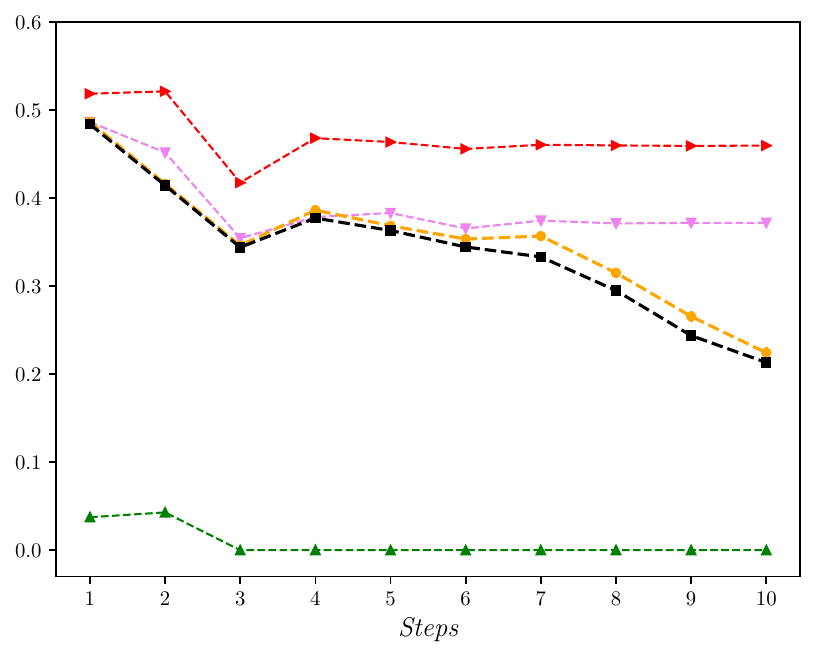} 
 \put(-115,185){\small (b)}
 \\
\includegraphics[width=0.45\textwidth]{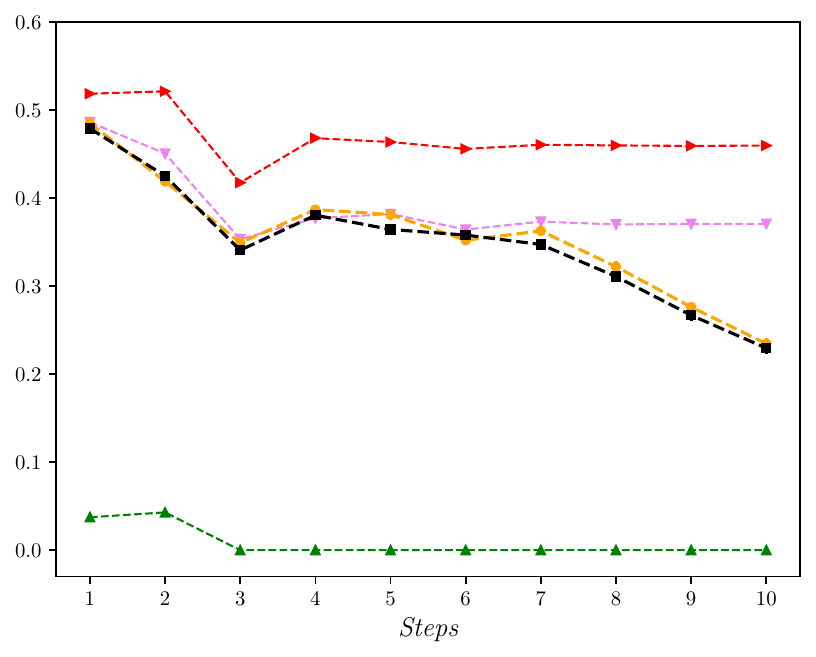}
 \put(-115,185){\small (c)}
\includegraphics[width=0.45\textwidth]{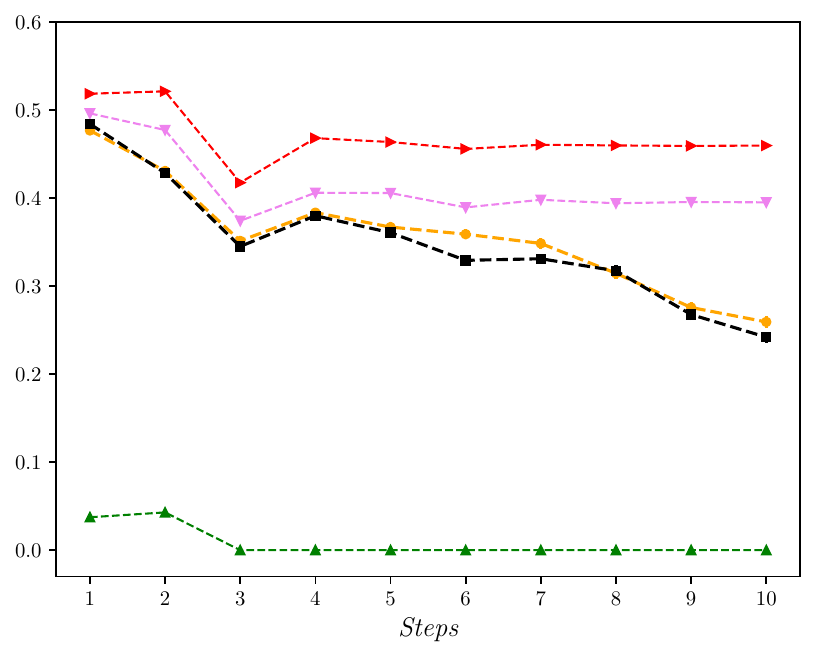}
 \put(-115,185){\small (d)}
\end{center}
\caption{We report additional plots for $\kappa=\alpha=1$, analogous to the one presented in Fig.~\ref{f:results2}, which is for completeness reported in panel (a). We refer to the caption of the same figure in the main text for the legend of the different curves. Other details about the physical parameters characterizing the experimental implementation on IBM quantum computers can be found in Table~\ref{table}.}
\label{f:additionalresults1}
\end{figure*}

\begin{figure*}[ht]
\begin{center}
\includegraphics[width=0.45\textwidth]{compare_Physical_noise_kappa_2_nolegend.pdf} 
 \put(-115,185){\small (a)}
\includegraphics[width=0.45\textwidth]{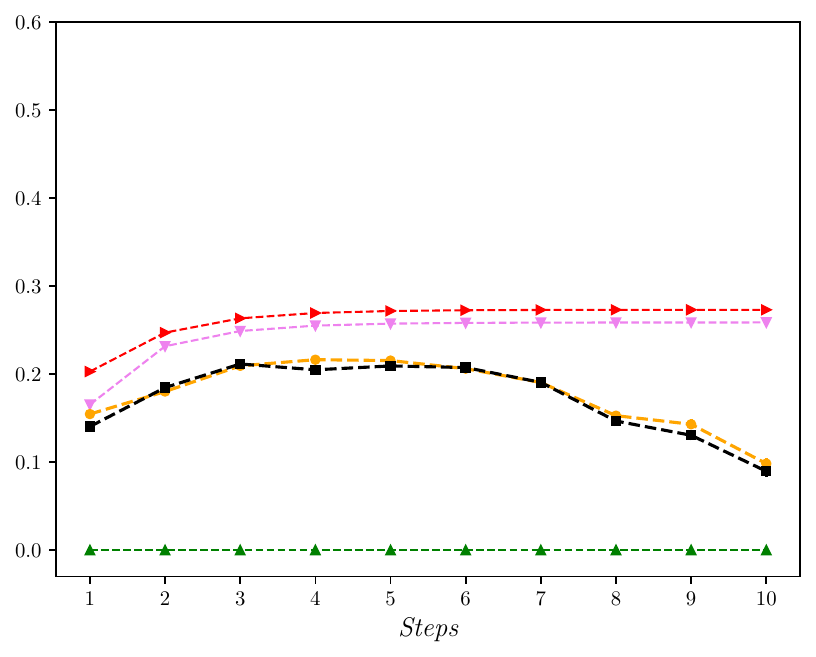} 
 \put(-115,185){\small (b)}
 \\
\includegraphics[width=0.45\textwidth]{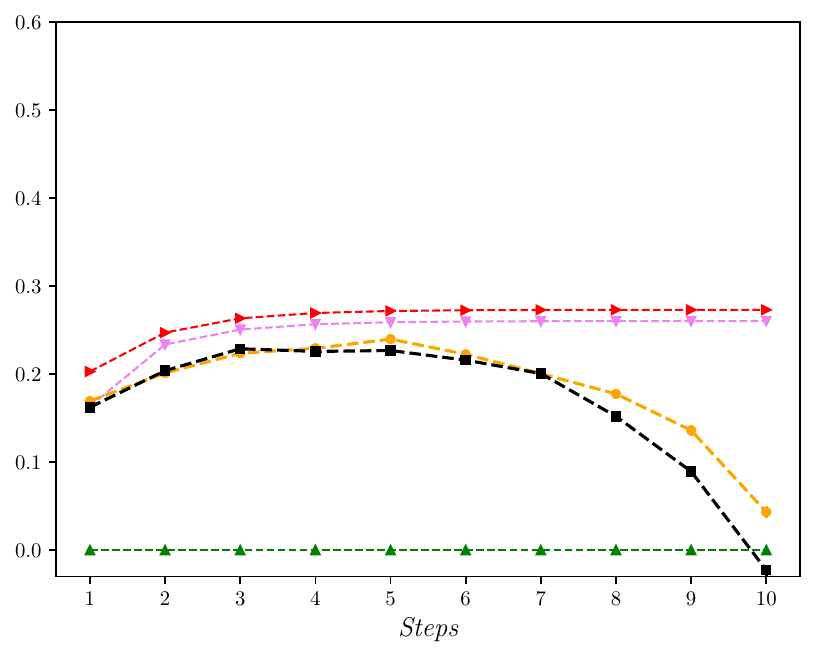}
 \put(-115,185){\small (c)}
\includegraphics[width=0.45\textwidth]{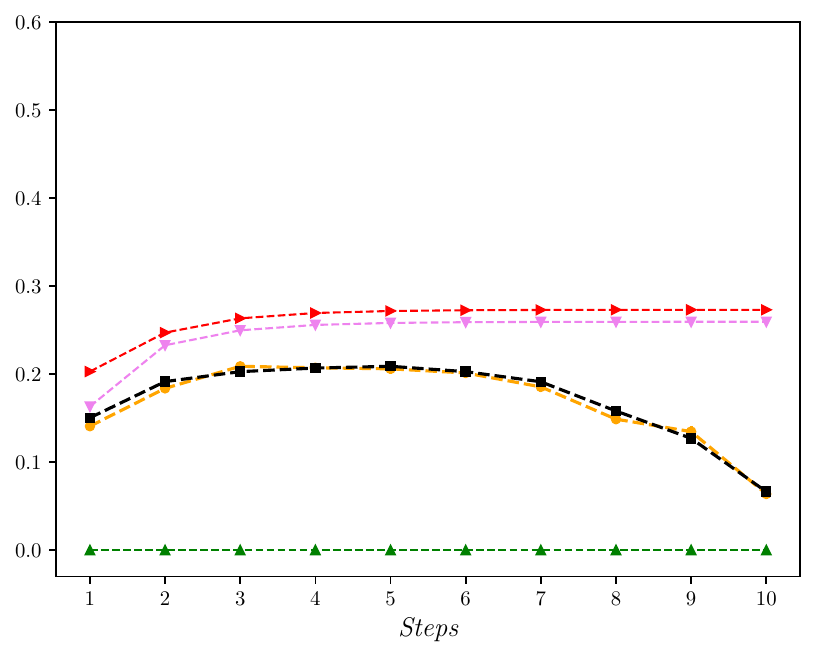} 
 \put(-115,185){\small (d)}
\end{center}
\caption{
We report additional plots for $\kappa=2\alpha=2$, analogous to the one presented in Fig.~\ref{f:results2}, which is for completeness reported in panel (a). We refer to the caption of the same figure in the main text for the legend of the different curves. Other details about the physical parameters characterizing the experimental implementation on IBM quantum computers can be found in Table~\ref{table}.}
\label{f:additionalresults2}
\end{figure*}

In this appendix we present the results of other experiments analogous to the ones presented in Figs.~\ref{f:results2} (for the sake of helping the reader, we report again the plots of Fig.~\ref{f:results2} in the top panel). These are reported in Fig.~\ref{f:additionalresults1} for the experiments obtained by fixing $\kappa=\alpha=1$, while in Fig.~\ref{f:additionalresults2} for the experiments obtained by fixing $\kappa=2\alpha=2$.

We observe how the results obtained for $\kappa=1$ are more consistent with each other: in all these examples the daemonic extracted work approaches the theoretical daemonic ergotropy evaluated according to the noisy model, and  the enhancement due to the work extraction unitaries optimized according to the noisy model is not really evident. On the other hand we observe that for $\kappa=2$ the results may differ a lot from one experiment to the other: in particular we observe how in several runs of the experiment the daemonic extracted work is not really close to the theoretical (noisy) daemonic ergotropy, and in particular in these cases it is also hard to observe a evident enhancement if one exploits the new optimized work extraction unitary. On the other hand, when, as in Fig.~\ref{f:results2} the noisy optimized daemonic extracted work approaches the corresponding theoretical ergotropy, the improvement with respect to the results obtained by employing the ideal noiseless model is clearly evident. This seems to indicate that when one implements the interaction gates $\hat{V}_j$ for $\kappa=2$, thus requiring a stronger interaction between the two qubits, additional noisy fluctuations play a non-negligible role in the different runs of the experiment that have not been taken into account in our simple noisy model. 
\end{document}